{}%disable hyper at arxiv
{}%use pdflatex at arxiv

\newif\ifpublic\publictrue
\newif\ifarxiv\arxivtrue

\documentclass[12pt,a4paper]{article}

\usepackage{amsmath,amssymb}
\ifpublic\else\usepackage{showkeys}\fi
\usepackage{graphicx}
\usepackage[bookmarks=true,hyperfigures=true]{hyperref}
\usepackage[usenames, dvipsnames]{color}
\usepackage[nosort]{cite}
\usepackage[bulletsep]{collref}
\usepackage[utf8]{inputenc}

\def\showkeysrefformat#1{{\normalfont\tiny\ttfamily#1}}
\makeatletter
\def\SK@@ref#1>#2\SK@{%
 {\@inlabelfalse\leavevmode\vbox to\z@{%
 \vss\SK@refcolor\rlap{\vrule\raise .75em%
  \hbox{\showkeysrefformat{#2}}}}}}
\makeatother

\usepackage[a4paper,text={160mm,247mm},centering]{geometry}
\RequirePackage{verbatim}

\makeatletter
\newwrite\bibinl@out
\newenvironment{bibtex}[1][\jobname]{%
  \immediate\openout\bibinl@out #1.bib
  \immediate\write\bibinl@out{\@percentchar generated from `\jobname' starting line \the\inputlineno^^J}%
  \def\verbatim@processline{\immediate\write\bibinl@out{\the\verbatim@line}}%
  \@bsphack\let\do\@makeother\dospecials\catcode`\^^M\active\verbatim@start
}%
{\immediate\closeout\bibinl@out\@esphack}
\makeatother

\allowdisplaybreaks[3]

\numberwithin{equation}{section}

\usepackage[font=small,labelfont=bf,width=0.85\textwidth]{caption}

\newcommand{\alignrel}[1][=]{\mathrel{\phantom{#1}}{}}
\makeatletter
\newcommand{\eqsep}{\@ifstar{\hspace{1em}}{\hspace{2em minus 1em}}}
\newcommand{\@eqjoin}[2]{\hspace{#1}#2\hspace{#1}}
\newcommand{\eqjoin}{\@ifstar{\@eqjoin{1em}}{\@eqjoin{2em minus 1em}}}
\makeatother

\renewcommand{\minalignsep}{2em}

\makeatletter
\expandafter\def\expandafter\calc@shift@align\expandafter
  {\expandafter\hook@align@sep\calc@shift@align}
\def\hook@align@sep{\ifnum\xatlevel@=\@ne
  \dimen@\displaywidth
  \advance\dimen@-\totwidth@
  \@tempcntb\maxfields@
  \divide\@tempcntb\tw@
  \@tempcnta\@tempcntb
  \advance\@tempcntb\m@ne
  \global\eqnshift@\dimen@
  \global\alignsep@\minalignsep\relax
  \global\advance\eqnshift@-\@tempcntb\alignsep@
  \global\divide\eqnshift@\tw@
  \ifdim\eqnshift@<\z@
   \global\eqnshift@\z@
  \fi
 \fi}
\makeatother

\makeatletter
\expandafter\def\expandafter\align\expandafter
  {\expandafter\hook@align@\align}
\expandafter\def\expandafter\math@cr@@@align\expandafter
  {\expandafter\hook@align@cr\math@cr@@@align}
\expandafter\def\expandafter\endalign\expandafter
  {\expandafter\hook@align@end\endalign}
\expandafter\def\expandafter\measure@\expandafter#\expandafter1\expandafter
  {\measure@{#1}\hook@align@}
\def\hook@align@{\gdef\hook@align@end{\donumber}%
  \gdef\hook@align@cr{\nonumber}}
\def\hook@align@cr{}
\def\hook@align@end{}

\newcommand{\donumber}{\gdef\hook@align@cr{\gdef\hook@align@cr{\nonumber}}}%
\newcommand{\numberhere}{\donumber\gdef\hook@align@end{}}%
\makeatother

\ifarxiv\PassOptionsToPackage{write=false,compile=false}{mpostinl}\fi

\RequirePackage{graphbox}
\RequirePackage[clean=false]{mpostinl}

\begin{mposttex}
\usepackage{amsmath,amssymb}
\usepackage[utf8]{inputenc}
\usepackage[usenames, dvipsnames]{color}
\end{mposttex}

\begin{mpostdef}
def pensize(expr s)=withpen pencircle scaled s enddef;
def fillshape(expr p,c)=
  fill p withcolor c;
  draw p
enddef;
def drawdot(expr z,s)=
  fill fullcircle scaled s shifted z
enddef;
def filldot(expr z,s,ci)=
  fillshape(fullcircle scaled s shifted z, ci)
enddef;
def drawcross(expr z,s,r,t,c)=
  draw ((-0.5,-0.5)--(+0.5,+0.5)) scaled s rotated r shifted z pensize(t) withcolor c;
  draw ((+0.5,-0.5)--(-0.5,+0.5)) scaled s rotated r shifted z pensize(t) withcolor c;
enddef;
\end{mpostdef}

\begin{mpostdef}
pair vpos[];
numeric nums[];
path paths[];
newinternal numeric xu;
xu:=1cm;
ahlength:=4pt;
\end{mpostdef}

\begin{mpostdef}
def midarrowperc (expr p, t) =
  fill arrowhead subpath(0,arctime(t*arclength(p)+0.5ahlength) of p) of p;
enddef;
def midarrow (expr p, t) =
  fill arrowhead subpath(0,arctime(arclength(subpath (0,t) of p)+0.5ahlength) of p) of p
enddef;
\end{mpostdef}

\begin{mposttex}[dual]
\usepackage{mathfixs}
\ProvideMathFix{autobold,greekcaps,frac,root}
\ProvideMathFix{multskip}
\end{mposttex}

\begin{mposttex}[dual]
\RequirePackage[extdef]{delimset}
\newcommand{\state}{\ket}

\end{mposttex}

\begin{mposttex}[dual]
\ProvideMathFix{vfrac,rfrac}
\newcommand{\iunit}{\mathring{\imath}}
\newcommand{\eunit}{\mathrm{e}}
\newcommand{\half}{\rfrac{1}{2}}
\newcommand{\ihalf}{\rfrac{\iunit}{2}}
\newcommand{\quarter}{\rfrac{1}{4}}
\end{mposttex}

\begin{mposttex}[dual]
\newcommand{\Real}{\mathbb{R}}
\newcommand{\Complex}{\mathbb{C}}
\newcommand{\Integer}{\mathbb{Z}}

\end{mposttex}

\newcommand{\vdot}{\mathord{\cdot}}

\newcommand{\vecsp}[1]{\vec{\mkern0mu#1\mkern1mu}}
\makeatletter
\def\vect{\@ifstar{\vecsp}{\vec}}
\makeatother

\begin{mposttex}[dual]
\newcommand{\alg}[1]{\mathfrak{#1}}
\newcommand{\grp}[1]{\mathrm{#1}}
\newcommand{\gen}[1]{\mathrm{#1}{}}

\newcommand{\envalg}{\grp{U}}

\newcommand{\genJ}{\gen{J}}
\newcommand{\genL}{\gen{L}}
\newcommand{\genM}{\gen{M}}
\newcommand{\genP}{\gen{P}}
\newcommand{\genRL}{\genL}
\newcommand{\genRH}{\gen{H}}
\newcommand{\genRP}{\genP}
\newcommand{\genQ}{\gen{Q}}

\newcommand{\permop}{\mathcal{P}}

\newcommand{\cybe}{\delimpair{[[}{[.],}{]]}} % not correct!
\newcommand{\liebr}{\comm}
\newcommand{\cobra}{\delta}

\newcommand{\dgen}{D\brk}

\newcommand{\casJJ}{\genJ^2}
\newcommand{\casLL}{\genL^2}
\newcommand{\casMM}{\genM^2}
\newcommand{\casMMa}{\genM_+^2}
\newcommand{\casMMp}{\genM_-^2}
\newcommand{\casPP}{\genP^2}
\newcommand{\casLP}{\genL\vdot\genP}
\newcommand{\casQQ}{\genQ^2}
\end{mposttex}

\begin{mposttex}[dual]
\newcommand{\mref}{{\hlcolor{red}\bar m}}
\newcommand{\rnorm}{{\hlcolor{red}\nu}}
\newcommand{\rtwist}{{\hlcolor{red}\xi}}
\newcommand{\bfnorm}{{\hlcolor{red}\zeta}}
\newcommand{\adsmass}{{\hlcolor{blue}\mu}}
\newcommand{\adsangmom}{{\hlcolor{blue}\kappa}}
\newcommand{\adseng}{{\hlcolor{blue}\omega}}
\newcommand{\adsbfnorm}{{\hlcolor{blue}\tilde\zeta}}
\newcommand{\redang}{{\hlcolor{red}\alpha}}
\newcommand{\redpar}{{\hlcolor{red}\beta}}
\newcommand{\redtmod}{{\hlcolor{red}h}}
\newcommand{\reppar}{{\hlcolor[rgb]{0,0.5,0}\gamma}}
\newcommand{\repang}{{\hlcolor[rgb]{0,0.5,0}\chi}}
\newcommand{\Sphere}{{\hlcolor{red}\mathrm{S}}}
\newcommand{\AdS}{{\hlcolor{red}\mathrm{AdS}}}
\end{mposttex}

\newcommand{\der}{\mathrm{d}}

\newcommand{\Order}{\mathcal{O}}
\newcommand{\spinup}{\uparrow}
\newcommand{\spindown}{\downarrow}
\newcommand{\algmid}{\mathrm{M}}
\newcommand{\algleft}{\mathrm{L}}
\newcommand{\algright}{\mathrm{R}}
\newcommand{\statebos}{\algright}
\newcommand{\stateferm}{\algleft}
\newcommand{\slpauli}{\tilde\sigma}
\newcommand{\repsinglet}{\mathbf{1}}
\newcommand{\repdoublet}{\mathbf{2}}

\def\[#1\]{\begin{equation}#1\end{equation}}

\providecommand{\href}[2]{#2}
\newcommand{\arxivlink}[1]{\href{http://arxiv.org/abs/#1}{arxiv:#1}}

\makeatletter
\def\mr@ignsp#1 {\ifx\:#1\@empty\else #1\expandafter\mr@ignsp\fi}%
\newcommand{\multiref}[1]{\begingroup%\let\protect\string%
\xdef\mr@no@sparg{\expandafter\mr@ignsp#1 \: }%
\def\mr@comma{}%
\@for\mr@refs:=\mr@no@sparg\do{\mr@comma\def\mr@comma{,}\ref{\mr@refs}}%
\endgroup}
\renewcommand{\eqref}[1]{(\multiref{#1})}
\makeatother

\makeatletter
\newcommand{\namedref}[2]{\hyperref[#2]{#1~\ref*{#2}}}
\newcommand{\secref}{\@ifstar{\namedref{Section}}{\namedref{Sec.}}}
\newcommand{\appref}{\@ifstar{\namedref{Appendix}}{\namedref{App.}}}
\newcommand{\tabref}{\@ifstar{\namedref{Table}}{\namedref{Tab.}}}
\newcommand{\figref}{\@ifstar{\namedref{Figure}}{\namedref{Fig.}}}
\makeatother

\let\oldbib=\thebibliography
\def\thebibliography{\phantomsection\addcontentsline{toc}{section}{\refname}\oldbib}

\let\oldtoc=\tableofcontents
\def\tableofcontents{\phantomsection\addcontentsline{toc}{section}{\contentsname}\oldtoc}

\providecommand{\hypersetup}[1]{}
\providecommand{\texorpdfstring}[2]{#1}

\hypersetup{plainpages=false}
\hypersetup{pdfpagemode=UseNone}
\hypersetup{bookmarksnumbered=true}
\hypersetup{pdfstartview=FitH}
\hypersetup{colorlinks=false}
\hypersetup{citebordercolor={.5 1 .5}}
\hypersetup{urlbordercolor={.5 1 1}}
\hypersetup{linkbordercolor={1 .7 .7}}
\usepackage{metastr}

\usepackage{ifpdf}
\ifpdf\else\RequirePackage[active]{srcltx}\fi
\RequirePackage{color}

\newcommand{\hlcolor}{\color}
\renewcommand{\hlcolor}[2][]{}
\ifpublic\renewcommand{\hlcolor}[2][]{}\fi
\begin{mposttex}
\newcommand{\hlcolor}[2][]{}
\end{mposttex}

\newcommand{\remark}[2][]{{\normalfont\sffamily\hspace{1ex}%
  \def\emph{\textsl}\def\textbullet{$\bullet$}
  \def\tmparga{#1}%
  \def\tmpargb{NB}\ifx\tmparga\tmpargb\color[rgb]{0,0,0.8}\fi%
  \def\tmpargb{EI}\ifx\tmparga\tmpargb\color[rgb]{0,0.5,0}\fi%
  \def\tmpargb{}\ifx\tmparga\tmpargb\color{red}\fi%
  \def\tmpargb{}\ifx\tmparga\tmpargb\else \textbf{#1:} \fi%
  #2\hspace{1ex}}}

\ifpublic\renewcommand{\remark}[2][]{}\fi

\metaset{title}{Classical Lie Bialgebras\texorpdfstring{\\}{ }%
for AdS/CFT Integrability\texorpdfstring{\\}{ }%
by Contraction and Reduction}
\metaset{author}{Niklas Beisert, Egor Im}

\begin{document}

\pdfbookmark[1]{Title Page}{title}
\thispagestyle{empty}

\begingroup\raggedleft\footnotesize\ttfamily
\arxivlink{2210.11150}
\par\endgroup

\vspace*{2cm}
\begin{center}%
\begingroup\Large\bfseries\metapick[print]{title}\par\endgroup
\vspace{1cm}

\begingroup\scshape
\metapick[print]{author}
\endgroup
\vspace{5mm}

\textit{Institut für Theoretische Physik,\\
Eidgenössische Technische Hochschule Zürich,\\
Wolfgang-Pauli-Strasse 27, 8093 Zürich, Switzerland}
\vspace{0.1cm}

\begingroup\ttfamily\small
\verb+{+nbeisert,egorim\verb+}+@itp.phys.ethz.ch\par
\endgroup
\vspace{5mm}

\vfill

\textbf{Abstract}\vspace{5mm}

\begin{minipage}{12.7cm}
Integrability of the one-dimensional Hubbard model 
and of the factorised scattering problem encountered on the worldsheet of AdS strings
can be expressed in terms of a peculiar quantum algebra. 
In this article, we derive the classical limit of these algebraic integrable structures 
based on established results for the exceptional simple Lie superalgebra $\alg{d}(2,1;\epsilon)$ 
along with standard $\alg{sl}(2)$
which form supersymmetric isometries on 3D AdS space.
The two major steps in this construction consist in
the contraction to a 3D Poincaré superalgebra
and a certain reduction to a deformation of the $\alg{u}(2|2)$ superalgebra.
We apply these steps to the integrable structure 
and obtain the desired Lie bialgebras with suitable classical r-matrices
of rational and trigonometric kind. 
We illustrate our findings in terms of representations for on-shell fields
on AdS and flat space.
\end{minipage}

\vspace*{4cm}

\end{center}

\ifpublic
\newpage
\tableofcontents
\fi

\newpage

%%%%%%%%%%%%%%%%%%%%%%%%%%%%%%%%%%%%%%%%%%%%%%%%%%%%%%%%%%%%%%%%%%%%%%%%%%%%%%%%
%%%%%%%%%%%%%%%%%%%%%%%%%%%%%%%%%%%%%%%%%%%%%%%%%%%%%%%%%%%%%%%%%%%%%%%%%%%%%%%%
\section{Introduction}
\label{sec:intro}

Integrable systems play an important role in theoretical physics,
especially for models of condensed matter physics as well as of high energy physics. 
Integrability of such models is often attributed to the existence of particular extended types of symmetry algebra.
In the case of many integrable quantum models, the relevant algebras 
are known as quantum groups and quantum algebras~\cite{Drinfel'd:1985, Drinfel'd:1988}. 
These describe not only the extended symmetries of a quantum model,
but also allow one to formulate the integrable structure purely in the algebraic language.
Quantum algebras thus give us a useful tool to describe, evaluate and study integrable systems. 

Two important, yet elaborate examples of integrable models are given by
the one-dimensional Hubbard model,
and by the planar limit of $\mathcal{N} = 4$ supersymmetric gauge theory, 
which is AdS/CFT dual to strings on $\AdS^{4,1}\times\Sphere^5$.
In fact, these two particular examples are not unrelated. 
The algebraic structures underlying integrability in the 1D Hubbard model
and factorised worldsheet scattering in the AdS/CFT models
are given by one and the same algebra.
Many features of this quantum algebra have already been worked out,
importantly, that it is based on certain extensions of the Lie superalgebra $\alg{u}(2|2)$
and that it is of some exceptional kind. 
The latter means that established standard constructions in quantum algebra based on
simple or semi-simple Lie algebras and superalgebras are not sufficient to describe it.
Let us elaborate on these achievements and on open questions.

The quantum R-matrix for the integrable structure of the 1D Hubbard model \cite{Hubbard:1963}, see \cite{Essler:2005aa},
has been proposed by Shastry \cite{Shastry:1986bb}. 
It is the result of an elaborate combination of two six-vertex models at the free fermion point using elliptic functions
and it was shown to satisfy the quantum Yang--Baxter equation.
A significant feature of this R-matrix is that it is not of a so-called difference form,
a feature that most of the known solutions to the Yang--Baxter share.
Much later, the R-matrix was reproduced \cite{Beisert:2006qh}
by a construction of the AdS/CFT worldsheet scattering matrix \cite{Beisert:2005tm},
see \cite{Beisert:2010jr},
which was based on a central extension of the Lie superalgebra $\alg{psu}(2|2)$ 
in combination with dynamics of excitations on the worldsheet. 
Quantum algebra structures were established for this system in
\cite{Gomez:2006va,Plefka:2006ze}, 
and the algebra was extended to an infinite-dimensional Yangian algebra in \cite{Beisert:2006fmy}.

Equipped with these algebraic tools, 
scattering matrices for some higher representations 
\cite{Dorey:2006dq,Beisert:2006qh,Chen:2006gp,Matsumoto:2014cka}
have been constructed \cite{Arutyunov:2008zt,deLeeuw:2008dp,Arutyunov:2009mi}.
Importantly, also the overall phase for the scattering matrix could be pinned down 
by consistency considerations of the quantum algebra together 
with considerations of the underlying physical system
\cite{Janik:2006dc,Hernandez:2006tk,Arutyunov:2006iu,Beisert:2006ib,Beisert:2006ez,Dorey:2007xn}.
All of this calls for the formulation of a universal R-matrix
which (in principle) could be evaluated in arbitrary representations 
in order to yield the corresponding scattering matrix along with a suitable overall phase.
Yet, our understanding of the quantum algebra is not complete, 
nor is the original Drinfel'd presentation well suited towards the construction of a universal R-matrix. 
Alternative presentations of the Yangian algebra have been formulated in
\cite{Spill:2008tp,Beisert:2014hya,Beisert:2016qei,Matsumoto:2022nrk}
with the aim of providing a complete formulation of this quantum algebra
from which all relevant properties of the algebra can be derived using established methods.

Progress towards a complete formulation of the quantum algebra
is compromised by an elevated complexity of the structures for this case.
In general, quantum algebras are highly non-linear objects,
but certain functions and structures have been established 
to formulate their objects in more convenient terms.
Among others, these are so-called q-deformations of 
group actions, exponential functions, logarithms, dilogarithms,
factorials, Gamma functions and Pochhammer symbols.
Unfortunately, in the present case, it is not yet known precisely 
how to compose these to formulate, e.g., the universal R-matrix.
Here, the classical limit comes along very handy,
where expressions are reduced to their leading order terms.
Consequently, the quantum algebra reduces to a Lie bialgebra
where most relations are linearised.
The classical limit for the R-matrix has been introduced in \cite{Klose:2006zd,Torrielli:2007mc,Moriyama:2007jt},
and a formulation as a Lie bialgebra was completed in \cite{Beisert:2007ty}.
The underlying algebra turned out to be 
a novel deformation of the $\alg{u}(2|2)$ loop superalgebra.
A corollary of this result was the discovery of an additional $\alg{u}(1)$ derivation
to extend the $\alg{psu}(2|2)$ symmetry algebra \cite{Matsumoto:2007rh}.

The article \cite{Beisert:2007ty} made an auxiliary proposition 
for the derivation of the deformed $\alg{u}(2|2)$ loop superalgebra
as a curious reduction of a maximally extended $\alg{sl}(2)\ltimes\alg{psu}(2|2)\ltimes\Real^{2,1}$ loop superalgebra.
The latter is a non-simple Lie superalgebra, yet its bialgebra structures take a standard form in the rational case.

A further clue in this direction was provided in the article \cite{Beisert:2017xqx},
where the above maximally extended $\alg{sl}(2)\ltimes\alg{psu}(2|2)\ltimes\Real^{2,1}$ superalgebra
was shown to be an algebraic contraction of the semi-simple algebra $\alg{d}(2,1;\epsilon)\times\alg{sl}(2)$
involving the exceptional Lie superalgebra $\alg{d}(2,1;\epsilon)$.
\unskip\footnote{The idea to involve the exceptional Lie superalgebra $\alg{d}(2,1;\epsilon)$
in the limit $\epsilon\to0$ appeared earlier in \cite{Matsumoto:2008ww}.}
This contraction follows along the lines of the contraction 
of the 3D AdS algebra $\alg{so}(2,2)=\alg{sl}(2)\times\alg{sl}(2)$
to the 3D Poincaré algebra $\alg{iso}(2,1)=\alg{sl}(2)\ltimes\Real^{2,1}$.
The latter contraction can be supersymmetrised by the replacement 
of one (or two) factors of $\alg{sl}(2)$ by $\alg{d}(2,1;\epsilon)$,
and by the introduction of one (or two) 
factors of the superalgebra $\alg{psu}(2|2)$ into the Poincaré algebra
$\alg{sl}(2)\ltimes\Real^{2,1}$.

\medskip

The combination of the latter two insights opens up a path 
towards a complete algebraic formulation of the classical integrable structures 
purely in terms of established elements of simple loop superalgebras and their Lie bialgebra structures.
In the present article, we carry out the full procedure 
in order to obtain the complete Lie bialgebra with its classical r-matrix. 
In particular, we will explore two concepts, \emph{contraction} and \emph{reduction}, 
that are essential in avoiding the complications mentioned above. 
Moreover, we will resort to the well-established representation theory of $\alg{sl}(2)$ 
together with analogous representations of $\alg{d}(2,1;\epsilon)$ 
to express a relevant class of representations for the resulting algebra.
This will allow us to express the classical r-matrix 
as the classical limit of a particle scattering matrix,
and it will generally illustrate some of the abstract results 
in more applied terms.

In this article we fill some of the missing steps of the above construction in the classical limit. 
In~\secref{sec:contraction} we start with the reduced case of the contraction
of the algebra $\alg{so}(2,2)$ of isometries of $\AdS^{2,1}$
to the 3D Poincaré algebra $\alg{iso}(2,1)$.
In particular, we describe on-shell field representations on $\AdS^{2,1}$ and on flat $\Real^{2,1}$,
and we show how to perform the contraction between the two.
Then we promote the discussion to loop algebras in~\secref{sec:loops} 
and establish the contraction of the r-matrix of rational type.
In~\secref{sec:reduction} we discuss a particular reduction of the algebras, their r-matrices and representations.
We then extend the receding construction from the rational to the trigonometric case in~\secref{sec:trig}.
Finally, the supersymmetric extension of the above constructions 
involving the exceptional superalgebra $\alg{d}(2,1;\epsilon)$ is discussed in~\secref{sec:susy}. 
Eventually, our construction yields the classical r-matrix which describes the classical limit 
of the 1D Hubbard model and of the AdS/CFT worldsheet S-matrix.

%%%%%%%%%%%%%%%%%%%%%%%%%%%%%%%%%%%%%%%%%%%%%%%%%%%%%%%%%%%%%%%%%%%%%%%%%%%%%%%%
%%%%%%%%%%%%%%%%%%%%%%%%%%%%%%%%%%%%%%%%%%%%%%%%%%%%%%%%%%%%%%%%%%%%%%%%%%%%%%%%
\section{Contraction}
\label{sec:contraction}

Our aim is to understand integrable structures of a physical model
with (an extension as well as a reduction of) Poincaré symmetry on flat Minkowski space $\Real^{2,1}$.
A difficulty is that the Lie algebra $\alg{iso}(2,1)=\alg{sl}(2)\ltimes\Real^{2,1}$ 
incorporating Poincaré symmetry is non-simple, 
whereas structures of integrability are best developed for simple and semi-simple algebras,
see \cite{Chari:1994pz}.
Moreover, unitary representation of this non-compact algebra are necessarily infinite-dimensional,
which complicates constructions in terms of physically relevant representations.
Our resolution to these problems is to resort to the fact that the Poincaré algebra is a contraction
of the Lie algebra $\alg{so}(2,2)=\alg{sl}(2)\times\alg{sl}(2)$ 
which incorporates the isometries of anti-de Sitter space $\AdS^{2,1}$.
The algebra factors $\alg{sl}(2)$ are simple, 
and their representation theory is well-understood and easy to handle.
We will then move, step by step, towards the originally intended situation
in the subsequent sections.

In order to set the stage for some more elaborate constructions in this work, 
we will review the algebra contraction, 
\[
\alg{so}(2,2)=\alg{sl}(2)\times\alg{sl}(2) 
\eqjoin*{\longrightarrow}
\alg{iso}(2,1)=\alg{sl}(2)\ltimes\Real^{2,1},
\]
first at the level of the algebra, then in terms of geometry
and finally at the level of infinite-dimensional representations.
This will also introduce the notation and 
relate the abstract mathematical considerations 
to physical fields on the symmetric space $\AdS^{2,1}$ 
and on flat Minkowski space $\Real^{2,1}$.

%%%%%%%%%%%%%%%%%%%%%%%%%%%%%%%%%%%%%%%%%%%%%%%%%%%%%%%%%%%%%%%%%%%%%%%%%%%%%%%%
\subsection{Algebra}
\label{sec:Algebra}

We start by introducing the above Lie algebras in terms of their generators,
see the summary in \tabref{tab:AlgebraGenerators},
and by describing the contraction that relates the two.

\begin{table}\centering
$\begin{array}{l|ll|l|l}
\text{algebra} & \text{generators} & \text{indices} & \alg{sl}(2) \text{ form} & a\in\set{0,\pm}
\\\hline
\alg{so}(3) & \genJ^k & 1,2,3
&
\alg{sl}(2)\text{ (generic)} & \genJ^a
\\
\alg{so}(2,2) & \genM^{\alpha\beta} & x,y;u,v
&
\alg{sl}(2)\times\alg{sl}(2) & \genM_1^a,\genM_2^a
\\
\alg{iso}(2,1) & \genL^{\mu\nu},\genP^\mu & x,y;t
&
\alg{sl}(2)\ltimes \Real^{2,1} & \genL^{a},\genP^{a}
\end{array}$
\caption{Generators of spacetime symmetries}
\label{tab:AlgebraGenerators}
\end{table}

%%%%%%%%%%%%%%%%%%%%%%%%%%%%%%%%%%%%%%%%
\paragraph{Spacetime Algebras.}

Here we present the generators of the relevant spacetime Lie algebras $\alg{so}(2,2)$ and $\alg{iso}(2,1)$
along with their Lie brackets and invariant quadratic forms.

The AdS algebra $\alg{so}(2,2)$ is spanned by a set of generators 
which we shall denote by $M^{\alpha\beta}=-M^{\beta\alpha}$ 
with the indices $\alpha,\beta\in\set{u,v,x,y}$.
Their Lie brackets are given by 
\unskip\footnote{Here and below, we follow the physics convention that
the generators for the real form of a Lie algebra
are typically assumed to be purely imaginary.}
\[
\label{eq:Mcomm}
\comm{\genM^{\alpha\beta}}{\genM^{\gamma\delta}}
= 
-\iunit\eta^{\beta\gamma}\genM^{\alpha\delta}
+\iunit\eta^{\alpha\gamma}\genM^{\beta\delta}
+\iunit\eta^{\beta\delta}\genM^{\alpha\gamma}
-\iunit\eta^{\alpha\delta}\genM^{\beta\gamma},
\]
where $\eta$ denotes a metric tensor of signature $(-,-,+,+)$
corresponding to the directions $(u,v,x,y)$. 
The algebra has two independent invariant quadratic forms
\[
\label{eq:Mcas}
\casMMa := -\half \eta_{\alpha\gamma}\eta_{\beta\delta}\.\genM^{\alpha\beta} \otimes \genM^{\gamma\delta},
\eqsep
\casMMp := \quarter \varepsilon_{\alpha\beta\gamma\delta}\.\genM^{\alpha\beta} \otimes \genM^{\gamma\delta},
\]
where $\varepsilon$ denotes totally anti-symmetric tensors
(we choose the normalisation $\varepsilon_{uvxy}=+1$).

The Lorentz algebra $\alg{so}(2,1)$ is spanned by the generators 
$\genL^{\mu\nu}=-\genL^{\nu\mu}$ 
with the indices $\mu,\nu\in\set{t,x,y}$ and the signature $(-,+,+)$.
Their Lie brackets take the same form as \eqref{eq:Mcomm}
but with indices $\mu,\nu\in\set{t,x,y}$
\[
\label{eq:Lcomm}
\comm{\genL^{\mu\nu}}{\genL^{\rho\sigma}}
= 
-\iunit\eta^{\nu\rho}\genL^{\mu\sigma}
+\iunit\eta^{\mu\rho}\genL^{\nu\sigma}
+\iunit\eta^{\nu\sigma}\genL^{\mu\rho}
-\iunit\eta^{\mu\sigma}\genL^{\nu\rho}.
\]
It has an invariant quadratic form 
$\casLL := -\half \eta_{\mu\rho}\eta_{\nu\sigma}\genL^{\mu\nu} \otimes \genL^{\rho\sigma}$
which is analogous to $\casMMa$ of $\alg{so}(2,2)$.

The Poincaré algebra $\alg{iso}(2,1)$ supplements 
the Lorentz generators $\genL^{\mu\nu}$ with the
momentum generators $\genP^\mu$, $\mu\in\set{t,x,y}$,
which obey the additional algebra relations
\[
\label{eq:LPcomm}
\comm{\genL^{\mu\nu}}{\genP^{\rho}}
= 
-\iunit\eta^{\nu\rho}\genP^{\mu}
+\iunit\eta^{\mu\rho}\genP^{\nu},
\eqsep
\comm{\genP^{\mu}}{\genP^{\nu}}=0.
\]
The Poincaré algebra has two invariant quadratic forms
\[
\label{eq:LPcas}
\casPP:=\eta_{\mu\nu}\.\genP^\mu\otimes\genP^\nu,
\eqsep
\casLP:=-\quarter\varepsilon_{\mu\nu\rho}(\genL^{\mu\nu}\otimes\genP^\rho+\genP^\rho\otimes\genL^{\mu\nu}).
\]
Note that the above invariant quadratic form $\casLL$ of the Lorentz algebra 
is not an invariant for the Poincaré algebra.

%%%%%%%%%%%%%%%%%%%%%%%%%%%%%%%%%%%%%%%%
\paragraph{$\alg{sl}(2)$ Forms.}

All of the above algebras are related to $\alg{sl}(2)$ in some way.
To make the relations between the various algebras more evident, 
we will use a common notation.
This will also streamline the contraction procedure.

We will typically denote the generators of an abstract complexified algebra $\alg{sl}(2)$ 
by $\genJ^a$ with index $a\in\set{0,\pm}$.
The generators obey the algebra relations
\[
\label{eq:Jcomm}
\comm{\genJ^0}{\genJ^\pm}=\pm\genJ^\pm,
\eqsep
\comm{\genJ^+}{\genJ^-}=-2\genJ^0
\eqjoin{\iff}
\comm{\genJ^a}{\genJ^b}=\iunit f^{ab}{}_c\genJ^c.
\]
The structure constants $f^{ab}{}_c$ of the latter universal form
are defined by the former explicit relations.
Finally, the quadratic invariant form of $\alg{sl}(2)$ reads
\[
\label{eq:Jcas}
\casJJ := -\genJ^0\otimes\genJ^0 + \half\genJ^+\otimes\genJ^- + \half\genJ^-\otimes\genJ^+
=c_{ab}(\genJ^a\otimes\genJ^b),
\]
with $c_{ab}$ denoting the coefficients in the basis $\genJ^a$.
Concretely, the above two sets of coefficients are given by 
\[ \label{eq:sl2Constants}
f^{0\pm}{}_\pm=\mp\iunit,
\eqsep*
f^{\pm\mp}{}_0=\pm 2\iunit,
\eqsep
c_{00}=-1,
\eqsep*
c_{\pm\mp}=\half.
\]

As an aside, note that the set of generators $\set{\genJ^{0,\pm}}$
maps almost trivially to the standard Cartan--Weyl basis 
$\set{\gen{H},\gen{E},\gen{F}}$ of $\alg{sl}(2)$ by
$(\genJ^0,\iunit\genJ^+,\iunit\genJ^-)=(\half\gen{H},\gen{E},\gen{F})$.
\unskip\footnote{For the real form $\alg{sl}(2,\Real)$, 
the generators $\genJ^0,\genJ^\pm$ or 
$\gen{H},\gen{E},\gen{F}$ can be taken to be real.}
It can also be cast as an imaginary basis $\genJ^k$, $k=1,2,3$ 
for the compact real form $\alg{so}(3)$ as
\unskip\footnote{For the real form $\alg{so}(3)$,
the generators $\genJ^0$ is imaginary and $\genJ^\pm$ obey $(\genJ^\pm)^*=\genJ^\mp$.}
\[
(\genJ^0,\iunit\genJ^\pm)=
(\genJ^3,\genJ^1\pm\iunit\genJ^2),
\eqsep
\comm{\genJ^j}{\genJ^k}=\iunit\varepsilon^{jkm}\genJ^m,
\eqsep
\casJJ=-\genJ^k\otimes\genJ^k.
\]

It is well-known that the AdS algebra $\alg{so}(2,2)$ is isomorphic to $\alg{sl}(2)\times\alg{sl}(2)$.
We use the following map for the generators $\genM_1^a,\genM_2^a$ 
of two copies of the algebra $\alg{sl}(2)$
\begin{align}
\genM_1^0&=-\half\genM^{uv}+\half\genM^{xy},
&
\genM_1^\pm&=\half(-\genM^{ux}+\genM^{vy})\mp\ihalf(+\genM^{uy}+\genM^{vx}),
\\
\genM_2^0&=+\half\genM^{uv}+\half\genM^{xy},
&
\genM_2^\pm&=\half(+\genM^{ux}+\genM^{vy})\mp\ihalf(-\genM^{uy}+\genM^{vx}).
\end{align}
Each set of generators $\genM_1^a$ and $\genM_2^a$ obeys the algebra relations
of one copy of $\alg{sl}(2)$ while the mixed Lie brackets are zero
\[
\label{eq:Mcommsl2}
\comm{\genM_1^a}{\genM_1^b}=\iunit f^{ab}{}_c\genM_1^c,
\eqsep
\comm{\genM_1^a}{\genM_2^b}=0,
\eqsep
\comm{\genM_2^a}{\genM_2^b}=\iunit f^{ab}{}_c\genM_2^c.
\]
The two quadratic invariant forms are related to the quadratic invariant
form \eqref{eq:Jcas} for the two copies of $\alg{sl}(2)$
\[
\label{eq:Mcassl2}
\casMMa=2\casMM_1+2\casMM_2,
\eqsep
\casMMp=2\casMM_1-2\casMM_2.
\]

The Lorentz algebra $\alg{so}(2,1)=\alg{sl}(2,\Real)$
can also be cast in the above $\alg{sl}(2)$ form.
The generators $\genL^{\mu\nu}$ are identified with $\genL^a$,
and it makes sense to express the Poincaré generators $\genP^\mu$
in the $\alg{sl}(2)$ basis $\genP^a$ as follows
\unskip\footnote{For the real form $\alg{so}(2,1)$,
the generators $\genL^0$ is imaginary and $\genL^\pm$ obey $(\genL^\pm)^*=-\genL^\mp$.}
\begin{align}
\genL^0&=\genL^{xy},
&
\genL^\pm&=\genL^{ty}\mp\iunit\genL^{tx},
&
\half\genL^++\half\genL^-&=\genL^{ty},
&
\ihalf\genL^+-\ihalf\genL^-&=\genL^{tx}.
\\
\genP^0&=\genP^t,
&
\genP^\pm&=\genP^{x}\pm\iunit\genP^{y},
&
\half\genP^+ +\half\genP^-&=\genP^{x},
&
-\ihalf\genP^+ -\ihalf\genP^-&=\genP^{y}.
\end{align}
Altogether, the algebra relations of $\alg{iso}(2,1)=\alg{sl}(2)\ltimes\Real^{2,1}$
then take the form 
\[
\label{eq:LPcommsl2}
\comm{\genL^a}{\genL^b}=\iunit f^{ab}{}_c\genL^c,
\eqsep
\comm{\genL^a}{\genP^b}=\iunit f^{ab}{}_c\genP^c,
\eqsep
\comm{\genP^a}{\genP^b}=0.
\]
The invariant quadratic forms match with their $\alg{sl}(2)$ counterparts
as follows
\[
\label{eq:LPcassl2}
\casPP=
c_{ab}(\genP^a\otimes\genP^b),
\eqsep
\casLP=
\half c_{ab}(\genL^a\otimes\genP^b+\genP^a\otimes\genL^b).
\]

%%%%%%%%%%%%%%%%%%%%%%%%%%%%%%%%%%%%%%%%
\paragraph{Algebra Contraction.}

The contraction $\alg{so}(2,2)\to\alg{iso}(2,1)$
is achieved by taking the limit $\epsilon\to0$ for the following identification
of generators
\unskip\footnote{Alternative choices of combinations for $\genP^a$ are conceivable,
e.g.\ 
$\genP^a=\half\epsilon\mref(\genM_1^a-\genM_2^a)$ or
$\genP^a=-\epsilon\mref\genM_2^a$ or
$\genP^a=c\epsilon\mref\genM_1^a$.
These lead to inessential modifications of the algebraic relations.}
\[ \label{eq:contraction}
\genL^a = \genM^a_1+\genM^a_2, 
\eqsep 
\genP^a = \epsilon\mref\genM_1^a.
\]
Here, we have introduced a supplementary reference mass constant $\mref$ 
in order to make appropriate mass dimensions manifest
when the limiting parameter $\epsilon$ is assumed to be dimensionless.
For finite $\epsilon$ this identification describes a bijective map
between $\genM_1^a,\genM_2^a$ and $\genL^a,\genP^a$
which becomes singular at $\epsilon=0$.
The algebraic relations of these generators at finite $\epsilon$ read
\[
\comm{\genL^a}{\genL^b}=\iunit f^{ab}{}_c\genL^c,
\eqsep
\comm{\genL^a}{\genP^b}=\iunit f^{ab}{}_c\genP^c,
\eqsep
\comm{\genP^a}{\genP^b}=\iunit \epsilon\mref f^{ab}{}_c\genP^c.
\]
In the limit $\epsilon\to 0$, the Lie bracket of generators $\genP$ 
becomes trivial due to $\comm{\genP}{\genP}=\Order(\epsilon)$,
and the algebra relations \eqref{eq:LPcommsl2} of $\alg{iso}(2,1)$ are recovered.

%%%%%%%%%%%%%%%%%%%%%%%%%%%%%%%%%%%%%%%%%%%%%%%%%%%%%%%%%%%%%%%%%%%%%%%%%%%%%%%%
\subsection{Geometry}

Our aim is to construct a representation of the AdS algebra $\alg{so}(2,2)$ 
which reduces to a field representation of the Poincaré algebra $\alg{iso}(2,1)$
under the contraction. Therefore it makes sense to analyse the
situation from a geometric point of view.

%%%%%%%%%%%%%%%%%%%%%%%%%%%%%%%%%%%%%%%%
\paragraph{Anti-de Sitter Space.}

The algebra $\alg{so}(2,2)$ acts canonically on the vector space $\Real^{2,2}$
for which we shall use linear coordinates $X^\alpha=(u,v,x,y)$ with signature $(-,-,+,+)$. 
The corresponding canonical action on a scalar field 
\unskip\footnote{We initially restrict to scalar fields for simplicity.
Later on, we will generalise our presentation to spinning fields.}
$F(u,v,x,y)$ 
by means of differential operators $\dgen{\genM}$ takes the form
\[
\dgen{\genM^{\alpha\beta}}=-\iunit X^\alpha\frac{\partial}{\partial X_\beta}+\iunit X^\beta\frac{\partial}{\partial X_\alpha}.
\]

This action is reducible thanks to the invariance of the square form $X^2:=\eta_{\alpha\beta}X^\alpha X^\beta$ under $\alg{so}(2,2)$
which can be used to restrict to submanifolds at constant $X^2$.
Correspondingly, $\alg{so}(2,2)$ has an action on the symmetric space $\AdS^{2,1}$ 
which can be embedded into $\Real^{2,2}$ as the submanifold at $X^2=-1$.
We shall use the embedding angular coordinates $(\tau,\psi,\phi)$
which are given by the relationship
\unskip\footnote{The non-periodicity of time $\tau$ implies that $\AdS^{2,1}$ covers 
the submanifold of $\Real^{2,2}$ infinitely often.
Conversely, the angle coordinate $\phi$ is $2\pi$-periodic,
and the domain of the radial coordinate $\psi$ is $0\leq \psi<\half\pi$.}
\[
X^\alpha=
\begin{pmatrix}
u\\v\\x\\y
\end{pmatrix}
=
\begin{pmatrix}
\sec(\psi)\cos(\tau)\\
\sec(\psi)\sin(\tau)\\
\tan(\psi)\cos(\phi)\\
\tan(\psi)\sin(\phi)
\end{pmatrix}.
\]
The metric of $\AdS^{2,1}$ in these coordinates reads
\[
\der s^2=
\sec^2(\psi) \brk!{-\der\tau^2+\der\psi^2+\sin^2(\psi)\.\der\phi^2}.
\]
The restriction of the above $\alg{so}(2,2)$ differential action 
to the submanifold $\AdS^{2,1}$ is given by
\begin{align}
\dgen{\genM^0_1}
&=
\frac{1}{2}\brk*{-\iunit\frac{\partial}{\partial\tau}
  -\iunit\frac{\partial}{\partial\phi}},
\\
\dgen{\genM^\pm_1}
&=
\frac{1}{2}\eunit^{\pm\iunit(\tau+\phi)}
\brk*{\mp\sin(\psi)\frac{\partial}{\partial\tau}
  +\iunit\cos(\psi)\frac{\partial}{\partial\psi}
  \mp\csc(\psi)\frac{\partial}{\partial\phi}
},
\\
\dgen{\genM^0_2}
&=
\frac{1}{2}\brk*{+\iunit\frac{\partial}{\partial\tau}
  -\iunit\frac{\partial}{\partial\phi}},
\\
\dgen{\genM^\pm_2}
&=
\frac{1}{2}\eunit^{\pm\iunit(-\tau+\phi)}
\brk*{\mp\sin(\psi)\frac{\partial}{\partial\tau}
  -\iunit\cos(\psi)\frac{\partial}{\partial\psi}
  \pm\csc(\psi)\frac{\partial}{\partial\phi}
}.
\end{align}
%

%%%%%%%%%%%%%%%%%%%%%%%%%%%%%%%%%%%%%%%%
\paragraph{Contraction.}

\begin{figure}
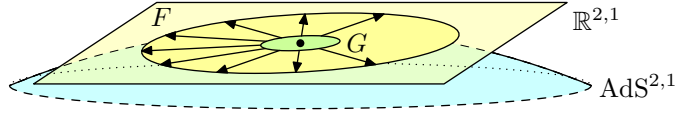

\centering
\begin{mpostfig}
paths[1]:=subpath (1.5,2.5) of fullcircle scaled 20xu;
paths[2]:=fullcircle scaled abs((point 0 of paths[1])-(point infinity of paths[1])) yscaled 0.08
  shifted 0.5[point 0 of paths[1],point infinity of paths[1]];
vpos[1]:=(point 0.5length(paths[1]) of paths[1]) + (0,-0.2xu);
paths[3]:=((-1xu,-1xu)--(1xu,-1xu)--(1xu,1xu)--(-1xu,1xu)--cycle) slanted 0.3 yscaled 0.2 scaled 2.7 shifted vpos[1];
paths[4]:=fullcircle scaled 1xu slanted 0.3 yscaled 0.2 shifted vpos[1];
paths[5]:=fullcircle scaled 4xu slanted 0.3 yscaled 0.2 shifted vpos[1];
fill paths[1]--(subpath (4,8) of paths[2])--cycle withcolor 0.2[white,blue+green];
draw paths[1] pensize(0.5pt);
draw paths[1] pensize(0.5pt) dashed evenly;
fill paths[3] withcolor 0.2[white,red+green];
fill buildcycle(subpath(2,6) of paths[3],paths[1]) withcolor 0.2[white,0.5red+green];
draw paths[1] pensize(0.1pt) dashed evenly;
draw paths[3];
draw subpath(0,4) of paths[2] dashed withdots scaled 0.5;
fillshape(paths[5],0.4[white,red+green]);
fillshape(paths[4],0.4[white,0.5red+green]);
drawdot(vpos[1], 3pt);
draw subpath(8,4) of paths[2] dashed evenly;
label.rt(btex $G$ etex, point 0 of paths[4]);
label.ulft(btex $F$ etex, point 3.5 of paths[5]);
for i:=2 upto 10: drawarrow (point ((i+0.5)*8/12) of paths[4])--(point ((i+0.5)*8/12) of paths[5]); endfor 
label.lrt(btex $\Real^{2,1}$ etex, point 2 of paths[3]);
label.rt(btex $\AdS^{2,1}$ etex, point 0 of paths[1]);
\end{mpostfig}
\caption{Mapping between a function on a small neighbourhood of a point on $\AdS^{2,1}$
and a function on the tangent space.}
\label{fig:tangent}
\end{figure}

The contraction can be thought of as 
the set of partially infinitesimal $\alg{so}(2,2)$ transformations
acting on a tangent space of $\AdS^{2,1}$ as follows
(see also \figref{fig:tangent} for an illustration):
Without loss of generality, we consider the tangent space 
at the marked point $\tau=\psi=0$ with arbitrary angle $\phi$.
We want to blow up a small neighbourhood of this point,
where the curvature of $\AdS^{2,1}$ becomes irrelevant,
to a finite neighbourhood of the origin in the tangent space.
The tangent space is $\Real^{2,1}$ and it has coordinates $(t,x,y)$
of signature $(-,+,+)$.
In the neighbourhood of the marked point,
we can map a function $F$ on the tangent space $\Real^{2,1}$ 
to a function $G$ on the manifold $\AdS^{2,1}$ 
according to the map
\[
F(t,x,y)\leftrightarrow G(\tau,\psi,\phi),
\eqsep
t=\frac{2}{\epsilon\mref}\tau,
\eqsep*
x=\frac{2}{\epsilon\mref}\psi\cos\phi,
\eqsep*
y=\frac{2}{\epsilon\mref}\psi\sin\phi,
\]
where the time and radial coordinates $\tau$ and $\psi$ need to be small in the limit $\epsilon\to 0$
while the angle coordinate $\phi$ should remain finite.
Note that the field $G$ fluctuates rapidly with the position
as far as $F$ is a smooth function.
Suitable transformations of $G$ then map 
to transformations of $F$ under the contraction.
Using the appropriate transformation of partial derivatives, 
\[
\frac{\partial}{\partial\tau}=
\frac{2}{\epsilon \mref}\partial_t,
\eqsep
\frac{\partial}{\partial\psi}=
\frac{2}{\epsilon \mref\sqrt{x^2+y^2}}(x\partial_x+y\partial_y),
\eqsep
\frac{\partial}{\partial\phi}=
x\partial_y-y\partial_x,
\]
we find the following contraction limit for the differential operators
using the prescriptions detailed in \secref{sec:Algebra}
\begin{align}
\dgen{\genL^0}
&= -\iunit(x\partial_y-y\partial_x),
&
\dgen{\genL^\pm}
&=
\mp(x\pm\iunit y)\partial_t
\mp t (\partial_x\pm\iunit\partial_y),
\\
\dgen{\genP^0}
&=
-\iunit\partial_t,
&
\dgen{\genP^\pm}
&=
\iunit(\partial_x\pm\iunit\partial_y)
.
\end{align}
The resulting expressions have no divergent terms,
and infinitesimal terms are discarded in the limit.
In the following, we demonstrate that these form a representation
of the limiting Poincaré algebra.

%%%%%%%%%%%%%%%%%%%%%%%%%%%%%%%%%%%%%%%%
\paragraph{Minkowski Space.}

These differential operators clearly agree with the canonical action of the Poincaré algebra $\alg{iso}(2,1)$
for scalar fields on $\Real^{2,1}$
\[
\dgen{\genL^{\mu\nu}}=-\iunit x^\mu \partial^\nu+\iunit x^\nu \partial^\mu,
\eqsep
\dgen{\genP^{\mu}}=\iunit \partial^\mu.
\]
Here $\state{x}$ is a position eigenstate on $\Real^{2,1}$
on which the Poincaré algebra acts as
\unskip\footnote{Note that a consistent representation
of some generator $\genJ$ on a position eigenstate $\state{x}$
requires the somewhat unintuitive identification $\genJ\state{x}=-\dgen{\genJ}\state{x}$
with the \emph{negative} differential operator $\dgen{\genJ}$.
For a functional eigenstate $\state{F}=\int \mathinner{\der x}F(x)\state{x}$
this leads to the desired positive sign upon acting on $F$ 
by integration by parts, namely $\genJ\state{F} = \state{\dgen{\genJ}F}$.
For this reason we shall explicitly distinguish between
a generator $\genJ$ and its associated differential operator $\dgen{\genJ}$.}
\[
\genL^{\mu\nu}\state{x} = -\dgen{\genL^{\mu\nu}}\state{x},
\eqsep
\genP^{\mu}\state{x} = -\dgen{\genP^{\mu}}\state{x}.
\]

Alternatively, fields on $\Real^{2,1}$ can be expressed in momentum space 
by means of the Fourier transformation 
\[
\state{p}=
\int (\der x)^3 
\exp(-\iunit p_\mu x^\mu)
\state{x},
\eqsep
\state{x}=
\int \frac{(\der p)^3}{(2\pi)^3} 
\exp(\iunit p_\mu x^\mu)
\state{p},
\]
where $\state{p}$ is a state of with definite momentum.
We will also use the representation in momentum space,
\[
\genL^{\mu\nu}\state{p}
=
\iunit p^\mu\frac{\partial}{\partial p_\nu}\state{p}
-\iunit p^\nu\frac{\partial}{\partial p_\mu}\state{p},
\eqsep
\genP^{\mu}\state{p}
=
p^\mu \state{p},
\]
which is obtained by Fourier transformation of the representation in position space.

Altogether, we have thus shown how to obtain the $\alg{iso}(2,1)$ momentum space representation on $\Real^{2,1}$
as a contraction limit of the $\alg{so}(2,2)$ representation acting on $\AdS^{2,1}$.

%%%%%%%%%%%%%%%%%%%%%%%%%%%%%%%%%%%%%%%%%%%%%%%%%%%%%%%%%%%%%%%%%%%%%%%%%%%%%%%%
\subsection{Irreducible Representations}
\label{sec:irreps}

The above field representations are off-shell, 
i.e.\ the fields are unconstrained by differential equations.
Such representations are reducible, 
and it makes sense to identify their irreducible components
and subsequently describe their contraction limit.

%%%%%%%%%%%%%%%%%%%%%%%%%%%%%%%%%%%%%%%%
\paragraph{Minkowski Space.}

Reducibility becomes most apparent for the above Poincaré representation in momentum space:
The orbit of the momentum $\state{p^\mu}$ under Lorentz transformations 
is a shell of common mass rather than all of $\Real^{2,1}$.
The mass $m$ and spin $s$ are identified as eigenvalues of the two quadratic invariants
\eqref{eq:LPcas} of $\alg{iso}(2,1)$
\[
\casPP\simeq p^2=-m^2,
\eqsep
\casLP\simeq -sm.
\]
For the above momentum space representation,
one finds that $\casLP\simeq 0$ which is in line with the fact that 
it describes a scalar field with $s=0$.
\unskip\footnote{Non-scalar fields require additional spin components 
for each momentum eigenstate on which some extra contributions to $\genL$ act.
For the sake of simplicity we do not discuss this more complicated case.
Instead we shall merely introduce a non-zero spin for irreps further below.}

The reduction of the above representation for a scalar field
to a mass shell can be performed by setting
\[
\state{\vect{p}}_m
:= \int \der e\. \theta(e)\.\delta(-e^2+\vect*{p}{}^2+m^2) \state{e,\vect{p}}
=\frac{1}{2e_m(\vect{p})}\state!{e_m(\vect{p}),\vect{p}}
\]
with the relativistic energy relation
\[
e_m(\vect{p}):=\sqrt{\vect*{p}{}^2+m^2}.
\]
This identification fixes the energy component $p^t$ to $e_m(\vect{p})$ 
and effectively makes all derivatives $\partial/\partial p^t$ drop out.
\unskip\footnote{More accurately, it leads to a derivative of the delta-function enforcing the mass shell condition.
It cancels in all combinations of derivatives that leave the mass shell condition invariant.}
%\[
%F(p_t,p_x,p_y)=\delta(p_x^2+p_y^2+m^2-p_t^2)\.f(p_x,p_y)
%\]
%\begin{align}
%\frac{\partial}{\partial p_t} F(p_t,p_x,p_y)
%&=-2p_t\delta'(p_x^2+p_y^2+m^2-p_t^2)\.f(p_x,p_y)
%\\
%\frac{\partial}{\partial p_x} F(p_t,p_x,p_y)
%&=2p_x\delta'(p_x^2+p_y^2+m^2-p_t^2)\.f(p_x,p_y)
%\\&\alignrel
%+\delta(p_x^2+p_y^2+m^2-p_t^2) \frac{\partial}{\partial p_x} f(p_x,p_y)
%\\
%\frac{\partial}{\partial p_y} F(p_t,p_x,p_y)
%&=2p_y\delta'(p_x^2+p_y^2+m^2-p_t^2)\.f(p_x,p_y)
%\\&\alignrel
%+\delta(p_x^2+p_y^2+m^2-p_t^2) \frac{\partial}{\partial p_y} f(p_x,p_y)
%\end{align}
%
Altogether the irreducible representation on an on-shell field
of mass $m$ and spin $s$ reads
\begin{align}
\label{eq:PoincareIrrep}
\genL^{xy}\state{\vect{p}}_{m,s}
&=\brk*{\iunit p_x\frac{\partial}{\partial p_y}-\iunit p_y\frac{\partial}{\partial p_x}+s
+p_x\frac{\partial\Theta}{\partial p_y}-p_y\frac{\partial\Theta}{\partial p_x}
}\state{\vect{p}}_{m,s},
\\
\genL^{ty}\state{\vect{p}}_{m,s}
&=\brk*{\iunit e_m(\vect{p})\frac{\partial}{\partial p_y}+\frac{sp_x}{e_m(\vect{p})+m}
+e_m(\vect{p})\frac{\partial\Theta}{\partial p_y}
}\state{\vect{p}}_{m,s},
\\
\genL^{tx}\state{\vect{p}}_{m,s}
&=\brk*{\iunit e_m(\vect{p})\frac{\partial}{\partial p_x}-\frac{sp_y}{e_m(\vect{p})+m}
+e_m(\vect{p})\frac{\partial\Theta}{\partial p_x}
}\state{\vect{p}}_{m,s},
\\
\genP^t\state{\vect{p}}_{m,s}
&=e_m(\vect{p})\state{\vect{p}}_{m,s},
\\
\genP^x\state{\vect{p}}_{m,s}
&=p_x\state{\vect{p}}_{m,s},
\\
\genP^y\state{\vect{p}}_{m,s}
&=p_y\state{\vect{p}}_{m,s}.
\end{align}
Here, we have generalised the above representation 
by a non-trivial spin $s$.
\unskip\footnote{The stabiliser of a massive momentum vector 
is $\alg{so}(2)$ whose irreducible representations are one-dimensional
and labelled by a spin $s\in\Integer$
(or $s\in\half\Integer$ for fields with bosonic and fermionic statistics).
Consequently, there are no additional spin degrees of freedom in this case.} 
Furthermore, the arbitrary function $\Theta=\Theta(\vect{p})$ 
incorporates the on-shell gauge freedom
\[
\label{eq:PoincareStateGauge}
\state{\vect{p}}\to\exp\brk!{\iunit\Theta(\vect{p})}\state{\vect{p}}.
\]
In position space, the mass shell and spin conditions turn 
into differential equations for the field.
One can formally solve these equations
by means of a Fourier transformation
which maps the representation to its momentum space counterpart.

%%%%%%%%%%%%%%%%%%%%%%%%%%%%%%%%%%%%%%%%
\paragraph{Anti-de Sitter Space.}

We would now like to formulate an irreducible representation of $\alg{so}(2,2)$ on $\AdS^{2,1}$
that limits to the above $\alg{iso}(2,1)$ representation upon contraction.
Unfortunately, the Fourier transformation does not apply to 
a curved manifold such as $\AdS^{2,1}$.
Hence it is not evident how to construct a momentum space representation,
but we can still impose a suitable differential equation.
The solutions of the differential equation then serve as the basis states
for an irreducible representation of $\alg{so}(2,2)$.
This representation will be infinite-dimensional.

The Laplacian for the $\AdS^{2,1}$ manifold reads
\[
\Delta=
-\cos^2(\psi)\frac{\partial^2}{\partial\tau^2}
+\cos^2(\psi)\frac{\partial^2}{\partial\psi^2}
+\cot(\psi)\frac{\partial}{\partial\psi}
+\cot^2(\psi)\frac{\partial^2}{\partial\phi^2}.
\]
The eigenvalue equation $\Delta \Upsilon=\lambda \Upsilon$
can be understood as an equation of motion for the 
free scalar field $\Upsilon$ where $\lambda=\adsmass^2-1$ describes its AdS mass $\adsmass$.
\unskip\footnote{The dimensionless AdS mass $\adsmass$ is measured in units of the inverse AdS radius.}
The eigenfunctions for eigenvalue $\lambda=\adsmass^2-1$
are given by the following hypergeometric functions
\unskip\footnote{Note that the two functions $\Upsilon_{\pm\adsmass,\adseng,\adsangmom}$
with opposite sign of $\adsmass$
serve as a basis for the two-dimensional space of solutions
of the second order differential equation
with otherwise equal dependency on $\tau$ and $\phi$.}
\begin{align}
\Upsilon_{\adsmass,\adseng,\adsangmom}(\tau,\psi,\phi)
&\sim
\eunit^{\iunit\adseng\tau+\iunit\adsangmom\phi}
\sin^\adsangmom(\psi)
\cos^{1+\adsmass}(\psi)
\\
&\qquad\cdot
\.\operatorname{{}_2F_1}\brk!{
\half(1+\adsangmom-\adseng+\adsmass),\half(1+\adsangmom+\adseng+\adsmass),1+\adsmass;\cos^2(\psi)
}.
\end{align}
The particular exponential dependencies of the eigenfunction $\Upsilon$ on $\tau$ and $\phi$ identify
$\adseng$ and $\adsangmom$ as the energy and angular momentum, respectively.

Let us understand how the eigenfunctions transform under the isometries $\alg{so}(2,2)$.
We find
\begin{align}
\label{eq:eigentrans}
\dgen{\genM_1^0} \Upsilon_{\adsmass,\adseng,\adsangmom}
&=
\half(\adsangmom+\adseng) \Upsilon_{\adsmass,\adseng,\adsangmom}
,\\
\dgen{\genM_1^\pm} \Upsilon_{\adsmass,\adseng,\adsangmom}
&=
\pm\ihalf(\adsangmom+\adseng\pm\adsmass\pm 1) \Upsilon_{\adsmass,\adseng\pm1,\adsangmom\pm1}
,\\
\dgen{\genM_2^0} \Upsilon_{\adsmass,\adseng,\adsangmom}
&=
\half(\adsangmom-\adseng) \Upsilon_{\adsmass,\adseng,\adsangmom}
,\\
\dgen{\genM_2^\pm} \Upsilon_{\adsmass,\adseng,\adsangmom}
&=
\mp\ihalf(\adsangmom-\adseng\pm\adsmass\pm 1) \Upsilon_{\adsmass,\adseng\mp1,\adsangmom\pm1}.
\end{align}
Here it makes sense to consider the action of invariant elements.
The two quadratic invariants for $\alg{so}(2,2)$ 
act as eigenvalues
\[
-\dgen{\casMMa} = \Delta\simeq \adsmass^2-1,
\eqsep
\dgen{\casMMp} \simeq 0.
\]
Moreover, we can compute the invariant group elements
\[ \label{eq:AdSInvGroupElems}
\exp\brk!{2\pi\iunit\dgen{\genM^{xy}}}\simeq \eunit^{2\pi\iunit\adsangmom},
\eqsep
\exp\brk!{2\pi\iunit\dgen{\genM^{uv}}}\simeq\eunit^{2\pi\iunit\adseng},
\]
which generate shifts of $\phi$ and $\tau$ by $2\pi$, respectively.
On $\AdS^{2,1}$ a shift $\phi\to\phi+2\pi$ must be trivial,
hence $\adsangmom\in\Integer$,
while the time $\tau$ is non-periodic,
and consequently $\adseng\in\Real$.

Finally, the eigenfunctions may or may not be normalisable
where the canonical square norm is given by
\[
\norm!{\Upsilon_{\adsmass,\adseng,\adsangmom}}^2:=
\int \der\phi\.\der\psi\.\tan(\psi)\abs!{\Upsilon_{\adsmass,\adseng,\adsangmom}(\tau,\psi,\phi)}^2.
\]
Normalisability is relevant for unitarity of the representation
which is a useful classification criterion in (quantum) physics.
Let us therefore discuss it:
A finite norm requires that the asymptotic behaviour at $\psi\to 0$ 
and at $\psi\to\half\pi$ must be benign, i.e.
\[
\lim_{\psi\to 0} \psi^{-\abs{\adsangmom}}\Upsilon_{\adsmass,\adseng,\adsangmom}(\tau,\psi,\phi)<\infty,
\eqsep
\lim_{\psi\to\pi/2}\Upsilon_{\adsmass,\adseng,\adsangmom}(\tau,\psi,\phi)=0.
\]
The function $\Upsilon_{\adsmass,\adseng,\adsangmom}$ 
is a linear combination of two basis functions with the leading behaviours $\sim\psi^{\pm\adsangmom}$ at $\psi\to 0$.
Clearly, one of these two is divergent and therefore undesirable. 
Within $\Upsilon_{\adsmass,\adseng,\adsangmom}$ it must have a vanishing coefficient
which happens to be achieved under the conditions $\adsangmom=n_1-n_2$ and $\abs{\adseng}=1+\adsmass+n_1+n_2$
with $n_1,n_2\in\Integer^+_0$ non-negative integers.
At $\psi\to\half\pi$ one finds the asymptotic behaviour $\Upsilon_{\adsmass,\adseng,\adsangmom}\sim(\half\pi-\psi)^{1+\adsmass}$
which implies a lower bound on $\adsmass$.
Altogether, an eigenfunction $\Upsilon_{\adsmass,\adseng,\adsangmom}$ is normalisable provided that
\[
\half\brk!{\abs{\adseng}-\abs{\adsangmom}-1-\adsmass}\in\Integer^+_0,
\eqsep
\adsmass>-1.
\]
%

%%%%%%%%%%%%%%%%%%%%%%%%%%%%%%%%%%%%%%%%
\paragraph{Principal Series Representations.}

The above representations \eqref{eq:eigentrans}
of $\genM_1$ and $\genM_2$ on the scalar eigenfunctions 
can be identified as infinite-dimensional principal series representations of 
the two copies of $\alg{sl}(2)$ in $\alg{so}(2,2)=\alg{sl}(2)\times\alg{sl}(2)$.
A principal series representation of $\genJ\in\alg{sl}(2)$ 
on the tower of states $\state{k}$, $k\in\Integer$, takes the form 
\begin{align} \label{eq:principalSeriesIrrep}
\genJ^0\state{k}_{\reppar,\repang}&=(k+\repang)\state{k}_{\reppar,\repang},
\\
\genJ^+\state{k}_{\reppar,\repang}&=\theta_k(k+\repang+\reppar+\half)\state{k+1}_{\reppar,\repang},
\\
\genJ^-\state{k}_{\reppar,\repang}&=\theta_{k-1}^{-1}(k+\repang-\reppar-\half)\state{k-1}_{\reppar,\repang}.
\end{align}
The parameter $\reppar$ describes the eigenvalue of the quadratic invariants \eqref{eq:Jcas}
whereas the non-integer part of the parameter $\repang$ 
describes the eigenvalue of the invariant group element $\exp(2\pi\iunit\genJ^0)$
\[
\casJJ \simeq \quarter-\reppar^2,
\eqsep
\exp(2\pi\iunit\genJ^0)\simeq \eunit^{2\pi\iunit\repang}.
\]
Furthermore, the parameters $\theta_k$ incorporate a gauge transformation 
$\state{k}\to\exp(\iunit\Theta_k)\state{k}$ acting on the states $\state{k}$
as the map $\theta_k\to\exp(\iunit\Theta_k-\iunit\Theta_{k-1})\theta_k$
which can be used to fix the $\theta_k$ to arbitrary values.
The states and parameters of the principal series representations
are matched with the scalar representation \eqref{eq:eigentrans} 
by the identifications $\adsangmom=k_1+k_2+\adsangmom_0$ and $\adseng=k_1-k_2+\adseng_0$
as well as $\reppar_{1,2}=\half\adsmass$ and $\repang_{1,2}=\half(\adsangmom_0\pm\adseng_0)$.
Note that this agrees with the relationship \eqref{eq:Mcassl2} 
between the quadratic invariants of $\alg{so}(2,2)$ and $\alg{sl}(2)\times\alg{sl}(2)$.

We want to generalise the above considerations 
to unitary irreps of $\alg{so}(2,2)$ with positive energy $\adseng$
and non-zero spin $s$:
This constrains the parameters $\reppar,\repang$ of the
$\alg{sl}(2)$ principal series representations somewhat:
the representation of $\genM_1$ needs be of lowest-weight type with
\[
\repang_1>0,
\eqsep
\reppar_1=\repang_1-\half,
\eqsep
\theta_{1;k}=\sqrt{\frac{k+1}{k+2\repang_1}},
\]
while the representation of $\genM_2$ needs to be of highest-weight type
with
\[
\repang_2<0,
\eqsep
\reppar_2=-\repang_2-\half,
\eqsep
\theta_{2;k}=\sqrt{\frac{-k-2\repang_2-1}{-k}}.
\]
A non-zero spin $s$ is achieved by choosing distinct values for $\repang_1$ and $-\repang_2$
\footnote{In the above discussion, 
this would correspond to certain tensor and/or spinor fields
which are eigenfunctions of both $2\casMM_1$ and $2\casMM_2$.}
\[ \label{eq:spinAdS}
\reppar_{1,2}=\half (\adsmass\pm s),
\eqsep
\repang_{1,2}=\half(s \pm (\adsmass + 1)).
\]
Unitarity then implies the bound $\adsmass>-1+\abs{s}$.

Altogether the representation for a field on $\AdS^{2,1}$ with spin $s$ 
is given by
\begin{align}
\label{eq:AdSIrrep}
\genM_1^0\state{k_1,k_2}_{\adsmass,s}
&=
(k_1+\half \adsmass+\half s+\half)\state{k_1,k_2}_{\adsmass,s}
,\\
\genM_1^+\state{k_1,k_2}_{\adsmass,s}
&=
\sqrt{(k_1+1)(k_1+\adsmass+s+1)}\state{k_1+1,k_2}_{\adsmass,s}
,\\
\genM_1^-\state{k_1,k_2}_{\adsmass,s}
&=
\sqrt{k_1(k_1+\adsmass+s)}\state{k_1-1,k_2}_{\adsmass,s}
,\\
\genM_2^0\state{k_1,k_2}_{\adsmass,s}
&=
-(k_2+\half\adsmass-\half s+\half)\state{k_1,k_2}_{\adsmass,s}
,\\
\genM_2^+\state{k_1,k_2}_{\adsmass,s}
&=
-\sqrt{k_2(k_2+\adsmass-s)}\state{k_1,k_2-1}_{\adsmass,s}
,\\
\genM_2^-\state{k_1,k_2}_{\adsmass,s}
&=
-\sqrt{(k_2+1)(k_2+\adsmass-s+1)}\state{k_1,k_2+1}_{\adsmass,s}
.
\end{align}
Here, the labels $k_2$ for states of the representation of $\genM_2$ 
have been flipped to make both numbers $k_1$ and $k_2$ non-negative integers.
The eigenvalues of the quadratic invariant forms and the full spatial rotation 
are given by
\[
\casMM_{1,2}=\quarter(1-\adsmass^2\mp 2s\adsmass-s^2),
\eqsep
\exp(2\pi\iunit\genM^{xy})\simeq\exp(2\pi\iunit s).
\]
The quadratic form eigenvalues identify $\adsmass$ and $s$ as mass and spin, respectively
\[
\casMMa\simeq 1-\adsmass^2-s^2,
\eqsep
\casMMp\simeq -2s\adsmass.
\]

The above representation for normalisable fields on $\AdS^{2,1}$
will serve as a principal starting point for many investigations of this article.

%%%%%%%%%%%%%%%%%%%%%%%%%%%%%%%%%%%%%%%%%%%%%%%%%%%%%%%%%%%%%%%%%%%%%%%%%%%%%%%%
\subsection{Irrep Contraction}
\label{sec:irrepcontr}

Next we would like to derive a Poincaré representation
from a unitary irrep of the AdS algebra. 
We have already constructed irreps for both algebras in 
\eqref{eq:AdSIrrep} and \eqref{eq:PoincareIrrep},
and they describe fields with certain properties
such as an AdS or Minkowski mass $\adsmass$ or $m$ and an intrinsic spin $s$. 
It is therefore conceivable that the AdS representation 
contracts to the Poincaré representation.
However, there is also a major difference between the states
of the representation: the AdS states are labelled by discrete numbers
whereas the Minkowski states are described by continuous momenta.
The contraction limit can indeed induce such a transmutation, 
but this is a singular process which requires some care. 
Let us therefore describe the limiting relationship of the representation in detail.

%%%%%%%%%%%%%%%%%%%%%%%%%%%%%%%%%%%%%%%%
\paragraph{Parameters.}

The contraction limit is performed via an identification of generators
\[
(\genL,\genP)
=(\genM_1+\genM_2,\mref\epsilon\genM_1)
\eqjoin*{\iff}
(\genM_1,\genM_2)=
(\epsilon^{-1}\mref^{-1}\genP,-\epsilon^{-1}\mref^{-1}\genP+\genL).
\]
We need to find a family of representations for $\genM_1,\genM_2$
so that the limiting representation for $\genL,\genP$ is finite. 
As the two representations superficially look rather unrelated,
a useful first step is to consider the eigenvalues of the algebra invariants.
These expressions are independent of the states, 
so the contraction limit will directly fix their relationship.

Altogether, we have the eigenvalue relations
\[
(\casPP,\casLP)\simeq(-m^2,-sm),
\eqsep
(\casMMa,\casMMp)\simeq(1-\adsmass^2-s^2,-2s\adsmass).
\]
The spin $s$ is a discrete quantity and it has to be identified directly
between the two representations. 
For the masses $m$ and $\adsmass$, 
we use the relationship of invariants
\[
\casMMa=\frac{4}{\epsilon^2\mref^2}\casPP+\Order(\epsilon^{-1}),
\eqsep
\casMMp=\frac{4}{\epsilon\mref}\casLP+\Order(\epsilon^{0})
\]
from which one obtains
\[ \label{eq:AdSmass}
\adsmass=\frac{2m}{\epsilon \mref}+\Order(\epsilon^0).
\]
We thus identify the parameters $\reppar_{1,2}$
of the $\alg{sl}(2)$ representations by
\[
\reppar_1=
\frac{m}{\epsilon \mref} +
\frac{s}{2},
\eqsep
\reppar_2=
\frac{m}{\epsilon \mref} -
\frac{s}{2}.
\]

%%%%%%%%%%%%%%%%%%%%%%%%%%%%%%%%%%%%%%%%
\paragraph{States.}

\begin{figure}
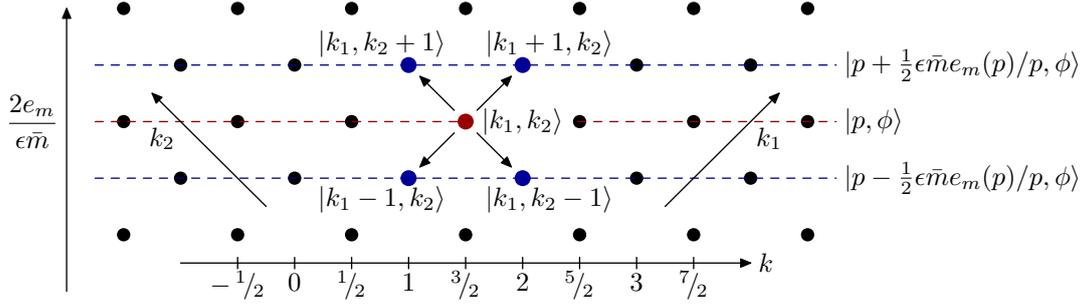

\centering
\begin{mpostfig}
interim xu:=0.75cm;
for j:=-2 step 2 until 2: for i:=-6 step 2 until 6:
drawdot((i*xu,j*xu),5pt);
endfor endfor
for j:=-1 step 2 until 1: for i:=-5 step 2 until 5:
drawdot((i*xu,j*xu),5pt);
endfor endfor
paths[5]:=((-6.5xu,0)--(+6.5xu,0)) shifted (0,-1xu);
paths[6]:=((-6.5xu,0)--(+6.5xu,0)) shifted (0,+0xu);
paths[7]:=((-6.5xu,0)--(+6.5xu,0)) shifted (0,+1xu);
draw paths[5] dashed evenly withcolor 0.6blue;
draw paths[6] dashed evenly withcolor 0.6red;
draw paths[7] dashed evenly withcolor 0.6blue;
label.rt(btex $\state{p-\half\epsilon \mref e_m(p)/p,\phi}$ etex, point 1 of paths[5]);
label.rt(btex $\state{p,\phi}$ etex, point 1 of paths[6]);
label.rt(btex $\state{p+\half\epsilon \mref e_m(p)/p,\phi}$ etex, point 1 of paths[7]);
drawdot((0xu,0xu),6pt) withcolor 0.6red;
drawdot((+1xu,+1xu),6pt) withcolor 0.6blue;
drawdot((-1xu,+1xu),6pt) withcolor 0.6blue;
drawdot((+1xu,-1xu),6pt) withcolor 0.6blue;
drawdot((-1xu,-1xu),6pt) withcolor 0.6blue;
paths[1]:=(+3.5xu,-1.5xu)--(+5.5xu,0.5xu);
paths[2]:=(-3.5xu,-1.5xu)--(-5.5xu,0.5xu);
unfill bbox(thelabel.rt(btex $(k_1,k_2)$ etex, (3pt,0xu)));
label.rt(btex $\state{k_1,k_2}$ etex, (3pt,0xu));
label.top(btex \qquad$\state{k_1+1,k_2}$ etex, (1xu,1xu));
label.top(btex $\state{k_1,k_2+1}$\qquad etex, (-1xu,1xu));
label.bot(btex \qquad$\state{k_1,k_2-1}$ etex, (1xu,-1xu));
label.bot(btex $\state{k_1-1,k_2}$\qquad etex, (-1xu,-1xu));
drawarrow paths[1];
drawarrow paths[2];
drawarrow subpath (0.2,0.8) of ((0,0)--(+1xu,+1xu));
drawarrow subpath (0.2,0.8) of ((0,0)--(-1xu,+1xu));
drawarrow subpath (0.2,0.8) of ((0,0)--(+1xu,-1xu));
drawarrow subpath (0.2,0.8) of ((0,0)--(-1xu,-1xu));
label.lrt(btex $k_1$ etex, point 0.75 of paths[1]);
label.llft(btex $k_2$ etex, point 0.75 of paths[2]);
paths[3]:=((-5xu,0xu)--(+5xu,0xu)) shifted (0,-2.5xu);
drawarrow paths[3];
label.rt(btex $k$ etex, point 1 of paths[3]);
for i:=-4 upto 4: 
draw ((0,-0.1xu)--(0,+0.1xu)) shifted (i*xu,-2.5xu);
endfor
label.bot(btex $-\vfrac{1}{2}$ etex, (-4xu,-2.5xu));
label.bot(btex $0\vphantom{0'}$ etex, (-3xu,-2.5xu));
label.bot(btex $\vfrac{1}{2}$ etex, (-2xu,-2.5xu));
label.bot(btex $1\vphantom{0'}$ etex, (-1xu,-2.5xu));
label.bot(btex $\vfrac{3}{2}$ etex, (0xu,-2.5xu));
label.bot(btex $2\vphantom{0'}$ etex, (1xu,-2.5xu));
label.bot(btex $\vfrac{5}{2}$ etex, (2xu,-2.5xu));
label.bot(btex $3\vphantom{0'}$ etex, (3xu,-2.5xu));
label.bot(btex $\vfrac{7}{2}$ etex, (4xu,-2.5xu));
paths[4]:=(-7xu,-3xu)--(-7xu,+2xu);
label.lft(btex $\displaystyle\frac{2e_m}{\epsilon \mref}$ etex, point 0.6 of paths[4]);
drawarrow paths[4];
\end{mpostfig}
\caption{Discrete states $\protect\state{k_1,k_2}$ of AdS irrep 
vs.\ continuous states $\protect\state{p,\phi}$ of Poincaré irrep.
Note that the lattice of permissible $k$ is alternating 
between integers and half-integers for consecutive energy levels.}
\label{fig:StateContraction}
\end{figure}

We can now approach the contraction limit for the representation.
The discrete states of the AdS representation
have to turn into continuous momentum states of the Poincaré representation
by some continuum limit involving states with large indices $k_1,k_2$.
For how to identify the indices precisely, 
we can take inspiration from their roles in the AdS representation.
We know that $\adseng=1+\adsmass+k_1+k_2$ represents the energy on AdS space
and in the contraction limit it must diverge.
Conversely, $\adsangmom=k_1-k_2$ is an angular momentum
which should remain finite in the limit.
These two considerations are taken into account
by the following prescription for a suitable state in the contraction limit
\[\label{eq:contrPrescription}
\state{p,\phi}_{m,s}:=
\sum_k\eunit^{-\iunit (k_1-k_2)\phi}\state{k_1,k_2}_{2m/\mref/\epsilon,s},
\eqsep
k_{1,2}:=\frac{e_m(p)-m}{\epsilon \mref}\pm k.
\]
Some remarks are in order:
The indices $k_{1,2}$ are not necessarily integers
for any given parameters $p,m,\epsilon$, but for sufficiently small $\epsilon$ 
one can always find a nearby momentum $p'$ such that $k_{1,2}$ become integers
while approximating the desired state well.
The bounds for the sum over $k$ are determined by non-negativity of $k_1$ and $k_2$,
and for small $\epsilon$ these diverge to $\pm\infty$.
Finally, note that the prescription assumes $k_1-k_2=2k$, 
and for $k_1$ and $k_2$ to be arbitrary non-negative integers,
we allow $k$ to be either an integer or a half-integer.
However, the sum over $k$ assumes a step size of 1,
so it is defined to be either over the integers or over the half-integers,
depending on where one starts. 
The situation is perhaps best illustrated by a figure, see \figref{fig:StateContraction}.

%%%%%%%%%%%%%%%%%%%%%%%%%%%%%%%%%%%%%%%%
\paragraph{Momentum Generators.}

Carrying out the contraction is most straight-forward for the energy generators $\genP^0$
\begin{align}
\genP^0\state{p,\phi}
&=
\epsilon \mref \sum_k\mathinner{\eunit}^{-\iunit (k_1-k_2)\phi}\.\genM_1^0\state{k_1,k_2}
\\
&=
\sum_k\eunit^{-\iunit (k_1-k_2)\phi}\.\epsilon \mref(k_1+\half \adsmass+\half s+\half)\state{k_1,k_2}
\\
&=
\sum_k\eunit^{-\iunit (k_1-k_2)\phi}\brk!{e_m(p)+\half\epsilon\mref(2k+s+1)}\state{k_1,k_2}
\\
&=e_m(p)\state{p,\phi}+\Order(\epsilon).
\end{align}
The result is the same state multiplied by the energy $e_m(p)$.
For the momentum generators $\genP^\pm$, some more work is needed, e.g.\
\begin{align}
\genP^+\state{p,\phi}
&=
%\epsilon \mref\genM_1^+\state{p,\phi}
%=
\sum_k\eunit^{-\iunit (k_1-k_2)\phi}\.\epsilon\mref \sqrt{(k_1+1)(k_1+\adsmass+s+1)}\state{k_1+1,k_2}
\\
%&=
%\sum_k\eunit^{-\iunit (k_1-k_2)\phi}\sqrt{e_m(p)^2-m^2}\state{k_1+1,k_2} + \Order(\epsilon)
%\\
&=
p\sum_k\eunit^{-\iunit (k_1-k_2)\phi}\state{k_1+1,k_2} + \Order(\epsilon).
\end{align}
Next we have to express the sum as a limiting state. 
For that we shift the summation index by $\vfrac{1}{2}$
to bring the sum closer to the original form. 
The resulting overall shift of $k_{1,2}$ by $\vfrac{1}{2}$ 
corresponds to a shift of energy by an elementary level
and thus to a shift of momentum 
\[
k_{1,2}'=k_{1,2}+\half,
\eqsep
e_m(p')=e_m(p)+\half\epsilon \mref,
\eqsep
p'=p+\half\epsilon\mref \frac{e_m(p)}{p}+\Order(\epsilon^2).
\]
According to \figref{fig:StateContraction} the shift of energy
fits nicely to the alternating lattices for the index $k$.
We thus find
\[
\sum_k\eunit^{-\iunit (k_1-k_2)\phi}\state{k_1+1,k_2}
=
\sum_{k-1/2}\eunit^{-\iunit (k_1-k_2-1)\phi}\state{k_1+\half,k_2+\half}
=
\eunit^{\iunit\phi}\state!{p+\half\epsilon \mref e_m(p)/p,\phi}.
\]
The shift of $k_1$ leads to a shift of energy and a phase $\eunit^{\iunit\phi}$.
Similar identities hold for shifts of $k_{1,2}$ by $\pm 1$.
%
\iffalse
\begin{align}
\sum_k\eunit^{-\iunit (k_1-k_2)\phi}\state{k_1+1,k_2}
&=
%\eunit^{\iunit\phi}\sum_{k-1/2}\eunit^{-\iunit (k_1-k_2)\phi}\state{k_1+\half,k_2+\half}
%=
\eunit^{\iunit\phi}\state{p+\half\epsilon\mref e_m(p)/p,\phi},
\\
\sum_k\eunit^{-\iunit (k_1-k_2)\phi}\state{k_1,k_2+1}
&=
%\eunit^{-\iunit\phi}\sum_{k+1/2}\eunit^{-\iunit (k_1-k_2)\phi}\state{k_1+\half,k_2+\half}
%=
\eunit^{-\iunit\phi}\state{p+\half\epsilon\mref e_m(p)/p,\phi},
\\
\sum_k\eunit^{-\iunit (k_1-k_2)\phi}\state{k_1-1,k_2}
&=
\eunit^{-\iunit\phi}\state{p-\half\epsilon\mref e_m(p)/p,\phi},
\\
\sum_k\eunit^{-\iunit (k_1-k_2)\phi}\state{k_1,k_2-1}
&=
\eunit^{\iunit\phi}\state{p-\half\epsilon\mref e_m(p)/p,\phi}
\end{align}
\fi
%
This concludes the evaluation of the contraction limit for the momentum generator
\[
\genP^+\state{p,\phi}
=
p\.\eunit^{+\iunit\phi}\state!{p+\half\epsilon\mref e_m(p)/p,\phi} + \Order(\epsilon)
=
\eunit^{+\iunit\phi}p\state{p,\phi} + \Order(\epsilon).
\]
In the limit, we end up with the same state we started with, 
therefore $\genP^+$ also acts by multiplication
with the momentum $\eunit^{\iunit\phi}p$.
The limit for the other momentum generator $\genP^-$ is analogous
and yields the eigenvalue $\eunit^{-\iunit\phi}p$.

%%%%%%%%%%%%%%%%%%%%%%%%%%%%%%%%%%%%%%%%
\paragraph{Lorentz Generators.}

We can now turn to the Lorentz generators $\genL$.
The rotation generator $\genL^0$ is easiest to handle,
we find
\begin{align}
\genL^0\state{p,\phi}
&=
\sum_k\eunit^{-\iunit (k_1-k_2)\phi}(\genM_1^0+\genM_2^0)\state{k_1,k_2}
=
\sum_k\eunit^{-\iunit (k_1-k_2)\phi}(k_1-k_2+s)\state{k_1,k_2}
\\
&=
\sum_k \brk*{\iunit\frac{\partial}{\partial\phi}+s}\eunit^{-\iunit (k_1-k_2)\phi}\state{k_1,k_2}
=
\brk*{\iunit\frac{\partial}{\partial\phi}+s}
\state{p,\phi}.
\end{align}
Here the appearance of a factor of $k_{1,2}$ in the Fourier sum 
translates to a derivative term in the conjugate variable as usual.
Finally, we compute the contraction limit for the remaining Lorentz generator $\genL^+$
where recycle most of the previously used relations
\begin{align}
\genL^+\state{p,\phi}
&=
\sum_k\eunit^{-\iunit (k_1-k_2)\phi}\sqrt{(k_1+1)(k_1+\adsmass+s+1)}\state{k_1+1,k_2}
\\
&\alignrel
-\sum_k\eunit^{-\iunit (k_1-k_2)\phi}\sqrt{k_2(k_2+\adsmass-s)}\state{k_1,k_2-1}
\\
&=
\sum_k\eunit^{-\iunit (k_1-k_2)\phi}
\brk*{
\frac{p}{\epsilon \mref}
+(k+1)\frac{e_m(p)}{p}
+\frac{1}{2}\frac{sp}{e_m(p)+m}
+\Order(\epsilon)
}\state{k_1+1,k_2}
\\
&\alignrel
-
\sum_k\eunit^{-\iunit (k_1-k_2)\phi}
\brk*{
\frac{p}{\epsilon\mref}
-k\frac{e_m(p)}{p}
-\frac{1}{2}\frac{sp}{e_m(p)+m}
+\Order(\epsilon)
}\state{k_1,k_2-1}
\\
&=
\eunit^{\iunit\phi}
\frac{p}{\epsilon\mref}
\brk!{
\state!{p+\half\epsilon \mref e_m(p)/p,\phi}
-\state!{p-\half\epsilon \mref e_m(p)/p,\phi}
}
\\
&\alignrel
+
\eunit^{\iunit\phi}
\brk*{
\iunit\frac{e_m(p)}{p}\frac{\partial}{\partial\phi}
+\frac{sp}{e_m(p)+m}
}
\state{p,\phi}
+\Order(\epsilon)
\\
&=
\eunit^{\iunit\phi}
\brk*{
e_m(p)\frac{\partial}{\partial p}
+\iunit\frac{e_m(p)}{p}\frac{\partial}{\partial\phi}
+\frac{sp}{e_m(p)+m}
}
\state{p,\phi}
+\Order(\epsilon).
\end{align}
In the final step, we have encountered the additional phenomenon
that a small shift of energy levels times a divergent factor of $1/\epsilon$ 
gives rise to a derivative with respect to momentum.

%%%%%%%%%%%%%%%%%%%%%%%%%%%%%%%%%%%%%%%%
\paragraph{Comparison.}

In summary, we have obtained the contraction limit of the irrep
\begin{align}
\label{eq:momrep}
\genL^0\state{p,\phi}_{m,s}
&=\brk*{\iunit \frac{\partial}{\partial\phi}+s}\state{p,\phi}_{m,s},
\\
\genL^\pm\state{p,\phi}_{m,s}
&=\eunit^{\pm\iunit\phi}\brk*{
\pm e_m(p)\frac{\partial}{\partial p} 
+\iunit\frac{e_m(p)}{p}\frac{\partial}{\partial\phi}
+\frac{sp}{e_m(p)+m}
}\state{p,\phi}_{m,s},
\\
\genP^0\state{p,\phi}_{m,s}
&=e_m(p)\state{p,\phi}_{m,s},
\\
\genP^\pm \state{p,\phi}_{m,s}
&=\eunit^{\pm\iunit\phi}p\state{p,\phi}_{m,s}.
\end{align}
One can convince oneself that these satisfy the relations of the Poincaré algebra.
In fact, the representation matches precisely with the Poincaré irrep \eqref{eq:PoincareIrrep}
with trivial gauge function $\Theta(\vect{p})=0$
provided we express the two-dimensional momenta $(p_x,p_y)$
in polar momentum coordinates $(p,\phi)$ as
\[\label{eq:momrepstate}
\state{p,\phi}_{m,s} = \state!{p\cos(\phi),p\sin(\phi)}_{m,s}.
\]
%

%\[
%\frac{\partial}{\partial\phi}
%=
%-p\sin(\phi)\frac{\partial}{\partial p_x}
%+p\cos(\phi)\frac{\partial}{\partial p_y}
%\eqsep
%\frac{\partial}{\partial p} 
%=
%\cos(\phi)\frac{\partial}{\partial p_x}
%+\sin(\phi)\frac{\partial}{\partial p_y}
%\]

%\[
%\frac{\partial}{\partial p_x}
%=
%\cos(\phi)\frac{\partial}{\partial p} 
%-\frac{1}{p}\sin(\phi)\frac{\partial}{\partial\phi}
%\eqsep
%\frac{\partial}{\partial p_y}
%=
%\sin(\phi)\frac{\partial}{\partial p} 
%+\frac{1}{p}\cos(\phi)\frac{\partial}{\partial\phi}
%\]

%%%%%%%%%%%%%%%%%%%%%%%%%%%%%%%%%%%%%%%%%%%%%%%%%%%%%%%%%%%%%%%%%%%%%%%%%%%%%%%%
%%%%%%%%%%%%%%%%%%%%%%%%%%%%%%%%%%%%%%%%%%%%%%%%%%%%%%%%%%%%%%%%%%%%%%%%%%%%%%%%
\section{Loop Algebras and r-Matrices}
\label{sec:loops}

The integrable structure of many classical physics models in $1+1$ dimensions 
are related to loop algebras based on some finite-dimensional Lie algebra
with a classical r-matrix of rational or trigonometric kind.
The framework for models with underlying Lie algebras of simple or semi-simple kind
is developed very well. The situation is not as fortunate for non-simple algebras.
Here we use the contraction limit of the AdS algebra $\alg{so}(2,2)$
to derive an algebraic integrability framework
based on the non-simple Poincaré algebra $\alg{iso}(2,1)$.

We start by reviewing some relevant elements of quasi-triangular loop algebras
of rational type for simple Lie algebras $\alg{g}$.
We then apply the contraction of \secref{sec:contraction} 
to derive the algebraic integrability structures for the Poincaré loop algebra. 
We also introduce the notion of integrable twists
that will be needed later on.

The results concerning quasi-triangular loop algebras 
based on non-simple Lie algebras (with invertible quadratic forms)
will turn out to be straight-forward and unsurprising generalisations 
of the case of simple Lie algebras. 
Nevertheless we will also use this section as an opportunity to introduce
the relevant algebraic framework 
and to derive their application to the quasi-triangular Poincaré loop algebra.

%%%%%%%%%%%%%%%%%%%%%%%%%%%%%%%%%%%%%%%%%%%%%%%%%%%%%%%%%%%%%%%%%%%%%%%%%%%%%%%%
\subsection{Simple Rational Case} 

In the following, we review rational r-matrices and loop algebras
for simple Lie algebras $\alg{g}$.

%%%%%%%%%%%%%%%%%%%%%%%%%%%%%%%%%%%%%%%%
\paragraph{Classical r-Matrix.}

A Lie algebra $\alg{g}$ can be supplemented by a classical r-matrix $r$
in order to turn it into a so-called quasi-triangular Lie bi-algebra.
Such an algebra provides relevant structures for integrability 
and it is a suitable starting point for quantisation~\cite{Drinfel'd:1988}.

A classical r-matrix is an element $r\in\alg{g} \otimes \alg{g}$
with two key properties. 
First, the symmetrisation $r+\permop(r)$ must be a quadratic invariant element of $\alg{g}$,
that is for any elements $\gen{X}\in\alg{g}$
\[
\liebr!{r+\permop(r)}{\gen{X}}
:=
\liebr!{r_{12}+r_{21}}{\gen{X}_1+\gen{X}_2}\stackrel{!}{=}0.
\]
Second, $r$ must satisfy the classical Yang--Baxter equation
\[
\cybe{r}{r}:=\liebr{r_{12}}{r_{13}} + \liebr{r_{12}}{r_{23}} + \liebr{r_{13}}{r_{23}}
\stackrel{!}{=}0.
\]
These two properties ensure that the Lie cobracket $\delta : \alg{g} \to \alg{g} \wedge \alg{g},$
defined by 
\[
\delta(\gen{X})
:=
-\liebr{r}{\gen{X}}
:=
-\liebr{r_{12}}{\gen{X}_1+\gen{X}_2},
\]
extends the Lie algebra to a proper Lie bi-algebra.

%%%%%%%%%%%%%%%%%%%%%%%%%%%%%%%%%%%%%%%%
\paragraph{Parametric r-Matrices and Loop Algebras.}

A classical r-matrix may also depend on a pair of so-called spectral parameters $u_1,u_2\in\Complex$.
The two spectral parameters are associated to the two tensor sites of the r-matrix
$r:\Complex\times\Complex \to  \alg{g} \otimes \alg{g}$
such that in its relations the latter typically extends according to the rule
\[
r_{12}\to r(u_1,u_2)_{12}.
\]
In the parametric case, $r(u_1,u_2)$ is usually anti-symmetric as follows
\[
\permop\brk!{r(u_2,u_1)}=-r(u_1,u_2)
\eqjoin{\iff}
r(u_1,u_2)_{12}=-r(u_2,u_1)_{21},
\]
so that its symmetric part is zero and thus trivially invariant.
The classical Yang--Baxter equation also extends canonically to
\[
\liebr!{r(u_1,u_2)_{12}}{r(u_1,u_3)_{13}} 
+ \liebr!{r(u_1,u_2)_{12}}{r(u_2,u_3)_{23}} 
+ \liebr!{r(u_1,u_3)_{13}}{r(u_2,u_3)_{23}}
=0.
\]

The dependency on spectral parameters $u_k$ can be encoded at the algebraic level
by lifting a given Lie algebra $\alg{g}$ to the infinite-dimensional loop extension $\alg{g}[u, u^{-1}]$ of $\alg{g}$.
This is achieved by adjoining $\alg{g}$ with Laurent polynomials in a formal parameter $u$:
\[
\alg{g}[u, u^{-1}] := \Complex[u, u^{-1}]\otimes\alg{g}.
\]
For a Lie algebra spanned by the generators $\genJ^a \in \alg{g}$ with $a = 1, \ldots, \dim\alg{g}$,
the corresponding loop algebra is spanned by $u^n\otimes\genJ^a =: \genJ^a_n$ with $n \in \Integer$. 
The Lie brackets are defined as 
\[
\comm{\genJ^a_n}{\genJ^b_m} = \iunit f^{ab}{}_{c} \genJ^c_{n+m}.
\]
For any representation $\rho$ of the original algebra $\alg{g}$ 
there is a corresponding one-parameter family of evaluation representations of the loop algebra 
$\rho_u$, $u\in\Complex$, defined by:
\[
\rho_u(\genJ^a_n)  = u^n \rho(\genJ^a).
\]
With these relations and identifications
(as well as an implicit reference to the evaluation representation)
the parametric r-matrix $r(u_1,u_2)\in\alg{g}\otimes\alg{g}$
becomes a plain r-matrix in the loop algebra $r\in\alg{g}[u,u^{-1}]\otimes\alg{g}[u,u^{-1}]$.
The cobracket in a quasi-triangular loop algebra thus becomes
\[
\delta(\genJ^a_n)
=
-
\liebr{r}{\genJ^a_n}
=
-
\liebr!{r(u_1,u_2)_{12}}{u_1^n\otimes\genJ_1^a+u_2^n\otimes\genJ_2^a}.
\]
%

%%%%%%%%%%%%%%%%%%%%%%%%%%%%%%%%%%%%%%%%
\paragraph{Rational r-Matrices.}

Many of the known parametric r-matrices obey the difference form $r(u_1,u_2)=r(u_1 - u_2)$.
Such r-matrices for simple Lie algebras are classified by their poles
which form regular lattices in the complex plane of dimension $0$, $1$ or $2$ \cite{Belavin:1982}.
These classes are respectively called rational, trigonometric and elliptic.
In this article we discuss rational and trigonometric r-matrices
for which standard expressions exist and are well-known.
In this and the following section we will deal only with rational r-matrices,
the generalisation to the trigonometric case is described in \secref{sec:trig}.

For a Lie algebra $\alg{g}$ with a quadratic invariant $\casJJ = c_{ab} \genJ^a \otimes \genJ^b$,
in particular for $\alg{g}=\alg{sl}(2)$, 
the standard parametric r-matrix of rational types takes the form
\[
\label{eq:GenericRationalrMat}
r(u_1,u_2) = \frac{-\rnorm\casJJ}{u_1 - u_2}.
\]
This r-matrix is anti-symmetric and it satisfies the classical Yang--Baxter equation.

%\[
% f^{0+}{}_+ = -\iunit  
%\eqsep
% f^{0+}{}_+ = +\iunit  
%\eqsep
% f^{+-}{}_0 = +2\iunit  
%\]
%\[
%c_{+-}=c_{-+}=\half,
%\eqsep
%c_{00}=-1
%\]
%\[
% f^{0}{}_{-+} = f^{0+}{}_{+}c_{+-} = -\ihalf
%\eqsep
% f^{+}{}_{+0} = f^{+-}{}_{0} c_{+-} = +\iunit  
%\eqsep
% f^{-}{}_{-0} = f^{-+}{}_0 c_{+-}= -\iunit  
%\]
%\[
%\half f^c{}_{ab} \genJ^a\wedge \genJ^b
%\eqsep
%-\iunit \genJ^0\wedge \genJ^+
%\eqsep
%+\iunit \genJ^0\wedge \genJ^-
%\eqsep
%+\ihalf \genJ^+\wedge \genJ^-
%\]

We can recast this r-matrix to the loop algebra form by identifying the spectral parameters $u_1$ and $u_2$ 
with the loop parameters of two copies of the loop algebra $\alg{g}[u,u^{-1}]$. 
We may expand the r-matrix around the point $u_1/u_2 = 0$ using a geometric series and obtain
\[
r = \rnorm c_{ab} \sum\limits_{k=0}^\infty \genJ^a_k \otimes \genJ^b_{-k-1}.
\]
This r-matrix satisfies the classical Yang--Baxter equation,
but notably it is not anti-symmetric 
in contradistinction to the corresponding parametric r-matrix $r(u_1,u_2)$.
\unskip\footnote{An alternative but equivalent form for $r$ 
is based on the expansion about the point $u_2/u_1=0$.
Here we fix the choice on one of the two alternative forms.}
The cobrackets based on the rational r-matrix read
\[
\cobra(\genJ^c_{n})
=
\ihalf \rnorm f^c{}_{ab} \sum_{k=0}^{n-1}
\genJ^a_k \wedge \genJ^b_{n-1-k}, 
\]
with the dual structure constants defined by $f^c{}_{ab}:=c_{da} f^{cd}{}_b$.

For $\alg{g}=\alg{sl}(2)$ we have concretely the rational r-matrix
\[
r_{\alg{sl}(2)} = \rnorm \sum\limits_{k=0}^\infty 
\brk!{
-\genJ^0_k \otimes \genJ^0_{-k-1}
+\half\genJ^+_k \otimes \genJ^-_{-k-1}
+\half\genJ^-_k \otimes \genJ^+_{-k-1}
}
\]
and the resulting cobrackets
\[
\cobra(\genJ^0_n)
=
-\half\rnorm \sum_{k=0}^{n-1}
\genJ^+_k \wedge \genJ^-_{n-1-k},
\eqsep
\cobra(\genJ^\pm_n)
=
\pm\rnorm \sum_{k=0}^{n-1}
\genJ^0_k \wedge \genJ^\pm_{n-1-k}.
\]

%%%%%%%%%%%%%%%%%%%%%%%%%%%%%%%%%%%%%%%%%%%%%%%%%%%%%%%%%%%%%%%%%%%%%%%%%%%%%%%%
\subsection{Contraction} \label{sec:contractionRationalCase}

Now we apply the contraction to the $\alg{so}(2,2)$ loop algebra with rational r-matrix, 
and we investigate several relevant aspects of it.

%%%%%%%%%%%%%%%%%%%%%%%%%%%%%%%%%%%%%%%%
\paragraph{Limit.}

As discussed above, the AdS algebra $\alg{so}(2,2)$ 
has two quadratic invariants $\casMM_1$ and $\casMM_2$. 
Therefore, there is a family of rational r-matrices
for this algebra
\[
\label{eq:rso22Rational}
r_{\alg{so}(2,2)}(u_{1,1},u_{2,1};u_{1,2},u_{2,2})
=
-\frac{\rnorm_1\casMM_1}{u_{1;1}-u_{1;2}}
-\frac{\rnorm_2\casMM_2}{u_{2;1}-u_{2;2}}
,
\]
with two constants $\rnorm_k$
and two pairs of spectral parameters $u_{k;j}$
which can all be chosen independently.
\unskip\footnote{Since the overall pre-factor of an r-matrix is usually of minor importance,
the r-matrix for this semi-simple algebra still depends 
on the ratio of $\rnorm_1$ and $\rnorm_2$ as one essential constant.}

Now let us see how the contraction limit of the $\alg{so}(2,2)$ r-matrix
leads to an r-matrix for $\alg{iso}(2,1)$.
We change variables from $\genM_{1,2}$ to $\genL,\genP$
according to \eqref{eq:contraction} to obtain
\[
r_{\alg{so}(2,2)}
%=\frac{-\rnorm_1\casMM_1-\rnorm_2\casMMp_2}{u_1-u_2}
=-\frac{
\rnorm_1 \mref^{-2}\epsilon^{-2}\casPP
}{u_{1;1}-u_{1;2}}
+\frac{
-\rnorm_2 \mref^{-2} \epsilon^{-2}\casPP
+2\rnorm_2 \mref^{-1}\epsilon^{-1}\casLP
-\rnorm_2\casLL
}{u_{2;1}-u_{2;2}}.
\]
We observe that this expression is divergent in the contraction limit $\epsilon \rightarrow 0$. 
In order for it to have a finite limit,
we need to transform the coefficients $\rnorm_{1,2}$ and spectral parameters $u_{1,2;j}$ accordingly 
such that $\rnorm_1+\rnorm_2=\Order(\epsilon^2)$ and $\rnorm_2=\Order(\epsilon)$
as well as $u_{1;j}-u_{2;j}=\Order(\epsilon)$.
This is easily achieved by the combinations
\[\label{eq:contractionCoefs}
\rnorm_{1,2}= \pm \rnorm\epsilon \mref + \half\rnorm'\epsilon^2 \mref^2+\Order(\epsilon^3),
\eqsep
u_{1,2;j}= u_j \pm \half\epsilon\mref v_j+\Order(\epsilon^2).
\]
Then we find in the contraction limit $\epsilon\to 0$
%%
%\[
%r_{\alg{so}(2,2)}(u_1,u_2)
%=\frac{
%-\rnorm'\casPP
%-2(\rnorm - \rnorm'\half\epsilon\mref) \casLP
%+2(\rnorm \half\epsilon\mref - \rnorm'\quarter\epsilon^2\mref^2)\casLL
%}{u_1-u_2}
%+\Order(\epsilon^2)
%\]
%%
%
\[
\label{eq:rationalContracted}
r_{\alg{so}(2,2)}
\to
r_{\alg{iso}(2,1)}
(u_1,v_1;u_2,v_2)
=
-\frac{2\rnorm\casLP}{u_1-u_2}
-\frac{\rnorm'\casPP}{u_1-u_2}
+\frac{\rnorm(v_1-v_2)\casPP}{(u_1-u_2)^2}
.
\]
The resulting expression up to the last term is of the generic rational form \eqref{eq:GenericRationalrMat}
for an algebra with two independent quadratic invariants $\casLP$ and $\casPP$.
Indeed, the construction of the rational r-matrix merely relies on a quadratic invariant form,
whose existence is guaranteed for semi-simple Lie algebras, 
but it may also exist in suitable non-simple Lie algebras.

Merely the last term proportional to the difference of the second spectral parameters $v_j$ is unusual. 
In fact, it is curious that the r-matrix admits a second spectral parameter $v_j$:
This spectral parameter appears in a very regular way 
thanks to its origin in the semi-simple loop algebra. 
For instance, the dependency on it can be reproduced 
easily by a suitable evaluation representation of the loop algebra
with two spectral parameters
\[
\rho_{u,v}(\genL^a_n) = u^n \rho(\genL^a) + nvu^{n-1} \rho(\genP^a),
\eqsep
\rho_{u,v}(\genP^a_n) = u^n \rho(\genP^a).
\]
The parametric r-matrix then follows as the evaluation representation 
of the loop algebra r-matrix
\[
r_{\alg{iso}(2,1)} 
= \sum\limits_{k=0}^\infty 
\brk!{
\rnorm c_{ab} \genL^a_k \otimes \genP^b_{-k-1}
+\rnorm c_{ab}\genP^a_k \otimes \genL^b_{-k-1}
+\rnorm' c_{ab}\genP^a_k \otimes \genP^b_{-k-1}
}.
\]
This expression agrees with the canonical form for rational r-matrices
with two independent quadratic invariants $\casLP$ and $\casPP$.
In fact, for many purposes, it would do to drop the second spectral parameter 
by setting it to zero, $v_j=0$. 
However, for reasons of generality, we shall keep it explicitly.

%%%%%%%%%%%%%%%%%%%%%%%%%%%%%%%%%%%%%%%%
\paragraph{Twists.}

Later on, we will need to apply a twist to the coalgebra structure. 
Let us introduce the twist deformation already here. 
A twist of some original r-matrix $r$ 
is obtained by adding some anti-symmetric combination of generators
\cite{Reshetikhin:1990ep}
\[
\tilde r=r+\rtwist\gen{X}\wedge \gen{Y}
\]
while ensuring that the classical Yang--Baxter equation remains valid.
Towards understanding the latter, we express the relevant combination 
of the classical Yang--Baxter equation in terms of the original cobracket $\cobra$ as
\[
\cybe{\tilde r}{\tilde r}
=
\cybe{r}{r}
-\rtwist\cobra(\gen{X})\wedge \gen{Y}
+\rtwist\cobra(\gen{Y})\wedge \gen{X}
+\rtwist^2(\gen{X}\wedge \gen{Y})\wedge\comm{\gen{X}}{\gen{Y}}
\stackrel{!}{=}0.
\]
In particular, if $\cobra(\gen{X})\wedge \gen{Y}$, $\cobra(\gen{Y})\wedge \gen{X}$
and $\comm{\gen{X}}{\gen{Y}}$ are all linear combinations of $\gen{X}$ and $\gen{Y}$,
the twist applies to arbitrary continuous values of the deformation parameter $\rtwist$.
The twist then correspondingly deforms the cobracket as follows
\[
\tilde\cobra(\gen{Z}) 
= \cobra(\gen{Z}) 
-\rtwist \gen{X}\wedge\comm{\gen{Y}}{\gen{Z}}
+\rtwist \gen{Y}\wedge\comm{\gen{X}}{\gen{Z}}.
\]

For example, we will later need to twist the rational $\alg{so}(2,2)$ r-matrix
as follows:
\[\label{eq:rattwist}
\tilde r_{\alg{so}(2,2)}=r_{\alg{so}(2,2)} + \rtwist (\genM_1^0+\genM_2^0)\wedge \genM_2^+.
\]
Defining the pair of level-0 generators 
$\gen{X}:=\genM_1^0+\genM_2^0$ and $\gen{Y}:=\genM_2^+$
we see that the twist is applicable for arbitrary $\rtwist$ because
\[
\delta(\gen{X})=\delta(\gen{Y})=0,
\eqsep
\comm{\gen{X}}{\gen{Y}}=\gen{Y}.
\]
%
%(on level zero only $\cobra(\genM_1^+)$ untwisted)
In the contraction limit with the scaling of parameter $\rtwist=\epsilon\mref\rtwist'$,
the twist yields the following r-matrix
\[
\label{eq:twistedrationaliso21}
\tilde r_{\alg{iso}(2,1)}
=
r_{\alg{iso}(2,1)}
+\rtwist' \genL^0\wedge \genP^+.
\]
This is in fact also a proper twist of $r_{\alg{iso}(2,1)}$ in \eqref{eq:rationalContracted}
because the generators $\gen{X}:=\genL^0$ and $\gen{Y}:=\genP^+$
obey the relations as described above.

%%%%%%%%%%%%%%%%%%%%%%%%%%%%%%%%%%%%%%%%
\paragraph{Momentum Representation.}

Now, we can evaluate the twisted rational r-matrix \eqref{eq:twistedrationaliso21} 
on the momentum representations which are the unitary irreps of the Poincaré algebra.
Upon subsequent reduction to be discussed in the following \secref{sec:reduction}, 
this representation of the r-matrix forms the diagonal part of the
tree level AdS/CFT S-matrix, as we will see in~\secref{sec:susy}.

The r-matrix acts on the tensor product of two states of the momentum representation \eqref{eq:momrepstate}:
\[
%\state{p_1, \phi_1; p_2,\phi_2} := 
\state{p_1, \phi_1}_{m_1, s_1} \otimes \state{p_2, \phi_2}_{m_2, s_2}
\]
as the differential operator
\[
\tilde r_{\alg{iso}(2,1)}(u_1,v_1;u_2,v_2)
\simeq
 \iunit A_{12}\frac{\partial}{\partial_{\phi_2}}
- \iunit A_{21}\frac{\partial}{\partial_{\phi_1}} 
+ \iunit B_{12}\frac{\partial}{\partial_{p_2}} 
- \iunit B_{21}\frac{\partial}{\partial_{p_1}}
+ C_{12}-C_{21},
\]
with the coefficient functions 
\begin{align}
A_{12} 
&=
\frac{\rnorm}{u_1 - u_2} \brk*{  e_{m_1}(p_1)  - \frac{p_1}{p_2} e_{m_2}(p_2) \cos(\phi_1 - \phi_2) }
- \rtwist' \mathinner{\eunit}^{\iunit \phi_1} p_1,
\\
B_{12}
&= 
\frac{\rnorm}{u_1 - u_2} p_1 e_{m_2}(p_2)\sin(\phi_1 - \phi_2),
\\
C_{12}
&=
\frac{\rnorm}{u_1 - u_2} s_2 \brk*{ e_{m_1}(p_1) - \frac{p_1 p_2}{e_{m_2}(p_2)+m_2} \cos(\phi_1 - \phi_2) } 
- \rtwist' s_2 \mathinner{\eunit}^{\iunit \phi_1} p_1
\\
&\alignrel
+ \brk[s]*{\frac{\half \rnorm'}{u_1 - u_2}-\frac{\half \rnorm(v_1-v_2)}{(u_1 - u_2)^2} }
\brk!{e_{m_1}(p_1)\. e_{m_2}(p_2) - p_1 p_2 \cos(\phi_1 - \phi_2)}
.
\end{align}
By construction, the above r-matrix differential operator acting on two-particle states 
obeys the classical Yang--Baxter equation.
We shall return to these expressions after having introduced
a reduction procedure for the algebra and for the states.

%%%%%%%%%%%%%%%%%%%%%%%%%%%%%%%%%%%%%%%%%%%%%%%%%%%%%%%%%%%%%%%%%%%%%%%%%%%%%%%%
%%%%%%%%%%%%%%%%%%%%%%%%%%%%%%%%%%%%%%%%%%%%%%%%%%%%%%%%%%%%%%%%%%%%%%%%%%%%%%%%
\section{Reduction}
\label{sec:reduction}

In this section we discuss a particular reduction 
of the $(3+3)$-dimensional Poincaré bi-algebra $\alg{sl}(2)\ltimes\Real^{2,1}$
to a $(1+1)$-dimensional bi-algebra $\alg{u}(1)\times\Real$.
This reduction was introduced in \cite{Beisert:2007ty} 
as a possibility to embed the classical algebra of the AdS/CFT worldsheet matrix 
in a larger but more conventional algebra.
It was motivated by the observation that 
particle momenta take values in a three-dimensional linear space
\cite{Beisert:2005tm,Arutyunov:2006ak,Hofman:2006xt},
but are further non-linearly constrained to one degree of freedom.
At the algebraic level, 
the reduction consists in restricting to a one-dimensional subalgebra of the Lorentz algebra $\alg{sl}(2)$
and dividing out a newly established two-dimensional ideal of the algebra of momentum generators $\Real^{2,1}$.
The two remaining generators represent the length of a momentum vector in three-dimensional space
as well as a Lorentz rotation about this vector.
The most interesting aspect of the reduction is that it incorporates a
dependency on the spectral parameter $u$ of the loop extension of the Poincaré algebra. 
In other words, the reduction is not homogeneous in the level of the loop algebra
because it relates generators of different loop levels. 
On the one hand, the resulting reduced algebra is abelian
and thus hardly exciting.
On the other hand, the reduction mechanism leaves the
coalgebra and r-matrix structures intact,
\unskip\footnote{The reduction mechanism is the same at the level of the coalgebra, 
but the roles of taking a subalgebra and dividing out an ideal are exchanged.} 
and we end up with a non-standard form for the r-matrix. 
This makes the overall construction 
extendable by $\alg{psu}(2|2)$ supersymmetry without further ado,
and has important applications within the AdS/CFT correspondence.
Also in practical terms it makes the reduction conveniently applicable to representations,
in particular to representations of the r-matrix.

%%%%%%%%%%%%%%%%%%%%%%%%%%%%%%%%%%%%%%%%
\paragraph{Subalgebra and Ideal.}

In \cite{Beisert:2007ty} the classical limit of the relevant representation 
for the AdS/CFT worldsheet matrix was discussed. 
This representation involves a $\Real^{2,1}$ momentum vector 
which was shown to be aligned along a common direction 
depending on the spectral parameter $u$.
This direction can be described by the two relations
\[
p^\pm = \frac{\eunit^{\pm\iunit \redang}\redpar}{u} p^0.
\]
Next to the spectral parameter $u$ it also depends on two constants $\redang,\redpar$.
These can be chosen arbitrarily; keeping them merely serves a more universal treatment. 
Furthermore, we can simply view the parameter $u$,
for the time being,
as just another adjustable parameter of the Poincaré representation,
as opposed to the spectral parameter for its loop algebra
which it ultimately becomes.

The above assignment of momentum components
is clearly not invariant under the full Lorentz algebra $\alg{sl}(2)$,
but there is a one-dimensional $\alg{u}(1)$ subalgebra which rotates about 
the direction of the above momentum.
This subalgebra is generated by the combination
\[ \label{eq:rationalgl1generator}
\genRL=\redpar^{-1}u\genL^0 - \half\eunit^{-\iunit \redang}\genL^+ - \half\eunit^{+\iunit \redang}\genL^- .
\]
As a one-dimensional subalgebra, the residual Lie bracket can only be trivial.
The algebra with the momentum generators is given by 
\begin{align}
\comm{\genRL}{\genP^0}
&= 
\half (\eunit^{-\iunit\redang} \genP^+ -\eunit^{+\iunit\redang} \genP^-)
=
\half (\eunit^{-\iunit\redang} \gen{I}^+ -\eunit^{+\iunit\redang} \gen{I}^-) ,
\\
\comm{\genRL}{\genP^\pm} 
&= 
\pm\redpar^{-1} u \genP^\pm
\mp \eunit^{\pm\iunit\redang} \genP^0 
=\pm\redpar^{-1} u \gen{I}^\pm.
\end{align}
Among the three resulting linear combinations of the momentum generators,
there are only two linearly independent directions 
\[
\gen{I}^\pm := \genP^\pm - \eunit^{\pm\iunit \redang}\redpar u^{-1} \genP^0.
\]
Consequently, these two span an ideal of $\alg{u}(1)\ltimes\Real^{2,1}$. 
Removing the ideal by the identifications
\unskip\footnote{Here we choose to identify $\genRP$ with $\genP^\pm$
such that the eigenvalue of $\genRP$ 
in a momentum representation is the magnitude of spatial momentum $p$. 
Alternatively, one might choose to identify $\genP^0=\genRH$
with a different generator $\genRH$ that measures the energy $p^0=e(p)$.
The identification between these two generators is $\redpar\genRH=u\genRP$.}
\[
\genP^0=\redpar^{-1}u\genRP,
\eqsep
\genP^\pm=\eunit^{\pm\iunit \redang}\genRP,
\]
where $\genRP$ is a single remaining momentum generator,
we reduce the algebra to $\alg{u}(1)\times\Real$
with the abelian relation 
\[
\comm{\genRL}{\genRP}=0.
\]
%

%%%%%%%%%%%%%%%%%%%%%%%%%%%%%%%%%%%%%%%%
\paragraph{r-Matrix.}

We can now address the reduction of the algebra to the coalgebra structures. 
Ordinarily, reducing an algebra to a subalgebra or removing an ideal 
is not compatible with the coalgebra structure 
because the latter typically leads out of the subalgebra
or does not respect the trivialisation of the ideal. 
Here, the combination of subalgebra and removal of ideal 
also ensures a proper coalgebra structure. 
First, removal of the ideal by the two relations
\[
\genP^\pm = \eunit^{\pm\iunit \redang}\redpar u^{-1} \genP^0
\]
can be understood as two linear relations on the dual generators 
which reduce it to a one-dimensional subcoalgebra.
Second, reduction to the subalgebra spanned by
\[
\genRL=\redpar^{-1}u\genL^0 - \half\eunit^{-\iunit \redang}\genL^+ - \half\eunit^{+\iunit \redang}\genL^-
\]
specifies a two-dimensional coideal of dual generators
which are annihilated by $\genRL$.

The compatibility of these reductions is more clearly seen in the r-matrix.
We first apply the removal of the ideal to the twisted r-matrix $\tilde r_{\alg{iso}(2,1)}$ in \eqref{eq:twistedrationaliso21} 
and find
\begin{align}
\tilde r_{\alg{iso}(2,1)}(u_1,u_2)
&\to
\frac{\rnorm}{u_1-u_2} \genRL\otimes \genRP
+\frac{\rnorm}{u_1-u_2}\genRP\otimes\genRL
\\
&\alignrel{}
+\brk[s]*{\frac{\rnorm'}{\redpar^2}-\frac{\rnorm(v_1-v_2)}{\redpar^2(u_1-u_2)}}\frac{u_1u_2-\redpar^2}{u_1-u_2} \genRP \otimes \genRP
\\
&\alignrel{}
+\brk*{-\frac{\rnorm}{\redpar}+\rtwist' \mathinner{\eunit}^{\iunit \redang}}\genL^0\wedge\genRP.
\end{align}
We note that almost all occurrences of the individual generators $\genL^a$ 
have been combined to the generator $\genRL$. 
Merely in the last term, there is a residual dependence 
on the generator $\genL^0$ 
which does not belong to the $\alg{u}(1)$ subalgebra.
This term can be removed by fixing the twist parameter
to
\unskip\footnote{The reduction presented in \cite{Beisert:2007ty}
used a slightly different twist of the original algebra
which respects the classical Yang--Baxter equation only upon reduction. 
Our twist resolves this issue by adding terms involving the ideal
so that the result is equivalent upon reduction.}
\[
\rtwist'=\frac{\rnorm}{\redpar}\eunit^{-\iunit \redang}.
\]
The twisted r-matrix thus reduces to
\[
r_{\alg{u}(1)\times\Real}
:=
\frac{\rnorm}{u_1-u_2} \genRL\otimes \genRP
+\frac{\rnorm}{u_1-u_2}\genRP\otimes\genRL
+\brk[s]*{\frac{\rnorm'}{\redpar^2}-\frac{\rnorm(v_1-v_2)}{\redpar^2(u_1-u_2)}}\frac{u_1u_2-\redpar^2}{u_1-u_2} \genRP \otimes \genRP,
\]
which now belongs completely to the reduced loop algebra $(\alg{u}(1)\times\Real)[u,u^{-1}]$.
Since the algebra is abelian, the classical Yang--Baxter equation is trivially
satisfied for any r-matrix.
 
Even though the classical Yang--Baxter equation is trivial in this case,
it is worth pointing out that it also holds by virtue of the reduction procedure alone.
We will work out this fact in \secref{sec:susy}
in order to conveniently demonstrate the applicability of the reduction procedure
to an extension by $\alg{psu}(2|2)$ supersymmetry.

In conclusion, we have constructed a consistent reduction 
of the Poincaré bi-algebra $\alg{sl}(2)\ltimes\Real^{2,1}$
to a quasi-triangular Lie bi-algebra $\alg{u}(1)\times\Real$.
In this construction, the role of the coalgebra twist remains somewhat unclear. 
Here we constructed it by the requirement that 
the r-matrix reduces consistently.
However, it would be interesting to understand whether there is some
(abstract) connection between the twist and the choice of the subalgebra or ideal.
Why is the twist necessary in the first place? 
Could one also start with a (suitable) twist, and construct the corresponding reduction?

%%%%%%%%%%%%%%%%%%%%%%%%%%%%%%%%%%%%%%%%
\paragraph{Loop Algebra Form.}

It makes sense to cast the reduction and the resulting relations in terms
of the loop algebra,
which we will occasionally need later on.
This reformulation makes explicit the non-standard nature of the resulting bi-algebra 
which is non-homogeneous in the loop levels.
Even more importantly, working in terms of the loop algebra makes the statements slightly more robust.
\unskip\footnote{For instance, the classical Yang--Baxter equation
can involve distributional terms in the dependence on the spectral parameter $u$
which have been discarded in the above treatment.}

The subalgebra $\alg{u}(1)[u,u^{-1}]$ of $\alg{sl}(2)[u,u^{-1}]$
is spanned by the generators
\[
\genRL_{n}:=\redpar^{-1}\genL^0_{n+1} - \half\eunit^{-\iunit \redang} \genL^+_n - \half\eunit^{+\iunit \redang}\genL^-_n .
\]
Similarly, the removal of the ideal is achieved by the relations
\[
\genP^0_n=\redpar^{-1}\genRP_{n+1},
\eqsep
\genP^\pm_n=\eunit^{\pm\iunit \redang}\genRP_{n}.
\]
Evidently, the resulting loop algebra is abelian,
$\comm{\genRL_n}{\genRP_m}=0$.
Finally, the reduced twisted r-matrix in loop algebra form reads
\begin{align} \label{eq:rmatrixgl1CRational}
\tilde r_{\alg{iso}(2,1)}&=
r_{\alg{iso}(2,1)}
+\frac{\rnorm}{\redpar}\eunit^{-\iunit \redang} \genL^0_0\wedge \genP^+_0
\to
r_{\alg{u}(1)\times\Real},
\\
r_{\alg{u}(1)\times\Real}
&:=
\rnorm\sum_{n=0}^\infty
\brk[s]*{-\genRL_n\otimes \genRP_{-n-1}-\genRP_n\otimes \genRL_{-n-1}}
\\
&\alignrel[:=]
+\rnorm'\sum_{n=0}^\infty
\brk[s]*{-\redpar^{-2}\genRP_{n+1} \otimes \genRP_{-n}+\genRP_n \otimes \genRP_{-n-1}}.
\end{align}

%%%%%%%%%%%%%%%%%%%%%%%%%%%%%%%%%%%%%%%%
\paragraph{Momentum Representation.}

Let us finally apply the reduction to representations.
In the momentum representation \eqref{eq:momrep},
the momentum generators are represented 
on a state $\state{u,p,\phi}_{m,s}$
by the eigenvalues 
\[
\genP^0_n\simeq u^n e_m(p),
\eqsep
\genP^\pm_n\simeq u^n\eunit^{\pm\iunit \phi}p.
\]
In order to trivialise the ideal, we need to impose the following relations
on the eigenvalues
\[
\eunit^{\pm\iunit \phi}p=\eunit^{\pm\iunit \redang}\redpar u^{-1} e_m(p).
\]
These are solved as functions of $u$ by 
\[
\phi=\redang,
\eqsep
p(u)=\frac{\redpar m}{\sqrt{u^2-\redpar^2}},
\eqsep
e(u)=\frac{mu}{\sqrt{u^2-\redpar^2}}.
\]

We thus define the reduced state as 
\[
\state{u,v}_{m,s}:=\state{u,v,p(u),\redang}_{m,s}.
\]
The remaining generators act on this state as eigenvalues 
\begin{align}
\genRL_n\state\state{u,v}_{m,s} 
&=
u^n \frac{sm}{p(u)}\state{u,v}_{m,s} 
+ v u^{n-1} \brk!{(n+1) u^2 \redpar^{-2} - n} p(u) \state\state{u,v}_{m,s} ,
\\
\genRP_n\state\state{u,v}_{m,s}
&=
u^n p(u) \state{u,v}_{m,s}.
\end{align}
Note that all the derivative operators in the representation of $\genRL$ 
have cancelled out on the particular state.

We can now act with the r-matrix on a tensor product state
$\state{u_1,v_1}_{m_1,s_1}\otimes\state{u_2,v_2}_{m_2,s_2}$.
Here we can either use the reduction of the r-matrix 
or apply the original twisted r-matrix to the reduced state,
in either case, the result is a pure scattering phase
\begin{align}
r_{\alg{u}(1)\times\Real}
&\simeq
\frac{\rnorm}{u_1-u_2} \frac{s_1m_1p_2}{p_1} 
+\frac{\rnorm}{u_1-u_2} \frac{s_2m_2p_1}{p_2} 
+\brk[s]*{\frac{\rnorm'}{\redpar^2}-\frac{\rnorm(v_1-v_2)}{\redpar^2(u_1-u_2)}}
\frac{u_1u_2-\redpar^2}{u_1-u_2} p_1p_2
\\
&=
\frac{\rnorm}{\redpar}\frac{s_1m_1p_2^2+s_2m_2p_1^2}{p_2e_1-p_1e_2} 
+\brk[s]*{\frac{\rnorm'}{\redpar}-\frac{\rnorm(v_1-v_2)p_1p_2}{\redpar^2(p_2e_1-p_1e_2)}}\frac{(e_1e_2-p_1p_2)p_1p_2}{p_2e_1-p_1e_2} .
\end{align}
As an aside, the second line displays the scattering phase
expressed purely in terms of energy and momentum variables.

%%%%%%%%%%%%%%%%%%%%%%%%%%%%%%%%%%%%%%%%%%%%%%%%%%%%%%%%%%%%%%%%%%%%%%%%%%%%%%%%
%%%%%%%%%%%%%%%%%%%%%%%%%%%%%%%%%%%%%%%%%%%%%%%%%%%%%%%%%%%%%%%%%%%%%%%%%%%%%%%%
\section{Trigonometric Case}
\label{sec:trig}

The rational classical r-matrix which we have discussed up to now 
has a trigonometric generalisation \cite{Beisert:2010kk}.
The corresponding quantum R-matrix was constructed in \cite{Beisert:2008tw}
and its quantum affine symmetry algebra was proposed in \cite{Beisert:2011wq}.
It provides the integrable structure for a quantum deformation 
of the one-dimensional Hubbard model, 
as well as for the worldsheet scattering matrix 
in quantum-deformed AdS/CFT \cite{Delduc:2013qra,Arutyunov:2013ega,Delduc:2014kha}.

In this section we will extend the discussion of the previous sections to the 
case of the trigonometric r-matrix. Whereas the contraction is performed precisely 
as in the rational case, the consistent reduction requires a different identification 
of the abelian subalgebra. 

%%%%%%%%%%%%%%%%%%%%%%%%%%%%%%%%%%%%%%%%%%%%%%%%%%%%%%%%%%%%%%%%%%%%%%%%%%%%%%%%
\subsection{Contraction} \label{sec:ContractionTrigCase}

We start with the trigonometric r-matrix of the algebra $\alg g  = \alg{sl}(2)$~\cite{Belavin:1982}. The standard form reads
with $z_i := \exp(u_i)$
%\iffalse
\begin{align}
\label{eq:rsl2trig}
r_{\alg{sl}(2)}(z_1,z_2) 
&= 
+ \half\rnorm\frac{z_1 + z_2}{z_1 - z_2} \genJ^0 \otimes \genJ^0 
- \frac{\half\rnorm z_1}{z_1-z_2} \genJ^+ \otimes \genJ^-
- \frac{\half\rnorm z_2}{z_1-z_2} \genJ^- \otimes \genJ^+ 
\\
&=
- \half \rnorm \frac{z_1 + z_2}{z_1 - z_2} \casJJ - \quarter \rnorm \genJ^+ \wedge \genJ^- .
\end{align}
 Similarly to the rational case, the trigonometric r-matrix can be expressed in terms of the loop algebra,
where $z_{1,2}$ (rather than $u_{1,2}$) serve as spectral parameters. 
The cobrackets take the form
\begin{align}
\cobra(\genJ^0_{n}) 
&= 
- \half \rnorm \sum_{k=1}^{n} \genJ^+_k \wedge \genJ^-_{n-k},
\\
\cobra(\genJ^+_{n}) 
&= 
+ \half \rnorm \genJ^0_0 \wedge \genJ^+_n
+ \rnorm  \sum_{k=1}^{n-1} \genJ^0_k \wedge \genJ^+_{n-k} , 
\\
\cobra(\genJ^-_{n}) 
&= 
- \half \rnorm \genJ^0_0 \wedge \genJ^-_n
- \rnorm  \sum_{k=1}^{n} \genJ^0_k \wedge \genJ^-_{n-k} .
\end{align}

%%%%%%%%%%%%%%%%%%%%%%%%%%%%%%%%%%%%%%%%
\paragraph{Contraction Limit.}

Similarly to the rational case, we construct the trigonometric r-matrix 
for the AdS algebra $\alg{so}(2,2)\simeq \alg{sl}(2)\times\alg{sl}(2)$ 
as a sum of two copies of the r-matrices above
\begin{align}
\label{eq:rso22trig}
r_{\alg{so}(2,2)}(z_{1;1},z_{2;1}; z_{1;2},z_{2;2}) 
&=
- \half \rnorm_1 \frac{z_{1;1} + z_{1;2}}{z_{1;1} - z_{1;2}} \casMM_1  - \quarter \rnorm_1 \genM_1^+ \wedge \genM_1^-
\\
&\alignrel 
- \half \rnorm_2 \frac{z_{2;1} + z_{2;2}}{z_{2;1} - z_{2;2}} \casMM_2 - \quarter \rnorm_2 \genM_2^+ \wedge \genM_2^-.
\end{align}

As the next step, we perform the contraction with the same prescription for the coefficients $\rnorm_{1,2}$ as we had in 
the rational case~\eqref{eq:contractionCoefs} and for the spectral parameters $z_{1,2;j} = z_j (1 \pm \half\epsilon \mref y_j) + \Order(\epsilon^2)$. 
This gives us a trigonometric r-matrix for the 3D Poincaré algebra
\begin{align} \label{eq:riso21trig}
r_{\alg{iso}(2,1)} (z_1, y_1; z_2, y_2)
&= 
- \rnorm \frac{z_1 + z_2}{z_1 - z_2} \casLP - \half \rnorm' \frac{z_1 + z_2}{z_1 - z_2} \casPP + \rnorm \frac{z_1 z_2 (y_1 - y_2)}{(z_1 - z_2)^2} \casPP
\\
&\alignrel
- \quarter \rnorm \genL^+ \wedge \genP^- + \quarter \rnorm \genL^- \wedge \genP^+ - \quarter \rnorm' \genP^+ \wedge \genP^-. 
\end{align}

This r-matrix can be obtained as the evaluation representation of the loop algebra r-matrix
\begin{align}
r_{\alg{iso}(2,1)} 
&= 
\sum_{k=0}^\infty  \brk*{
\rnorm c_{ab} \genL^a_{k} \otimes \genP^b_{-k} 
+ \rnorm c_{ab} \genP^a_{k} \otimes \genL^b_{-k} 
+ \rnorm' c_{ab} \genP^a_{k} \otimes \genP^b_{-k} }
\\
&\alignrel
- \rnorm \casLP - \half \rnorm' \casPP 
- \quarter \rnorm \genL^+_0 \wedge \genP^-_0 + 
\quarter \rnorm \genL^-_0 \wedge \genP^+_0 - 
\quarter \rnorm' \genP^+_0 \wedge \genP^-_0,
\end{align}
with the evaluation representation being defined as
\[
\rho_{z,y}(\genL^a_n) = z^n \rho(\genL^a) + n y z^n \rho(\genP^a),
\eqsep
\rho_{z,y}(\genP^a_n) = z^n \rho(\genP^a).
\]
%

%%%%%%%%%%%%%%%%%%%%%%%%%%%%%%%%%%%%%%%%
\paragraph{Twist.} 

Similarly to the rational case, there exists a twist with an arbitrary parameter $\rtwist$ 
that preserves the classical Yang--Baxter equation
\[
\tilde{r}_{\alg{so}(2,2)} = r_{\alg{so}(2,2)} + \half \rnorm_2 \genM_1^0 \wedge \genM_2^0  
+ \rtwist \brk{\genM_1^0 + \genM_2^0 } \wedge \genM_2^+.
\]
After the contraction one obtains with $\rtwist = -\epsilon\mref\rtwist'$
\[\label{eq:riso21trigtwist}
\tilde{r}_{\alg{iso}(2,1)} = r_{\alg{iso}(2,1)} + \half \rnorm \genL^0 \wedge \genP^0 
+  \rtwist'  \genL^0 \wedge \genP^+ .
\]
We will use this twist later on.

%%%%%%%%%%%%%%%%%%%%%%%%%%%%%%%%%%%%%%%%
\paragraph{Momentum Representation.}

Finally, we also compute the momentum representation of the twisted trigonometric 
r-matrix on two-particle states. This yields the following differential operator
\begin{align}
r_{\alg{iso}(2,1)}(z_1, y_1; z_2, y_2) \simeq  
\iunit A_{12} \frac{\partial}{\partial\phi_2}
-\iunit  A_{21} \frac{\partial}{\partial\phi_1}
+\iunit  B_{12} \frac{\partial}{\partial p_2} 
-\iunit  B_{21} \frac{\partial}{\partial p_1} 
+ C_{12} - C_{21},
\end{align}
where we now define the coefficient functions as
\begin{align}
A_{12}
&= 
\half \rnorm \frac{ z_1 + z_2 }{z_1 - z_2} \brk*{ e_{m_1}(p_1)  - e_{m_2}(p_2) \frac{p_1}{p_2}  \cos(\phi_1 - \phi_2) } 
\\
&\alignrel
- \ihalf \rnorm e_{m_2}(p_2) \frac{p_1}{p_2}  \sin(\phi_1 - \phi_2) 
- \rnorm e_{m_1}(p_1) 
- \rtwist' p_1 \eunit^{\iunit \phi_1}, 
\\
B_{12}
&= 
\half \rnorm \frac{z_1 + z_2 }{z_1 - z_2} p_1 e_{m_1}(p_2) \sin(\phi_1 - \phi_2) 
- \ihalf \rnorm p_1 e_{m_2}(p_2) \cos(\phi_1 -\phi_2),
\\
C_{12} 
&=
\half \rnorm \frac{z_1 + z_2}{z_1 - z_2} s_2 \brk*{ e_{m_1}(p_1) - 
\frac{p_1 p_2 \cos(\phi_1 - \phi_2)}{m_2 + e_{m_2}(p_2)}} 
- \ihalf \rnorm \frac{p_1 p_2 s_2}{m_2 + e_{m_2}(p_2)} \sin(\phi_1 - \phi_2)
\\
&\alignrel
+ \brk*{ \quarter \rnorm' \frac{z_1 + z_2}{z_1 - z_2} 
- \half \rnorm \frac{z_1 z_2 (y_1 - y_2)}{(z_1 - z_2)^2}} \brk!{e_{m_1}(p_1) e_{m_2}(p_2) - p_1 p_2 \cos(\phi_1 - \phi_2)} 
\\
&\alignrel
- \rfrac{\iunit}{4} \rnorm' p_1 p_2 \sin(\phi_1 - \phi_2) - \half \rnorm e_{m_1}(p_1) s_2 - \rtwist' p_1 s_2 \eunit^{\iunit \phi_1}.
\end{align}

%%%%%%%%%%%%%%%%%%%%%%%%%%%%%%%%%%%%%%%%%%%%%%%%%%%%%%%%%%%%%%%%%%%%%%%%%%%%%%%%
\subsection{Reduction}

Now, we introduce a different reduction procedure 
to reduce the trigonometric 3D Poincaré bi-algebra to the bi-algebra $\alg{u}(1)\times\Real$. 
The alternative reduction consists in a different choice for the one-dimensional subalgebra of $\alg{sl}(2)$ 
and subsequent ideal subalgebra. Whereas the difference is inessential for the resulting abelian algebras,
in the supersymmetric extension to be discussed in \secref{sec:susy},
the novel reduction will yield the trigonometric bi-algebra
for the classical limit of the quantum-deformed Hubbard model \cite{Beisert:2010kk}.

%%%%%%%%%%%%%%%%%%%%%%%%%%%%%%%%%%%%%%%%
\paragraph{Subalgebra and Ideal.} 

The new $\alg{u}(1)$ subalgebra generator now is chosen as
\[
\genL =  
\half \redtmod^{-1} z \genL^0 - \half \redtmod^{-1} \genL^0 
+ \ihalf \eunit^{\iunit \redang} \genL^- + \ihalf \eunit^{- \iunit \redang} z \genL^+,
\]
where $z$ will later be identified with the parameter of the evaluation representation.
We also introduce two global parameters $\redtmod$ and $\redang$ 
that take on the roles of $\redpar$ and $\redang$ in the rational case. 
This generator acts on the momentum generators according to
\begin{align}
\comm{\genL}{\genP^0} 
&=
\ihalf \eunit^{\iunit \redang} \genP^- 
- \ihalf \eunit^{-\iunit \redang} z \genP^+ 
=: \gen{I}^0,
\\
\comm{\genL}{\genP^+} 
&= 
\iunit \eunit^{\iunit \redang} \genP^0
- \half \redtmod^{-1} \genP^+ + \half \redtmod^{-1} z \genP^+ 
=:  \gen{I}^+,
\\
\comm{\genL}{\genP^-} 
&= -\iunit \eunit^{-\iunit \redang} z \genP^0
+ \half \redtmod^{-1} \genP^- 
- \half \redtmod^{-1} z \genP^-
\\
&= - \eunit^{ - 2 \iunit \redang} z \gen{I}^+ 
- \iunit \eunit^{- \iunit \redang} \redtmod^{-1} \gen{I}^0
+ \iunit \eunit^{- \iunit \redang} \redtmod^{-1} z \gen{I}^0.
\end{align}
We see that there are only two linear independent brackets out of three between $\genL$ and $\genP^{0, \pm}$. 
This implies that $\{\gen{I}^{0, +}_n\}_{n\in \Integer}$ span an ideal subalgebra for $\alg{u}(1)\ltimes\Real^{2,1}$. 
By dividing the ideal out we recover an abelian algebra 
spanned by $\genL$ and $\genP$ with the identifications:
\[ \label{eq:reductiontrigonometric}
\genP^0 \simeq \half \redtmod^{-1} z \genP - \half \redtmod^{-1} \genP, 
\eqsep
\genP^+ \simeq - \iunit \eunit^{ \iunit \redang} \genP, 
\eqsep
\genP^- \simeq - \iunit \eunit^{ -\iunit \redang} z \genP.
\]

The relations~\eqref{eq:reductiontrigonometric} can be implemented at the level of the momentum representation 
forming constraints on the eigenvalues of the momentum generators. 
However, it will prove useful to consider an alternative parametrisation, in which the constraints are expressed as
\[
\label{eq:relationsfrompaper2}
e_m(p) = (z-1)\frac{qm}{\redtmod},
\eqsep
p \mathinner{\eunit}^{\iunit \phi} = -2 \iunit \mathinner{\eunit}^{\iunit\redang}qm,
\eqsep
p \mathinner{\eunit}^{- \iunit \phi} = -2\iunit \mathinner{\eunit}^{-\iunit\redang}qzm.
\]
For the time being we consider the spectral parameter $z$ as a usual representation variable.
The parameter $q$ depends on $z$, but in order to resolve square roots in $q(z)$ we introduce a new parameter $x$
such that
\[
z=\frac{\iunit x}{(\redtmod'x-\iunit \redtmod)(\redtmod x+\iunit \redtmod')},
\eqsep
q=\frac{(\redtmod'x-\iunit \redtmod)(\redtmod x+\iunit \redtmod')}{\redtmod'(x^2-1)},
\eqsep
\redtmod'=\sqrt{1-\redtmod^2}.
\]
The on-shell condition is explicitly satisfied in the new variables
\[
e^2-p^2-m^2
=\frac{m^2}{\redtmod^2} \brk!{z^2q^2+2zq^2(2\redtmod^2-1)+q^2-\redtmod^2}
=0.
\]

%%%%%%%%%%%%%%%%%%%%%%%%%%%%%%%%%%%%%%%%
\paragraph{r-Matrix.}

As we did for the rational case, let us apply the reduction relations~\eqref{eq:reductiontrigonometric} to the
twisted trigonometric r-matrix~\ref{eq:riso21trigtwist}:
\begin{align} 
r_{\alg{iso}(2,1)} 
&\rightarrow
\rnorm \frac{z_2}{z_1 - z_2} \genL \otimes \genP + \rnorm \frac{z_1 }{z_1 - z_2} \genP \otimes \genL
\\
&\alignrel
+  \rnorm' \frac{\rfrac{1}{8} \redtmod^{-2} (z_1 + z_2) (z_1 - 1) (z_2 - 1) + z_1 z_2}{z_1 - z_2} \genP \otimes \genP
\\
&\alignrel
- \rnorm \frac{z_1 z_2 (y_1 - y_2)}{z_1 - z_2} \frac{\half (z_1 + z_2) + \quarter \redtmod^{-2} (z_1 - 1) (z_2 - 1) }{z_1 - z_2} \genP \otimes \genP.
\end{align}
As before, we observe that all the terms which do not belong to the reduced subalgebra cancel thanks to the twist and 
we obtain a trigonometric classical r-matrix for $\alg{u}(1)\times\Real$.

%%%%%%%%%%%%%%%%%%%%%%%%%%%%%%%%%%%%%%%%
\paragraph{Loop Form.}

For slightly more rigorous treatment we can reformulate the reduction and the r-matrix in the loop algebra form by 
identifying $z$ with the spectral parameter.
The $\alg{u}(1)[z, z^{-1}]$ generators then read
\[
\genL_n =  
\half \redtmod^{-1} \genL^0_{n + 1} - \half \redtmod^{-1}  \genL^0_{n} 
+ \ihalf \eunit^{\iunit \redang} \genL^-_{n} + \ihalf \eunit^{- \iunit \redang} \genL^+_{n + 1},
\]
and the following identifications between $\Real^{2,1}[z, z^{-1}]$ generators remove the ideal
\[ 
\genP^0_n \simeq \half \redtmod^{-1} \genP_{n+1} - \half \redtmod^{-1} \genP_n, \eqsep
\genP^+_n \simeq - \iunit \eunit^{ \iunit \redang} \genP_n, \eqsep
\genP^-_n \simeq \iunit \eunit^{ -\iunit \redang} \genP_{n+1}.
\]

In the loop form, the resulting reduced trigonometric r-matrix reads
\begin{align} \label{eq:rmatrixgl1CTrigonometric}
r_{\alg{u}(1)\times\Real} 
&= 
- \rnorm \sum_{k=0}^\infty \brk[s]{\genL_k \otimes \genP_{-k} + \genP_{k+1} \otimes \genL_{-k-1} } 
\\
&\alignrel
- \quarter \rnorm' \redtmod^{-2} \sum_{k=0}^\infty \brk[s]{\genP_{k+1}\otimes \genP_{-k+1} + \genP_{k+1}\otimes \genP_{-k-1} + (4 \redtmod^2 - 2) \genP_{k+1}\otimes \genP_{-k}}
\\
&\alignrel
- \rfrac{1}{8} \rnorm' \redtmod^{-2} \brk{\genP_{0}\otimes \genP_{0} - \genP_{1}\otimes \genP_{1} + \genP_{1} \wedge \genP_{0} }.
\end{align}

%%%%%%%%%%%%%%%%%%%%%%%%%%%%%%%%%%%%%%%%
\paragraph{Rational Limit.} 

The trigonometric and rational reductions are actually related by the so-called rational limit.
We identify
\[ \label{eq:rationallimt}
z_i = \eunit^{\lambda u_i}, 
\eqsep 
y_i = \lambda v_i,
\eqsep 
\redtmod = \ihalf \lambda \redpar ,
\]
and consider the leading-order terms in $\lambda$ in all expressions as $\lambda \rightarrow 0$.
In the evaluation representation  the leading terms of the generator $\genL$ 
coincide with the rational generator~\eqref{eq:rationalgl1generator} (up to a factor of $\iunit$)
and the leading terms of $\gen{I}^{0,+}$ span the same ideal as in the rational case.
We can also verify that the obtained trigonometric twist is consistent with the 
rational one in the limiting sense. 
We apply the rational limit relations~\eqref{eq:rationallimt} to the r-matrix 
and only keep track of leading order terms in $\lambda$:
\begin{align}
r^{\text{trig}}_{\alg{iso}(2,1)}(z_1, y_1; z_2, y_2) 
&=
- \frac{\rnorm}{\lambda} \frac{2 \casLP}{u_1 - u_2} 
- \frac{\rnorm'}{\lambda} \frac{\casPP}{u_1 - u_2} 
+ \frac{\rnorm}{\lambda} \frac{(v_1 - v_2) \casPP }{(u_1 - u_2)^2} 
\\
&\alignrel
+ \frac{\rnorm}{\lambda} \frac{\eunit^{- \iunit \redang}}{\redpar} \genL^0 \wedge \genP^+ + \Order(\lambda^0)
\\
&=
\lambda^{-1} r^{\text{rat}}_{\alg{iso}(2,1)}(u_1, v_1; u_2, v_2) 
+\Order(\lambda^0).
\end{align}
The singular contribution in the limit is $\lambda \rightarrow 0$ the resulting r-matrix rescaled by $\lambda^{-1}$.
We observe that it exactly coincides with the rational twisted r-matrix~\eqref{eq:twistedrationaliso21}:
\[
r^{\text{rat}}_{\alg{iso}(2,1)}
(u_1,v_1;u_2,v_2)
=
-\frac{2\rnorm\casLP}{u_1-u_2}
-\frac{\rnorm'\casPP}{u_1-u_2}
+\frac{\rnorm(v_1-v_2)\casPP}{(u_1-u_2)^2}
+\frac{\rnorm}{\redpar}\eunit^{-\iunit \redang} \genL^0\wedge \genP^+.
\]

%%%%%%%%%%%%%%%%%%%%%%%%%%%%%%%%%%%%%%%%
\paragraph{Momentum Representation.} 

Relations~\eqref{eq:relationsfrompaper2} define a reduced state that depends on the variable $z$
\[
\state{z}_{m,s} := \state!{p(z), \phi(z)}_{m, s}.
\]
The reduced generators then act on this state according to
\begin{align}
\genL_n \state{z}_{m,s} 
&= 
\frac{s z^n}{2 q(z)} \state{z}_{m,s} 
+ \frac{2 m q(z) y z^n}{(2h)^2}  
\brk!{(n+1) z^2 - (2 \iunit \redtmod^2 + 1) (2n + 1) z + n} \state{z}_{m,s}, 
\\
\genP_n \state{z}_{m,s} &= 2m q(z) z^n \state{z}_{m,s}.
\end{align}
Altogether, the representation of the r-matrix reads (with $q_{1,2} = q(z_{1,2}))$
\begin{align}
r_{\alg{u}(1)\times\Real}(z_1, y_1; z_2, y_2) 
&\simeq \frac{\rnorm}{z_1 - z_2} \brk*{\frac{q_2}{q_1} z_2 m_2 s_1 
+ \frac{q_1}{q_2} z_1 m_1 s_2} 
\\
&\alignrel
+ \frac{\rnorm'}{z_1 - z_2} \brk*{z_1 z_2 
+ \rfrac{1}{8} \redtmod^{-2}  (z_1 + z_2) (z_1 - 1) (z_2 - 1) } q_1 q_2 
\\
&\alignrel
- \frac{\rnorm ( y_1 - y_2)}{(z_1 - z_2)^2} \brk*{ \half (z_1 + z_2)  
+ \quarter \redtmod^{-2}  (z_1 - 1) (z_2 - 1) }z_1 z_2 q_1 q_2.
\end{align}
%

%%%%%%%%%%%%%%%%%%%%%%%%%%%%%%%%%%%%%%%%%%%%%%%%%%%%%%%%%%%%%%%%%%%%%%%%%%%%%%%%
%%%%%%%%%%%%%%%%%%%%%%%%%%%%%%%%%%%%%%%%%%%%%%%%%%%%%%%%%%%%%%%%%%%%%%%%%%%%%%%%
\section{Supersymmetric Extension}
\label{sec:susy}

In this section we consider supersymmetric extensions of the algebras we discussed so far.
Namely, we extend the AdS algebra $\alg{so}(2,2)$ to a semi-simple superalgebra $\alg{d}(2,1;\epsilon)\times\alg{sl}(2)$,
\unskip\footnote{Throughout this article, we consider the real form of $\alg{d}(2,1;\epsilon)$
whose even part is $\alg{sl}(2)\times\alg{su}(2)\times\alg{su}(2)$.}
where $\alg{so}(2,2)=\alg{sl}(2)\times\alg{sl}(2)$ 
is realised as the combination of the external $\alg{sl}(2)$ and the $\alg{sl}(2)$ subalgebra inside $\alg{d}(2,1;\epsilon)$. 
On the other hand, the Poincaré algebra $\alg{iso}(2,1)$ is lifted to 
$\alg{sl}(2)\ltimes\alg{psu}(2|2)\ltimes\Real^{2,1}$ 
where the Lorentz and momentum generators of $\alg{iso}(2,1)=\alg{sl}(2)\ltimes\Real^{2,1}$
act as three derivations and three central charges, respectively, to the simple super-algebra $\alg{psu}(2|2)$.
The extension of the contraction and reduction is rather plain 
and we obtain a (deformed) loop $\alg{u}(2|2)$ algebra and classical r-matrices for it
that are relevant for AdS/CFT integrability and for the one-dimensional Hubbard 
model~\cite{Klose:2006zd, Torrielli:2007mc, Beisert:2007ty, Beisert:2010kk}.

In the following we shall introduce the above superalgebras and some relevant representations.
We will then discuss how they are related by contractions, generalise their reduction
and investigate r-matrices of rational and trigonometric form. 
Finally, we discuss how a second superalgebra $\alg{d}(2,1;\epsilon')$
can be used in place of a plain $\alg{sl}(2)$ algebra
in order to double the amount of supersymmetry.

%%%%%%%%%%%%%%%%%%%%%%%%%%%%%%%%%%%%%%%%%%%%%%%%%%%%%%%%%%%%%%%%%%%%%%%%%%%%%%%%
\subsection{Superalgebras}

Let us briefly introduce the relevant superalgebras, their Lie brackets and explain how the contraction works in the 
supersymmetric setting.

%%%%%%%%%%%%%%%%%%%%%%%%%%%%%%%%%%%%%%%%
\paragraph{Exceptional Lie Superalgebra $\alg{d}(2,1;\epsilon)$.} 

The exceptional Lie superalgebra $\alg{d}(2,1;\epsilon)$ is spanned by 
a set of $\alg{sl}(2)$ generators $\genJ_\algmid^{0,\pm}$,
two copies of $\alg{su}(2)$ generators $\genJ_\algleft^{0,\pm}$, $\genJ_\algright^{0,\pm}$ 
and eight supercharges $\set{\genQ^{j,l r}}{}_{j,l,r=\spinup,\spindown}$. 
The algebra relations for the $\alg{sl}(2)$ or $\alg{su}(2)$ subalgebras are described in \secref{sec:Algebra},
whereas the supercharges transform in spin-$\vfrac{1}{2}$ representations of $\alg{sl}(2)_{\algmid}$ and $\alg{su}(2)_{\algleft,\algright}$ 
in each index, respectively. 
In other words, the Lie brackets read
\begin{align}
\comm{\genJ_\algmid^a}{\genQ^{j, l r}} 
&= 
- \half c^{ab} (\slpauli_b)^j{}_k \genQ^{k, l r}, 
\\
\comm{\genJ_\algleft^a}{\genQ^{j, l r}} 
&= 
- \half c^{ab} (\slpauli_b)^l{}_m \genQ^{j, m r}, 
\\
\comm{\genJ_\algright^a}{\genQ^{j, l r}} 
&= 
- \half c^{ab} (\slpauli_b)^r{}_p \genQ^{j, l p}, 
\eqsep
\end{align}
where we introduce a set of Pauli matrices $\slpauli$ adjusted to 
the signature of $\alg{sl}(2)$
as well as the anti-symmetric tensor $\varepsilon$
\[
\slpauli_0 = \begin{pmatrix} 1 & 0 \\ 0 & -1 \end{pmatrix}, \eqsep
\slpauli_+ = \begin{pmatrix} 0 & \iunit\\ 0 & 0 \end{pmatrix}, \eqsep
\slpauli_- = \begin{pmatrix} 0 & 0\\ \iunit & 0 \end{pmatrix}, \eqsep
\varepsilon = \begin{pmatrix} 0 & +1 \\ -1 & 0 \end{pmatrix}.
\]
The anti-symmetric Lie bracket between the supercharges is
\[
\acomm{\genQ^{j, l r}}{\genQ^{k, m p}} 
= -s_\algmid (\slpauli_a \varepsilon)^{jk} \varepsilon^{l m} \varepsilon^{r p} \.\genJ_\algmid^a
- s_\algleft \varepsilon^{jk} (\slpauli_a \varepsilon)^{lm} \varepsilon^{r p} \.\genJ_\algleft^a
- s_\algright \varepsilon^{jk} \varepsilon^{l m} (\slpauli_a \varepsilon)^{rp} \.\genJ_\algright^a.
\]
It depends on three parameters $s_{\algmid,\algleft,\algright}$ subject to two considerations:
First, the Jacobi identity requires the linear constraint $s_\algmid + s_\algleft + s_\algright = 0$.
Second, the possibility to rescale the supercharges removes another degree of freedom. 
Thus, there is effectively only one independent parameter that we label as $\epsilon$. 
For further purposes we fix the parameters to 
\[
s_\algmid = \epsilon, 
\eqsep 
s_\algleft = 1 - \epsilon, 
\eqsep 
s_\algright = -1.
\]
The superalgebra has a quadratic invariant form given by
\[\label{eq:d21invariant}
\casJJ = - \epsilon \casJJ_\algmid - (1 - \epsilon) \casJJ_\algleft + \casJJ_\algright +\casQQ,
\]
where $\casJJ_{\algmid,\algleft,\algright}$ denote the invariant quadratic forms 
for the even subalgebras $\alg{sl}(2)_\algmid$ and $\alg{su}_{\algleft,\algright}(2)$, respectively,
and where we have defined $\casQQ$ as the following quadratic combination of the supercharges
\[
\casQQ := \half \varepsilon_{jk} \varepsilon_{lm} \varepsilon_{rp} \.\genQ^{j, lr}\otimes \genQ^{k, mp}.
\]

%%%%%%%%%%%%%%%%%%%%%%%%%%%%%%%%%%%%%%%%
\paragraph{AdS Supersymmetry.} 

The simple superalgebra $\alg{d}(2,1;\epsilon)$ can be supplemented by a factor of $\alg{sl}(2)$ to 
a supersymmetry algebra $\alg{d}(2,1;\epsilon)\times\alg{sl}(2)$ for $\AdS^{2,1}$ space.
In that case, we relabel the generators $\genJ_\algmid^{0,\pm}$ of the subalgebra $\alg{sl}(2)_\algmid$ as $\genM_1^{0,\pm}$
of the subalgebra $\alg{sl}(2)_1$.
The additional algebra $\alg{sl}(2)_2$ is spanned by the triplet of generators $\genM_2^{0,\pm}$
which have trivial algebra relations with $\alg{d}(2,1;\epsilon)$.

%%%%%%%%%%%%%%%%%%%%%%%%%%%%%%%%%%%%%%%%
\paragraph{Poincaré Supersymmetry.}

The other relevant superalgebra is maximally extended $\alg{psu}(2|2)$ algebra
which can also be expressed as $\alg{sl}(2)\ltimes\alg{psu}(2|2)\ltimes\Real^{2,1}$
\cite{Nahm:1977tg}.
We would like to obtain it as a contraction limit of the algebra above.
Therefore, we write the algebra starting from the above basis 
where the triplet of generators $\genJ_\algmid^a$
is replaced by the triplet of Lorentz generators $\genL^a$, 
and an additional triplet of momentum generators $\genP^a$
is introduced.
The action of the $\alg{sl}(2)$ and $\alg{su}(2)_{\algleft,\algright}$ generators on the supercharges 
is the same as in the algebra $\alg{d}(2,1;\epsilon)$
where $\genL^a$ replaces $\genJ_\algmid^a$.
The Lie bracket between the supercharges is
\[
\acomm{\genQ^{j, l r}}{\genQ^{k, m p}} 
= - \frac{1}{\mref}(\slpauli_a \varepsilon)^{jk} \varepsilon^{lm} \varepsilon^{rp} \.\genP^a  
- \varepsilon^{jk} (\slpauli_a \varepsilon)^{lm} \varepsilon^{rp} \.\genJ^a_\algleft 
+ \varepsilon^{jk} \varepsilon^{lm} (\slpauli_a \varepsilon)^{rp} \.\genJ^a_\algright.
\]
Here, we have introduced a reference mass scale $\mref$, 
which could be absorbed by rescaling the momentum generators.
Nevertheless, we would like to keep it explicit to manifestly make the momenta have the dimension of a mass. 
The algebra has two invariant quadratic forms 
\[\label{eq:superpoincareinvariant}
2\casLP - \mref(\casQQ - \casJJ_{\algleft} + \casJJ_{\algright}),
\eqsep
\casPP.
\]

%%%%%%%%%%%%%%%%%%%%%%%%%%%%%%%%%%%%%%%%
\paragraph{Contraction.} 

We start with the semi-simple AdS superalgebra $\alg{d}(2,1;\epsilon)\times\alg{sl}(2)$
with the AdS generators $\genM_1^{0,\pm}=\genJ_\algmid^{0,\pm}$
from $\alg{d}(2,1;\epsilon)$ and $\genM_2^{0,\pm}$ from $\alg{sl}(2)$.
One can easily check that the contraction \eqref{eq:contraction} prescribed by the identifications
\[
\genL^a = \genM^a_1+\genM^a_2, 
\eqsep 
\genP^a = \epsilon \mref \genM_1^a,
\]
leads to the algebra $\alg{sl}(2)\ltimes\alg{psu}(2|2)\ltimes\Real^{2,1}$ 
in the limit $\epsilon\to 0$. 
Note that the limit now involves both the identification of generators for the
contraction as well as the structure of the algebra $\alg{d}(2,1;\epsilon)$ itself.
For the even subalgebra all the relations hold precisely the same as before. 
For the odd subalgebra the contraction is manifest, since 
$\comm{\genM_2^a}{ \genQ^{j,lr}} = 0$ and the Lie brackets between supercharges 
coincide if we replace $\epsilon \genM_1^a \rightarrow \genP^a/\mref$ and then 
take the limit $\epsilon \rightarrow 0$.
Note that the limits of the quadratic invariants read
\[
\epsilon \mref^2\casJJ \to - \casPP ,
\eqsep
\epsilon^2 \mref^2\casMM_2 \to \casPP ,
\eqsep
\mref\casJJ+\epsilon\mref\casMM_2 \to - 2\casLP+\mref(\casQQ-\casJJ_\algleft + \casJJ_\algright).
\]

%%%%%%%%%%%%%%%%%%%%%%%%%%%%%%%%%%%%%%%%%%%%%%%%%%%%%%%%%%%%%%%%%%%%%%%%%%%%%%%%
\subsection{Irreducible Representations}

Now, we would like to discuss a particular type of irreducible representation for the algebras above
that appear in the context of AdS integrability \cite{Beisert:2005tm}.
Since these representations are intended for physics applications, 
we use unitarity considerations to select appropriate representation parameters.

%%%%%%%%%%%%%%%%%%%%%%%%%%%%%%%%%%%%%%%%
\paragraph{Poincaré Supersymmetry.} 

Asymptotic particles on the string worldsheet 
or spin states of the one-dimensional Hubbard model 
transform in a four-dimensional representation
of the even subalgebra $\alg{su}(2)_\algleft\times\alg{su}(2)_\algright$.
The representation space consists of two two-dimensional components `$\stateferm$' and `$\statebos$' as follows:
On the components `$\stateferm$', the $\alg{su}(2)_\algleft$ generators 
act in the fundamental representation $\repdoublet$ 
and the $\alg{su}(2)_\algright$ generators act in the trivial representation $\repsinglet$.
On the other component `$\statebos$', the roles of the two $\alg{su}(2)$'s are exchanged.
We denote the states by $\state{\statebos^{\spinup,\spindown}}$, $\state{\stateferm^{\spinup,\spindown}}$
and they transform canonically in the representations 
$(\repsinglet, \repdoublet)$ and $(\repdoublet, \repsinglet)$, respectively, 
under $\alg{su}(2)_\algleft\times\alg{su}(2)_\algright$.
Furthermore, these states must obey opposite bosonic and fermionic statistics.

For the Poincaré subalgebra $\alg{iso}(2,1)$ we assume 
a field representation in momentum space whose states $\state{p, \phi}_{m,s}$
are labelled as in~\eqref{eq:momrepstate}
by the two spatial momentum components $p$ and $\phi$ (in radial coordinates), 
the mass $m$ and the spin $s$.
Altogether, the basis states of the representation are given by
\[
\state{p, \phi; \statebos^{\spinup,\spindown}}_{m,s_{\statebos}},
\eqsep*
\state{p, \phi; \stateferm^{\spinup,\spindown}}_{m,s_{\stateferm}},
\]
where we allow the two components
to have different spins $s_\statebos$ and $s_\stateferm$
while the same mass $m$ applies to both. 
The even generators act as in \eqref{eq:momrep}.
The supersymmetry generators act on the components according to
\begin{align} \label{eq:maxsl22SuperChargeRep}
\genQ^{\spinup, l r} \state{p, \phi; \statebos^p }_{s}
&= 
\bfnorm\mref^{-1/2} \sqrt{e_m(p) + m} \varepsilon^{r p} \state{p, \phi; \stateferm^l }_{s}, 
\\
\genQ^{\spindown, l r} \state{p, \phi; \statebos^p }_{s}
&= 
\iunit \bfnorm\mref^{-1/2} \eunit^{- \iunit \phi} \sqrt{e_m(p) - m} \varepsilon^{r p} \state{p, \phi; \stateferm^l }_{s},
\\
\genQ^{\spinup, l r} \state{p, \phi; \stateferm^m }_{s} 
&= 
- \iunit \bfnorm^{-1}\mref^{-1/2} \eunit^{\iunit \phi} \sqrt{e_m(p) - m} \varepsilon^{l m} \state{p, \phi; \statebos^r }_{s},
\\
\genQ^{\spindown, l r} \state{p, \phi; \stateferm^m }_{s} 
&= 
\bfnorm^{-1}\mref^{-1/2} \sqrt{e_m(p) + m} \varepsilon^{l m} \state{p, \phi; \statebos^r }_{s},
\end{align}
where the constant $\bfnorm$ governs the relative normalisation of states $\statebos$ vs.\ $\stateferm$.
The algebra relations then imply the following mass and spin constraints
\unskip\footnote{In fact, the consistency conditions require $m^2 = \quarter \mref^2$. 
Depending on the concrete root chosen, the lowering supercharges annihilate either 
$\statebos$ or $\stateferm$ states for $p=0$ in~\eqref{eq:maxsl22SuperChargeRep},
and in addition the difference between spins in \eqref{eq:massAndSpinContsr} is flipped.} 
\[ \label{eq:massAndSpinContsr}
m = \half \mref, 
\eqsep 
s := s_\statebos = s_\stateferm - \rfrac{1}{2}.
\]
The invariant quadratic forms \eqref{eq:superpoincareinvariant} take the following eigenvalues 
on the momentum representation
\[ \label{eq:superpoincareinvariantrep}
2\casLP- \mref(\casQQ - \casJJ_{\algleft} + \casJJ_{\algright})\simeq
- \mref (s + \quarter), 
\eqsep 
\casPP\simeq
- \quarter \mref^2.
\]
% 

%%%%%%%%%%%%%%%%%%%%%%%%%%%%%%%%%%%%%%%%
\paragraph{Ultra-Short Representation of $\alg{d}(2,1;\epsilon)$.}

We would now like to identify an irreducible representation of $\alg{d}(2,1;\epsilon)$
that could lead to the above representation upon contraction. The representation theory of 
$\alg{d}(2,1;\epsilon)$ is given in \cite{VanderJeugt:1985} 
(see also \cite{OhlssonSax:2011ms, Eberhardt:2019niq} in the context of AdS$_3$/CFT$_2$).
In particular, the configuration of $\alg{su}(2)_\algleft\times\alg{su}(2)_\algright$ representations 
will serve as a convenient starting point.

Irreducible representations of the superalgebra $\alg{d}(2,1;\epsilon)$ 
are composed from irreps of the even subalgebra $\alg{sl}(2)_\algmid\times\alg{su}(2)_\algleft\times\alg{su}(2)_\algright$.
These could be principal series, highest-weight, lowest-weight or fixed spin representations.
The interaction with supercharges organises a collection of different irreps of $\alg{sl}(2)$ into a supermultiplet.
These typically have a large number of components, 
but for particular choices of the $\alg{sl}(2)$ representations, they may be substantially shorter.
For instance, if we take the representation of the compact subalgebras $\alg{su}(2)_\algleft$ and $\alg{su}(2)_\algright$
to be $\repsinglet$ or $\repdoublet$, we obtain the so-called ultra-short representation. 
Just as the above representation for the Poincaré supersymmetry algebra, 
this representation consists of two components, 
on which $\alg{su}(2)_\algleft\times\alg{su}(2)_\algright$ 
act in $(\repsinglet, \repdoublet)$ and $(\repdoublet, \repsinglet)$, respectively.
We denote the corresponding states as $\state{\statebos^{\spinup,\spindown}}$, $\state{\stateferm^{\spinup,\spindown}}$,
and they need to respect opposite bosonic and fermionic statistics.
The third, non-compact $\alg{sl}(2)_\algmid$ acts on the two components
in the principal series representation~\eqref{eq:principalSeriesIrrep} 
with the parameters $(\reppar_{\statebos,\stateferm},\repang_{\statebos,\stateferm})$.
It turns out that a consistent action of the supercharges on the ultra-short multiplet
implies the following constraints on the representation parameters
\[ \label{eq:fermVsBosConstraintD21a}
(\reppar_{\statebos}, \repang_{\statebos})=(\half\epsilon^{-1} - \half, \repang),
\eqsep
(\reppar_{\stateferm}, \repang_{\stateferm})=(\half\epsilon^{-1}, \repang+\half),
\]
which leaves just a single adjustable representation parameter $\repang$.

Here we assume that the representation space is spanned by the vectors 
\[\label{eq:d21states}
\state{k; \statebos^{\spinup,\spindown}}_{\repang},
\eqsep*
\state{k + \half; \stateferm^{\spinup,\spindown}}_{\repang}
\eqjoin*{\text{with}} 
k \in \Integer,
\]
where the label $k$ enumerates states of the principal series representation. 
Notice that we label the $\stateferm$ states by a half-integer index.
The even generators then act as in \eqref{eq:principalSeriesIrrep}
with a shift of $\stateferm$ labels by $\vfrac{1}{2}$,
\begin{align} \label{eq:bosRepD21a}
\genJ_\algmid^0\state{k;\statebos^r}_{\repang}
&=
(k+\repang)\state{k;\statebos^r}_{\repang},
\\
\genJ_\algmid^+\state{k;\statebos^r}_{\repang}
&=
\theta^\statebos_k(k+\repang+\half\epsilon^{-1})\state{k+1;\statebos^r}_{\repang},
\\
\genJ_\algmid^-\state{k;\statebos^r}_{\repang}
&=
(\theta^\statebos_{k-1})^{-1}(k+\repang-\half\epsilon^{-1})\state{k-1;\statebos^r}_{\repang},
\\
\genJ_\algmid^0\state{k+\half;\stateferm^l}_{\repang}
&=
(k+\repang+\half)\state{k+\half;\stateferm^l}_{\repang},
\\
\genJ_\algmid^+\state{k+\half;\stateferm^l}_{\repang}
&=
\theta^\stateferm_k(k+\repang+1+\half\epsilon^{-1})\state{k+\rfrac{3}{2};\stateferm^l}_{\repang},
\\
\genJ_\algmid^-\state{k+\half;\stateferm^l}_{\repang}
&=
(\theta^\stateferm_{k-1})^{-1}(k+\repang-\half\epsilon^{-1})\state{k-\half;\stateferm^l}_{\repang}.
\end{align}
The supercharges act on the components according to
\begin{align} \label{eq:superchargesRepD21a}
\genQ^{\spinup, l r} \state{k; \statebos^p}_{\repang} 
&= 
(k+\repang + \half \epsilon^{-1} ) 
\theta^\statebos_k \eta_k^{-1}\. \varepsilon^{rp} \state{k+\half; \stateferm^l}_{\repang},
\\
\genQ^{\spindown, l r} \state{k; \statebos^p}_{\repang} 
&= 
\iunit(k+\repang-\half\epsilon^{-1}) \eta_{k-1}^{-1}\. \varepsilon^{rp} \state{k-\half; \stateferm^l}_{\repang} ,
\\
\genQ^{\spinup, l r} \state{k+\half ; \stateferm^m}_{\repang} 
&=
-\iunit\epsilon \eta_k\. \varepsilon^{lm} \state{k+1; \statebos^r}_{\repang},
\\
\genQ^{\spindown, l r} \state{k+\half ; \stateferm^m}_{\repang} 
&=
\epsilon \eta_k (\theta^\statebos_k)^{-1}\. \varepsilon^{lm} \state{k; \statebos^r}_{\repang}.
\end{align}
Here, the normalisation coefficients $\eta_k$ must satisfy the difference equation
\[
\eta_{k+1} \theta^\stateferm_{k} = \eta_k \theta^\statebos_{k+1}
\]
and $\theta^{\statebos,\stateferm}_k$ are the gauge degrees of freedom of the principal series representations 
for the $\statebos$ and $\stateferm$ components, respectively.
The representation is characterised by the eigenvalues of the 
quadratic invariant form \eqref{eq:d21invariant} 
and of the group invariant
\[
\casJJ\simeq\quarter\epsilon^{-1}-\quarter,
\eqsep
\exp(2\pi\iunit\genJ_\algmid^0+2\pi\iunit \genJ_{\algleft}^0)\simeq \eunit^{2\pi\iunit\repang}.
\]

In what follows we will need the representation to be of the unitary semi-infinite type. 
This is achieved by fixing the presentation parameter and the gauge as follows
\[ 
\label{eq:AdSSuperIrrepGauge}
\repang = \half\epsilon^{-1},
\eqsep
\theta^\statebos_k = \sqrt{\frac{k+1}{k+\epsilon^{-1}}},
\eqsep
\theta^\stateferm_k = \sqrt{\frac{k+1}{k+\epsilon^{-1}+1}},
\eqsep 
\eta_k = \frac{\sqrt{k+1}}{\adsbfnorm \sqrt{\epsilon}}.
\]
The normalisation $\adsbfnorm$ needs to be a pure phase, $\abs{\adsbfnorm}=1$.
The lowest-weight representation closes on the states with half-integer label $k\geq 0$
and it is unitary for $\epsilon>0$.
The highest-weight representation closes on the states with label $k\leq -\half$
and it is unitary for $\epsilon<0$ or $\epsilon>1$.
%
\iffalse
\begin{align} 
\label{eq:superchargesRepD21aUni}
\genJ^0\state{k;\statebos^r}
&=
(k+\half\epsilon^{-1})\state{k;\statebos^r},
\\
\genJ^+\state{k;\statebos^r}
&=
\sqrt{k+1}\sqrt{k+\epsilon^{-1}}\state{k+1;\statebos^r},
\\
\genJ^-\state{k;\statebos^r}
&=
\sqrt{k}\sqrt{k+\epsilon^{-1}-1}\state{k-1;\statebos^r},
\\
\genJ^0\state{k+\half;\stateferm^l}
&=
(k+\half\epsilon^{-1}+\half)\state{k+\half;\stateferm^l},
\\
\genJ^+\state{k+\half;\stateferm^l}
&=
\sqrt{k + 1}\sqrt{k+\epsilon^{-1}+1}\state{k+\rfrac{3}{2};\stateferm^l},
\\
\genJ^-\state{k+\half;\stateferm^l}
&=
\sqrt{k}\sqrt{k+\epsilon^{-1}}\state{k-\half;\stateferm^l}.
\\
\genQ^{\spinup, l r} \state{k; \statebos^p} 
&= 
\adsbfnorm \sqrt{\epsilon}\sqrt{k+\epsilon^{-1}} \. \varepsilon^{rp} \state{k+\half; \stateferm^l},
\\
\genQ^{\spindown, l r} \state{k; \statebos^p} 
&= 
\iunit \adsbfnorm\sqrt{\epsilon} \sqrt{k} \. \varepsilon^{rp} \state{k-\half; \stateferm^l},
\\
\genQ^{\spinup, l r} \state{k+\half; \stateferm^m}
&=
-\iunit\adsbfnorm^{-1}\sqrt{\epsilon} \sqrt{k+1}\. \varepsilon^{lm} \state{k+1; \statebos^r},
\\
\genQ^{\spindown, l r} \state{k+\half; \stateferm^m}
&=
\adsbfnorm^{-1} \sqrt{\epsilon} \sqrt{k+\epsilon^{-1}} \. \varepsilon^{lm} \state{k; \statebos^r}.
\end{align}
\fi
%
An alternative pair of unitary semi-infinite representations 
is obtained by setting
\[
\label{eq:AdSSuperIrrepGaugeAlt}
\repang = -\half\epsilon^{-1},
\eqsep
\theta^\statebos_k = \sqrt{\frac{k-\epsilon^{-1}+1}{k}},
\eqsep
\theta^\stateferm_k = \sqrt{\frac{k-\epsilon^{-1}+1}{k+1}},
\eqsep 
\eta_k = \frac{\sqrt{k-\epsilon^{-1}+1}}{\adsbfnorm \sqrt{\epsilon}}.
\]
Here, the lowest-weight representation closes on states with label $k\geq \half$, 
and it is unitary for $\epsilon<0$ or $\epsilon>1$.
The highest-weight representation closes on the states with label $k\leq 0$
and it is unitary for $\epsilon>0$.
%
\iffalse
\begin{align}
\genJ^0\state{k;\statebos^r}
&=
(k-\half\epsilon^{-1})\state{k;\statebos^r},
\\
\genJ^+\state{k;\statebos^r}
&=
\sqrt{k}\sqrt{k-\epsilon^{-1}+1} \state{k+1;\statebos^r},
\\
\genJ^-\state{k;\statebos^r}
&=
\sqrt{k-1}\sqrt{k-\epsilon^{-1}}\state{k-1;\statebos^r},
\\
\genJ^0\state{k+\half;\stateferm^l}
&=
(k-\half\epsilon^{-1}+\half)\state{k+\half;\stateferm^l},
\\
\genJ^+\state{k+\half;\stateferm^l}
&=
\sqrt{k+1}\sqrt{k-\epsilon^{-1}+1} \state{k+\rfrac{3}{2};\stateferm^l},
\\
\genJ^-\state{k+\half;\stateferm^l}
&=
\sqrt{k}\sqrt{k-\epsilon^{-1}} \state{k-\half;\stateferm^l},
\\
\genQ^{\spinup, l r} \state{k; \statebos^p} 
&= 
\adsbfnorm \sqrt{\epsilon} \sqrt{k} \varepsilon^{rp} \state{k+\half; \stateferm^l},
\\
\genQ^{\spindown, l r} \state{k; \statebos^p} 
&= 
\iunit \adsbfnorm \sqrt{\epsilon} \sqrt{k-\epsilon^{-1}}  \varepsilon^{rp} \state{k-\half; \stateferm^l} ,
\\
\genQ^{\spinup, l r} \state{k+\half ; \stateferm^m} 
&=
-\iunit\adsbfnorm^{-1}\sqrt{\epsilon} \sqrt{k-\epsilon^{-1}+1} \varepsilon^{lm} \state{k+1; \statebos^r},
\\
\genQ^{\spindown, l r} \state{k+\half ; \stateferm^m} 
&=
\adsbfnorm^{-1}\sqrt{\epsilon} \sqrt{k} \varepsilon^{lm} \state{k; \statebos^r}.
\end{align}
\fi
%
These four semi-infinite representations mainly differ in the lowest-weight or highest-weight states
being either type $\statebos$ or $\stateferm$.

\paragraph{AdS Supersymmetry.} 

Now we supplement the $\alg{d}(2,1;\epsilon)$ algebra with another set of $\alg{sl}(2)$ generators $\genM_2^a$
to a supersymmetry algebra for $\AdS^{2,1}$. 
The additional non-compact $\alg{sl}(2)$ acts on the states
in a principal series representation \eqref{eq:principalSeriesIrrep},
and we thus add an additional integer label $k_2$ to the states \eqref{eq:d21states}.

In~\secref{sec:irreps} we identified the pair of principal series representations 
with the (normalisable) (tensor) fields on AdS space. 
In the supersymmetric case we have two types of on-shell fields, $\statebos$ and $\stateferm$.
Therefore, we would like to identify both states in~\eqref{eq:superAdSStates} 
with AdS fields of, possibly, different masses $\adsmass_{\statebos, \stateferm}$ 
and spins $s_{\statebos, \stateferm}$. 
In order to ensure a positive energy, as in \secref{sec:irreps},
we turn the states of the representation of $\alg{sl}(2)_2$ around
and use a highest-weight representation instead,
while the representation of $\alg{d}(2,1;\epsilon)$ remains of lowest-weight type.

As we have already seen in the previous paragraph, the two components must have different parameters 
$(\reppar_{\statebos,\stateferm}, \repang_{\statebos,\stateferm})$
for the representations of the $\alg{sl}(2)_1$ subalgebra~\eqref{eq:fermVsBosConstraintD21a}.
On the other hand, due to the overall algebra being a direct product, the commutator between supercharges and $\genM_2^{a}$ 
must be trivial. Thus, the parameters $(\reppar_2, \repang_2)$ of the principal series representation 
must be the same for $\statebos$ and $\stateferm$ states.

As pointed out in \secref{sec:irreps}, the mass and spin are encoded into 
$\reppar_1=\reppar_{\statebos,\stateferm}$ and $\reppar_2$ according to~\eqref{eq:spinAdS}
resulting in 
$\adsmass_{\statebos,\stateferm}=\reppar_{\statebos,\stateferm} +\reppar_2$ and
$s_{\statebos,\stateferm}=\reppar_{\statebos,\stateferm} -\reppar_2$. 
The above constraint~\eqref{eq:fermVsBosConstraintD21a} 
then relates the mass and spin parameters for $\statebos$ and $\stateferm$ states
according to
\[\label{eq:superadsconstraints}
(\adsmass_\statebos,s_\statebos)=(\epsilon^{-1} - 1 - s,s),
\eqsep
(\adsmass_\stateferm,s_\stateferm) = (\epsilon^{-1} - \half - s, s+\half),
\]
where the overall spin parameter $s$ 
is one remaining degree of freedom for the representation.
\unskip\footnote{We will see in \secref{sec:doublesusy} 
that doubling the amount of supersymmetry singles out $s=-\quarter$ 
as the most symmetric choice.}
Finally, we impose the lowest-weight relation
$\repang = \half\epsilon^{-1}$ for $\alg{d}(2,1;\epsilon)$ 
and the highest-weight relation $\repang_2 = - ( \reppar_2 + \half )$ for $\alg{sl}(2)_2$.
This constrains the labels $k_1$ and $k_2$ to be non-negative 
and the representation for both $\statebos$ and $\stateferm$ components 
is given by~\eqref{eq:AdSIrrep} with masses and spins 
$(\adsmass_{\statebos,\stateferm},s_{\statebos,\stateferm})$.
Unitarity imposes the bounds $2s<\epsilon^{-1}$ and $\epsilon>0$
on the algebra parameter $\epsilon$ and on the spin $s$.
The resulting states
\[ \label{eq:superAdSStates}
\state{k_1, k_2; \statebos^{\spinup,\spindown}}_s,
\eqsep*
\state{k_1 + \half, k_2; \stateferm^{\spinup,\spindown}}_s
\eqjoin*{\text{with}} 
k_1,k_2 \in \Integer^+_0.
\]
transform under the AdS representation \eqref{eq:AdSIrrep}
supplemented by the supercharge actions in \eqref{eq:superchargesRepD21a,eq:AdSSuperIrrepGauge}.

%%%%%%%%%%%%%%%%%%%%%%%%%%%%%%%%%%%%%%%%
\paragraph{Irrep Contraction.}

The contraction of the representation of $\alg{d}(2,1;\epsilon)\times\alg{sl}(2)$ discussed above 
follows the same lines as for the case of $\alg{so}(2,2)$ in~\secref{sec:irrepcontr} with some minor adjustments.

First of all, we note that the constraints \eqref{eq:superadsconstraints}
can be expressed due to the mass constraint $m=\half \mref$ as
\[
\adsmass_{\statebos,\stateferm} = \frac{2m}{\epsilon \mref} +\Order(\epsilon^0).
\]
This relation agrees with the relation \eqref{eq:AdSmass}
required for a proper contraction of the representation.
We then generalise the prescription introduced in \eqref{eq:contrPrescription} 
by supplementing the identifications for the $\stateferm$ states 
\[
\state{p, \phi; \stateferm^j}_{s} := \sum_k \eunit^{- \iunit (k_1 - k_2) \phi } \state{k_1 + \half, k_2; \stateferm^j}_{s},
\eqsep
k_{1,2} := \frac{e_m(p) - m}{\epsilon \mref} \pm k.
\]
Furthermore we identify the normalisations 
$\adsbfnorm$ in~\eqref{eq:AdSSuperIrrepGauge} and $\bfnorm$ in~\eqref{eq:maxsl22SuperChargeRep}
as $\adsbfnorm=\bfnorm$.

Contraction of the momentum and Lorentz generators precisely repeats the calculation given in \secref{sec:irrepcontr}. 
The last step is to consider the contraction of the supercharge generators. 
Let us perform the computation explicitly only for $\genQ^{\spinup, l r} \state{p, \phi; \statebos^p}_{s} $ 
as other cases are treated analogously:
\begin{align}
\gen{Q}^{\spinup, l r} \state{p,\phi; \statebos^p}_{s} 
&= 
\adsbfnorm \sum_k \mathinner{\eunit}^{- \iunit (k_1 - k_2) \phi } 
\sqrt{\epsilon}\sqrt{k_1 + \epsilon^{-1}} \varepsilon^{rp} \state{k_1 + \half, k_2; \stateferm^l}_{s} 
\\
&= \bfnorm\mref^{-1/2} \sqrt{e_m(p) + m} \varepsilon^{rp} \state{p, \phi; \stateferm^l}_{s},
\end{align}
which precisely reproduces the action of the supercharge 
on the $\statebos$ state in the conventions~\eqref{eq:maxsl22SuperChargeRep}.
One may easily check that the representation of other supercharges is contracted consistently as well.

%%%%%%%%%%%%%%%%%%%%%%%%%%%%%%%%%%%%%%%%%%%%%%%%%%%%%%%%%%%%%%%%%%%%%%%%%%%%%%%%
\subsection{Reduction}

In the full $\alg{sl}(2)\ltimes\alg{psu}(2|2)\ltimes\Real^{2,1}$ superalgebra 
we only apply the reduction procedure to the $\alg{sl}(2)\ltimes\Real^{2,1}$ subalgebra.
Thus, the supersymmetry generators as well as the left and right $\alg{su}(2)_{\algleft,\algright}$ generators
remain unchanged within the reduced subalgebra. The resulting set of generators corresponds to those
of $\alg{u}(2|2)$ superalgebra. However, the Lie brackets between the supercharges and the $\alg{u}(1)$ generator
are non-standard and depend on the reduction scheme. Let us compute them explicitly.

%%%%%%%%%%%%%%%%%%%%%%%%%%%%%%%%%%%%%%%%
\paragraph{Rational Case.} 

We straight-forwardly compute the Lie bracket between $\genL$ and $\genQ^{j, lr}$
using the identification~\eqref{eq:rationalgl1generator} 
\begin{align}
\comm{\genL}{\genQ^{j, lr}} 
&=
W_{\text{rat}}(u)^{j}{}_k \genQ^{k, lr},
\end{align}
where we introduce the traceless matrix $W_{\text{rat}}(u)$ 
\[
W_{\text{rat}}(u) 
= \frac{1}{2} \begin{pmatrix}
\redpar^{-1} u & \iunit \eunit^{\iunit \redang}
\\
\iunit \eunit^{-\iunit \redang} & - \redpar^{-1} u
\end{pmatrix}.
\]
This matrix also allows us to write the anti-symmetric Lie brackets between supercharges compactly
\[
\acomm{\genQ^{j, l r}}{\genQ^{k, m p}} 
= - \frac{2}{\mref} \brk!{W_{\text{rat}}(u)\varepsilon}{}^{jk} \varepsilon^{l m} \varepsilon^{r p} \.\genP 
- \varepsilon^{jk} (\slpauli_a\varepsilon)^{l m} \varepsilon^{r p} \.\genJ^a_\algleft 
+ \varepsilon^{jk} \varepsilon^{l m} (\slpauli_a\varepsilon)^{r p} \.\genJ^a_\algright.
\]
We observe that now the Lie brackets involve the spectral parameter $u$. Therefore, we cannot
interpret the resulting algebra as a finite-dimensional $\alg{u}(2|2)$. Rather, we treat the relations 
above as if they were computed in the evaluation representation of a loop algebra. Therefore, we 
call the algebra a deformation of the loop algebra $\alg{u}(2|2)[u, u^{-1}]$. In fact, we here merely 
reproduce the result obtained in~\cite{Beisert:2007ty} (upon some rescalings and shift of levels in the generators).

%%%%%%%%%%%%%%%%%%%%%%%%%%%%%%%%%%%%%%%%
\paragraph{Trigonometric Case.}

The trigonometric reduction, naturally, give the same form for the brackets with
the supercharges except for replacing the matrix $W_{\text{rat}}(u)$ with
\[
W_{\text{trig}}(z) 
= \frac{1}{2} \begin{pmatrix}
\half \redtmod^{-1} (z-1) & \eunit^{\iunit \redang} \\
z \eunit^{- \iunit \redang} & - \half \redtmod^{-1} (z-1) 
\end{pmatrix}.
\]
As before, the algebra relations depend on the spectral parameter $z$ and we interpret
the resulting algebra as a (different) deformation of the loop algebra $\alg{u}(2|2)[z, z^{-1}]$.
The commutation relations coincide with those found in~\cite{Beisert:2010kk}.

%%%%%%%%%%%%%%%%%%%%%%%%%%%%%%%%%%%%%%%%%%%%%%%%%%%%%%%%%%%%%%%%%%%%%%%%%%%%%%%%
\subsection{r-Matrix}

The supersymmetric extension of the r-matrices is rather straight-forward. Let us briefly discuss the additional 
structures appearing in this case. 

%%%%%%%%%%%%%%%%%%%%%%%%%%%%%%%%%%%%%%%%
\paragraph{AdS Supersymmetry.}

In order to construct the rational and trigonometric r-matrices for the superalgebra 
$\alg{d}(2,1;\epsilon)\times\alg{sl}(2)$, we simply add terms corresponding to the left and right 
$\alg{sl}(2)$ and the supercharges to the r-matrices introduced earlier in~\secref{sec:contractionRationalCase} 
and~\secref{sec:ContractionTrigCase}. 
Namely, the rational r-matrix reads
\[
r_{\alg{d}(2,1;\epsilon)\times\alg{sl}(2)}^{\text{rat}}
= \frac{\rnorm_1}{\epsilon} 
\frac{\casQQ - \epsilon \casMM_1 - (1-\epsilon)\casJJ_\algleft + \casJJ_\algright }{u_{1;1} - u_{1;2}}
-\rnorm_2 \frac{\casMM_2}{u_{2;1}-u_{2;2}},
\]
where the numerators are given by the invariant quadratic forms 
of $\alg{d}(2,1;\epsilon)$ in \eqref{eq:d21invariant} and of $\alg{sl}(2)_2$.

Similarly, in the trigonometric case we have
\begin{align}
r_{\alg{d}(2,1;\epsilon)\times\alg{sl}(2)}^{\text{trig}}
&=
\frac{\rnorm_1}{\epsilon \rnorm} \brk*{
 \epsilon \. r_{1}^{\text{trig}}(z_{1;1}, z_{1;2})
+ (1 - \epsilon) \. r_{\algleft}^{\text{trig}}(z_{1;1}, z_{1;2}) 
-   r_{\algright}^{\text{trig}}(z_{1;1}, z_{1;2})}
\\
&\alignrel
-\frac{\rnorm_1}{\epsilon \rnorm} 
r_{\genQ}^{\text{trig}}(z_{1;1}, z_{1;2})
+ \frac{\rnorm_2}{\rnorm} r_{2}^{\text{trig}}(z_{2;1}, z_{2;2})
\end{align}
Here, $r_{1,2,\algleft,\algright}^{\text{trig}}(z_1, z_2)$ denote
the trigonometric r-matrices of the subalgebras $\alg{sl}(2)_{1,2}$ and $\alg{su}(2)_{\algleft,\algright}$, respectively,
and (with some abuse of notation) $r_{\genQ}^{\text{trig}}(z_1, z_2)$
describes the supercharge contribution to the r-matrix given by
\[
r_{\genQ}^{\text{trig}}(z_1, z_2)
:=
 -\frac{\half\rnorm z_1}{z_1 - z_2} \varepsilon_{lm}\varepsilon_{rp} \.\genQ^{\spinup, l r} \otimes\genQ^{\spindown, m p} 
+ \frac{\half\rnorm z_2}{z_1 - z_2} \varepsilon_{lm}\varepsilon_{rp} \.\genQ^{\spindown, l r} \otimes\genQ^{\spinup, m p}.
\]
%

%%%%%%%%%%%%%%%%%%%%%%%%%%%%%%%%%%%%%%%%
\paragraph{Contraction.}

As we discussed at the beginning of this section, the supercharges and additional $\alg{su}(2)$'s 
do not interfere with the contraction procedure. 
Therefore, one can straight-forwardly exploit the contraction prescription defined in~\secref{sec:loops} 
and obtain corresponding r-matrices for the maximally extended $\alg{sl}(2|2)$. 
The rational r-matrix then takes the form
\[
r_{\alg{sl}(2)\ltimes\alg{psu}(2|2)\ltimes\Real^{2,1}}^{\text{rat}}
= 
- \rnorm \frac{2\casLP - \mref(\casQQ - \casJJ_{\algleft} + \casJJ_{\algright})}{u_1 - u_2}
-\frac{\rnorm'\casPP}{u_1-u_2}
+\frac{\rnorm(v_1-v_2)\casPP}{(u_1-u_2)^2},
\]
and the trigonometric one
%
%\begin{align}
%r_{\alg{sl}(2)\ltimes\alg{psu}(2|2)\ltimes\Real^{2,1}}^{\text{trig}}
%&= 
%- \rnorm \frac{z_1 + z_2}{z_1 - z_2} \casLP 
%- \quarter \rnorm \genL^+ \wedge \genP^-
% + \quarter \rnorm \genL^- \wedge \genP^+ 
%\\
%&\alignrel
%+ \half \rnorm \mref \frac{z_1 + z_2}{z_1 - z_2} \casJJ_\algleft 
%+ \quarter \rnorm \mref \genJ_\algleft^+ \wedge \genJ_\algleft^-
%- \half \rnorm \mref \frac{z_1 + z_2}{z_1 - z_2} \casJJ_\algright 
%- \quarter \rnorm \mref \genJ_\algright^+ \wedge \genJ_\algright^-
%\\
%&\alignrel
%- \half \rnorm \mref \frac{z_1+z_2}{z_1 - z_2} \casQQ
%- \quarter \rnorm \mref \varepsilon_{lm}\varepsilon_{rp}\. \genQ^{\spinup, l r} \wedge\genQ^{\spindown, m p} 
%\\
%&\alignrel
%- \half \rnorm' \frac{z_1 + z_2}{z_1 - z_2} \casPP 
%- \quarter \rnorm' \genP^+ \wedge \genP^-
%+ \rnorm \frac{z_1 z_2 (y_1 - y_2)}{(z_1 - z_2)^2} \casPP
%\\
%&=
%r_{\alg{iso}(2,1)}^{\text{trig}}
%+ \mref (
%r_{\genQ}^{\text{trig}}
%- r_{\algleft}^{\text{trig}}
%+ r_{\algright}^{\text{trig}}
%)
%. 
%\end{align}
%
\[
r_{\alg{sl}(2)\ltimes\alg{psu}(2|2)\ltimes\Real^{2,1}}^{\text{trig}}
=
r_{\alg{iso}(2,1)}^{\text{trig}}
- \mref \brk!{r_{\genQ}^{\text{trig}} - r_{\algleft}^{\text{trig}} + r_{\algright}^{\text{trig}}}. 
\]

%%%%%%%%%%%%%%%%%%%%%%%%%%%%%%%%%%%%%%%%
\paragraph{Reduction.} 

The possible twists of the r-matrices introduced earlier also work without any 
changes in the supersymmetric case. Since the reduction is not affected by the 
supercharges, no new structures appear in the reduced r-matrices. In the end 
one obtains the r-matrices of the deformed loop $\alg{u}(2|2)$ algebra that 
coincide with those obtained in~\cite{Beisert:2007ty,Beisert:2010kk}.

Here, we notice that unlike $\alg{u}(1)\times\Real$ from~\secref{sec:reduction},
the full supersymmetric reduced algebra $\alg{u}(2|2)$ is not abelian. 
Evidently, the r-matrices~\eqref{eq:rmatrixgl1CRational} and~\eqref{eq:rmatrixgl1CTrigonometric} 
for an abelian algebra satisfy the classical Yang--Baxter equation,
but it is non-trivial that their supersymmetric counterparts do so. 
The latter happens due to the consistency in choosing the ideal subalgebra 
of the reduction and the twist of the r-matrix as follows: 

Concretely, let us describe the reduction of the original algebra 
$\alg{g}=\alg{sl}(2)\ltimes\alg{psu}(2|2)\ltimes\Real^{2,1}$ 
as
\[
\alg{g}\to\alg{h}/\alg{i}:
\eqsep
\alg{g}=\alg{k}\oplus\alg{h},
\eqsep*
\liebr{\alg{h}}{\alg{h}}\subset\alg{h},
\eqsep
\alg{i}\subset\alg{h},
\eqsep*
\liebr{\alg{h}}{\alg{i}}\subset\alg{i},
\]
where $\alg{k}$ denotes the subspace of $\alg{g}$ 
which does not belong to the subalgebra $\alg{h}=\alg{u}(1)\ltimes\alg{psu}(2|2)\ltimes\Real^{2,1}$,
and $\alg{i}=\Real^{2}$ denotes the ideal of $\alg{h}$.
We know that the r-matrix reduces to elements of the reduced algebra
upon removing elements of the ideal. 
In other words, the reduced r-matrix deviates from the original r-matrix
by pairings of the ideal with arbitrary elements of the original algebra
\[
r\in \alg{h}\otimes\alg{h}
 + \alg{i}\otimes\alg{g}
 + \alg{g}\otimes\alg{i}.
\]
In particular, generators not belonging to the subalgebra can only be paired 
with elements of the ideal
\[
r\in (\alg{h}\otimes\alg{h})
 \oplus (\alg{i}\otimes\alg{k})
 \oplus (\alg{k}\otimes\alg{i}).
\]
We now decompose the terms $\cybe{r}{r}$ 
of the original classical Yang--Baxter equation to these subspaces as follows
\[
\cybe{r}{r}\in
\liebr{\alg{h}}{\alg{h}}\wedge\alg{h}\wedge\alg{h}
+
\liebr{\alg{h}}{\alg{i}}\wedge\alg{h}\wedge\alg{k}
+
\liebr{\alg{i}}{\alg{i}}\wedge\alg{k}\wedge\alg{k}
+
\liebr{\alg{h}}{\alg{k}}\wedge\alg{h}\wedge\alg{i}
+
\liebr{\alg{i}}{\alg{k}}\wedge\alg{k}\wedge\alg{i}
+
\liebr{\alg{k}}{\alg{k}}\wedge\alg{i}\wedge\alg{i}.
\]
The original classical Yang--Baxter equation $\cybe{r}{r}=0$ holds by assumption,
and it thus also holds upon factoring out the ideal 
by setting $\alg{i}=0$.
This eliminates all potential terms from the listed subspaces but the first one
because there is an explicit tensor factor from $\alg{i}$
or because the Lie bracket $\liebr{\alg{h}}{\alg{i}}$ belongs to $\alg{i}$.
The remaining terms can only belong to the subspace
$\liebr{\alg{h}}{\alg{h}}\wedge\alg{h}\wedge\alg{h}$
or equivalently 
$\liebr{\alg{h}/\alg{i}}{\alg{h}/\alg{i}}\wedge\alg{h}/\alg{i}\wedge\alg{h}/\alg{i}$.
In particular, the residual terms rely on Lie brackets from the subalgebra only, 
and thus they constitute the validity of the classical Yang--Baxter equation
for the reduced r-matrix.

%%%%%%%%%%%%%%%%%%%%%%%%%%%%%%%%%%%%%%%%
\paragraph{Representation.}

For completeness, let us write down the representation of the resulting rational and 
trigonometric r-matrices after the reduction. We would like to recover the same classical r-matrices as 
in~\cite{Beisert:2007ty,Beisert:2010kk}. Therefore, in the rational case we adopt the following change of variables 
that resolves the square roots in various relations 
\[
u(x):=\frac{\redpar}{2}\frac{x^2+1}{x}.
\]
Then the momentum and energy eigenvalues read
\[
p(x)= \mref \frac{x}{x^2-1},
\eqsep
e(x)= \frac{\mref}{2} \frac{x^2+1}{x^2-1}.
\]
For the explicit matching of all the coefficients we also fix the normalisation of the supercharge representation to be
\[ \label{eq:bosonVsFermionNormFix}
\bfnorm = \bfnorm(x) := \mref^{-1/2}\tilde{\gamma}(x)\sqrt{1-x^{-2}}.
\]
The unitarity now requires $\tilde{\gamma}(x) = \sqrt{\mref}/\sqrt{1-x^{-2}}$.
Notice that we naturally define the action of the r-matrix on the tensor product such that 
it obeys the fermionic statistics when a supercharge is interchanged with a fermionic state
\[
\brk{\gen{A} \otimes \gen{B}}\brk!{ \state{u_1, \mathrm{X}}_{m, s} \otimes  \state{u_2, \mathrm{Y}}_{m, s}} 
:= (-1)^{\delim||{\gen{B}} \delim||{\mathrm{X}}} \gen{A} \state{u_1, \mathrm{X}}_{m, s}  \otimes \gen{B} \state{u_2, \mathrm{Y}}_{m, s},
\]
where $\delim||{\ldots} = 0,1$ describes the grading of a generator or a state.
These identifications lead us exactly to the classical rational r-matrix from~\cite{Torrielli:2007mc, Beisert:2007ty}
(with the normalisation $\abs{\mref}=1$) 
plus an additional term proportional to the unit matrix. 
The coefficient of the latter is given by the scalar phase function
\[ \label{eq:phaseRational}
P_{12} =
\rnorm \frac{\mref}{2} \frac{p_2 (s_1 + \quarter)}{p_1 (u_1 - u_2)} + \rnorm \frac{\mref}{2} \frac{p_1 (s_2 + \quarter)}{p_2 (u_1 - u_2)} + 
\brk*{\rnorm' - \rnorm \frac{v_1 - v_2}{u_1 - u_2} } \frac{e_1 e_2 - p_1 p_2}{u_1 - u_2}.
\]
For a perfect matching with \cite{Beisert:2007ty}, we need a trivial phase $P_{12}=0$:
We find that this is obtained by 
fixing the two spin parameters $s_{1,2}$, 
the secondary spectral parameters $v_{1,2}$,
as well as the secondary r-matrix normalisation $\rnorm'$ 
as follows
\[ \label{eq:phasematching}
s_{1,2}=-\frac{1}{4},
\eqsep
v_{1,2}=0,
\eqsep
\rnorm'=0.
\]
These assignments are in perfect agreement with the
algebra representation proposed in \cite{Beisert:2007ty}. Notably, this particular choice of spin 
leads to the trivial representation of one of the quadratic invariants in~\eqref{eq:superpoincareinvariantrep}.

It is interesting to contemplate different assignments for the parameters $s_{1,2}$, $v_{1,2}$ and $\rnorm'$
or to add some other twist in $\genL \otimes \genP$,
all of which merely affect the phase function $P_{12}$. 
For instance, one could reproduce the phase factor obtained in~\cite{Arutyunov:2004vx} 
at the classical level~\cite{Beisert:2007ty}
\[
P_{12} = \half \rnorm \eta (u_2 - u_1) p_1 p_2
\]
with $\eta$ some (dimensionful) normalisation constant.
This phase was originally obtained by twisting the r-matrix with terms non-homogeneous in loop level
\[
\half \rnorm \eta \genP_0 \wedge \genP_1 \simeq P_{12},
\]
which obviously preserves the classical Yang--Baxter equation as the generators $\genP_0$ and $\genP_1$ are central. 
Curiously, the same effect at the representation level can be achieved by merely fixing the parameters as follows
\[
s_{1,2} = - \frac{1}{4} + \frac{p_{1,2}^2}{\mref} \eta (u_{1,2}^2 - \redpar^2) 
= \frac{\eta \mref \beta^2 - 1}{4},
\eqsep 
v_{1,2} = \eta \redpar^2 u_{1,2},
\eqsep
\rnorm' = 0.
\]
However, it is not clear whether such manipulations are permissible,
in particular once the affine extension is taken into consideration~\cite{BeisertIm2}. 
It is interesting to pursue this question further in order to better understand the phase of the r-matrix. 

Similarly, in the trigonometric case we also obtain the fundamental r-matrix 
displayed in~\cite{Beisert:2010kk} with an additional phase 
factor 
\begin{align}
P_{12} &= 
\rnorm \frac{\mref}{2} \frac{z_2 q_2 (s_1 + \quarter)}{ q_1 (z_1 - z_2)} 
+ \rnorm \frac{\mref}{2} \frac{z_1 q_1 (s_2 + \quarter)}{ q_2 (z_1 - z_2)}
+ \rnorm' \mref^2 \frac{z_1 z_2 q_1 q_2}{z_1 - z_2} 
+ \rnorm' \frac{e_1 e_2}{z_1 - z_2} \frac{z_1 + z_2}{2} 
\\
&\alignrel
- \rnorm \frac{y_1 - y_2}{z_1 - z_2} \frac{z_1 z_2 e_1 e_2 }{z_1 - z_2}
- \rnorm \mref^2 \frac{y_1 - y_2}{z_1 - z_2} \frac{z_1 z_2 q_1 q_2 }{z_1 - z_2} \frac{z_1 + z_2}{2}.
\end{align}
Again, by varying the parameters $s_{1,2}, y_{1,2}$ or adding a twist we can shift the phase of the r-matrix.

%%%%%%%%%%%%%%%%%%%%%%%%%%%%%%%%%%%%%%%%%%%%%%%%%%%%%%%%%%%%%%%%%%%%%%%%%%%%%%%%
\subsection{Double Supersymmetry}
\label{sec:doublesusy}

The symmetry relevant to worldsheet scattering in the AdS/CFT correspondence 
requires twice as many supercharges and is
attributed to the algebra \cite{Beisert:2004ry}
\[
\alg{u}(1)\ltimes\alg{psu}(2|2)^2\ltimes\Real. 
\]
This algebra can be obtained by means of the contraction and reduction discussed in this article 
from the extended $\AdS$ supersymmetry algebra
\[
\alg{d}(2,1;\epsilon_1)_1\times\alg{d}(2,1;\epsilon_2)_2,
\]
where the corresponding $\alg{so}(2,2)$ subalgebra is composed of the two $\alg{sl}(2)_{1,2}$ subalgebras
taken from each $\alg{d}(2,1;\epsilon_{1,2})_{1,2}$ factor. The consistent contraction limit requires
$\epsilon_1 \sim \epsilon_2\to 0$, thus we set
\[
\epsilon_1\mref_1=\epsilon_2\mref_2 + \Order(\epsilon_{1,2}^2).
\]
In this way one obtains an enhanced Poincaré superalgebra with two distinct mass scales $\mref_{1,2}$.
Note also that the quadratic invariants are now contracted according to
\[
\epsilon_1 \mref_1^2\casJJ_1 \to - \casPP ,
\eqsep
\epsilon_2 \mref_2^2\casJJ_2 \to - \casPP ,
\eqsep
\mref_1\casJJ_1-\mref_2\casJJ_2 \to - 2\casLP+\ldots.
\]

Next, we apply the representation theory that was developed earlier in this section. 
We have two copies of the exceptional algebra, thus, the representations space is spanned by 
a tensor product of two states from~\eqref{eq:d21states}. Therefore, each state is labelled by 
two (half) integers and four possible pairs of the letters $\statebos$ and $\stateferm$.
We also exchange the roles of the states $\stateferm$ and $\statebos$ in the second $\alg{d}(2,1;\epsilon_2)_2$ 
compared to $\alg{d}(2,1;\epsilon_1)_1$ for the reason explained shortly.
Recall, that we had to choose the principal series representation of the external $\alg{sl}(2)$ algebra 
to be of the highest-weight type in order to contract the representations.
Now, we redefine the representation of the $\alg{sl}(2)$ algebra within $\alg{d}(2,1;\epsilon_2)_2$ 
to be of highest-weight kind, which, according to~\eqref{eq:AdSSuperIrrepGaugeAlt}, leads to the
following constraints on the parameters of the representation 
\[
(\reppar_{2,\statebos}, \repang_{2,\statebos}) 
= 
(\half \epsilon_2^{-1}, \half \epsilon_2^{-1} - \half),
\eqsep
(\reppar_{2,\stateferm}, \repang_{2,\stateferm}) 
= 
(\half \epsilon_2^{-1} - \half, \half \epsilon_2^{-1} - 1).
\]
Note that we invert the sign of $k_2$ in~\eqref{eq:bosRepD21a} and shift it by $-1$
so that one of the charge relations takes the form
\[
\genM_2^0\state{k_2;\stateferm^r}_{\repang}
=
(-k_2 + \repang_{2,\stateferm})\state{k_2;\stateferm^r}_{\repang},
\]
and so on for other generators. Now, we have the highest-weight representation with 
$k_2$ bounded from below as $k_2 \ge -\half$ and unitarity holds if $\epsilon_2<0$ or $\epsilon_2>1$.
For the parameters of $\alg{d}(2,1;\epsilon_1)_1$ we have the analogous relations from~\eqref{eq:fermVsBosConstraintD21a} and 
the unitarity condition $\epsilon_1>0$.
This representation has fixed masses $\adsmass_{(\statebos,\stateferm)(\statebos,\stateferm)} 
= \repang_{1,(\statebos,\stateferm)} - \repang_{2,(\statebos,\stateferm)} - 1$ and spins
$s_{(\statebos,\stateferm)(\statebos,\stateferm)} = \repang_{1,(\statebos,\stateferm)}+\repang_{2,(\statebos,\stateferm)}$.

A relevant observation is that a proper contraction limit of the above ultra-short representations 
requires $\epsilon_2=-\epsilon_1+\Order(\epsilon_{1,2}^2)$
for the resulting spins to be finite.
This also implies equal reference masses, $\mref_1=-\mref_2$, for the limiting Poincaré superalgebra.
\unskip\footnote{A Poincaré superalgebra with unequal reference masses is perfectly conceivable,
however, it does not support doubly ultra-short irreps (of equal kinds).}
Incidentally, a proper combination of two ultra-short representation of the Poincaré superalgebra 
needs $\abs{\mref_1}=\abs{\mref_2}$ because the Poincaré mass $m$ is constrained to both $\mref_1$ and $\mref_2$.
Furthermore, for unitary representations, 
we need to assume that $\epsilon_1$ approaches $0$ from above, 
consequently, $\epsilon_2$ will approach $0$ from below,
which is consistent with the above assignments.

In the end, we find the following finite spin configuration 
for the full representation
\[
s_{\statebos\stateferm} 
= 
\half \xi - 1 + \Order(\epsilon_{1,2}^2),
\eqsep
s_{\statebos\statebos} 
= 
s_{\stateferm\stateferm} 
= 
\half \xi - \half  + \Order(\epsilon_{1,2}^2),
\eqsep
s_{\stateferm\statebos} 
= 
\half \xi  + \Order(\epsilon_{1,2}^2),
\]
where $\xi$ is a parameter that relates $\epsilon_1$ and $\epsilon_2$ at sub-leading orders 
according to $\epsilon_2=-\epsilon_1-\xi\epsilon_1^2+\Order(\epsilon_{1,2}^3)$.
We observe that although all the parameters of the representation are constrained by the algebra parameters $\epsilon_{1,2}$, 
the spin $s$ appears to be unconstrained in the limit because $\xi$ disappears from the algebra relations. 
This leads us to a rather curious situation: 
before taking the limit the spin is determined by the remaining algebra parameter $\xi$, 
whereas in the limiting Poincaré superalgebra the spin is rather a representation parameter $s$.
\unskip\footnote{In particular, there is only a discrete choice of
representations before the limit, whereas the representations in the limit appear to have a continuous parameter.
Consequently, two Poincaré superalgebra representations with different spins
should not be obtainable simultaneously as the contraction from the same AdS superalgebra.
We emphasise that this feature is perhaps curious but not self-contradictory.}
At this point, one would be tempted to assign $\xi = 1$ in order to achieve a symmetric distribution 
of spins 
\[ \label{eq:naturalspindoublesuperalgebra}
s_{\statebos\stateferm} 
= 
-\half,
\eqsep
s_{\statebos\statebos} 
= 
s_{\stateferm\stateferm} 
= 
0,
\eqsep
s_{\stateferm\statebos} 
= 
+\half,
\]
and a natural interpretation of $\statebos\statebos$, $\stateferm\stateferm$ being bosons
and $\stateferm\statebos$, $\statebos\stateferm$ being fermions.
In fact, such an assignment can be motivated by two observations: On the one hand, the physical 
phase of the classical r-matrix required $s_{1,\stateferm} = -s_{1,\statebos} = \quarter$ (see~\eqref{eq:phasematching}).
Since for the second exceptional algebra we exchanged the roles of $\statebos$ and $\stateferm$, we also 
have $s_{2,\statebos} = -s_{2,\stateferm} = \quarter$. Altogether, this consideration fixes spins to be
as in~\eqref{eq:naturalspindoublesuperalgebra}.
On the other hand, we may use one of the 6 equivalences of the $\alg{d}(2,1;\epsilon)$ algebra
to define $\epsilon_2$ in terms of $\epsilon_1$ as 
\[
\epsilon_2 = \frac{- \epsilon_1}{1 - \epsilon_1} = -\epsilon_1-\epsilon_1^2+\Order(\epsilon_1^3).
\]
This equally singles out the value $\xi=1$.
Furthermore, this equivalence exchanges the left and right $\alg{su}(2)_{\algleft}$ and $\alg{su}(2)_{\algright}$,
hence it motivates the exchange of the roles $\statebos$ and $\stateferm$ in $\alg{d}(2,1;\epsilon_2)_2$.

As a final remark, in~\cite{Beisert:2017xqx} the contraction of the q-deformed algebras 
$\envalg_{h_1}(\alg{d}(2,1;\epsilon_1))\times\envalg_{h_2}(\alg{d}(2,1;\epsilon_2)) $ is investigated.
The consistent algebra contraction imposes a relation between the deformation parameters $h_{1,2}$ and algebra
parameters $\epsilon_{1,2}$
\[
h_1 \epsilon_1 + h_2 \epsilon_2 = \Order(\epsilon_{1,2}^2), 
\]
Our brief analysis of the representation contraction seems to suggest that 
one has to set $h_1 = h_2$ and
fix the next to the leading order term to be $-\epsilon_1/h_1$ in the quantum case. 
Therefore, it is interesting to investigate the discussed representation
for the quantum algebras and understand all the interaction between deformation, algebra 
and representation parameters.

%%%%%%%%%%%%%%%%%%%%%%%%%%%%%%%%%%%%%%%%%%%%%%%%%%%%%%%%%%%%%%%%%%%%%%%%%%%%%%%%
%%%%%%%%%%%%%%%%%%%%%%%%%%%%%%%%%%%%%%%%%%%%%%%%%%%%%%%%%%%%%%%%%%%%%%%%%%%%%%%%
\section{Conclusions and Outlook}
\label{sec:conclusions}

In this article, we have constructed the classical algebra 
relevant to integrability of the one-dimensional Hubbard model
and worldsheet scattering in the AdS/CFT correspondence
by means of contracting the semi-simple loop superalgebra based on
$\alg{d}(2,1;\epsilon)\times\alg{sl}(2)$ to the loop algebra of the 3D Poincaré superalgebra 
$\alg{sl}(2)\ltimes\alg{psu}(2|2)\ltimes\Real^{2,1}$ and by subsequently reducing it
to a deformation of the $\alg{u}(2|2)$ loop algebra. 
We have explicitly constructed infinite-dimensional unitary irreducible representations 
compatible with the contraction and reduction 
that result in the physically relevant representations for the Hubbard model and AdS/CFT models. 
We have also determined the rational and trigonometric r-matrices for these algebras 
and deduced that they are consistent with the contraction and reduction 
so that they satisfy the classical Yang--Baxter equation at each step. 
Evaluating the obtained r-matrices on the physically relevant representation
produces the tree-level scattering matrix for the string worldsheet within the AdS/CFT correspondence as expected. 
This constitutes a deductive derivation of the classical r-matrix 
for the one-dimensional Hubbard model and for AdS/CFT worldsheet scattering.

The natural future goal is to lift our construction to the level of quantum algebra. 
The feasibility of the contraction of the finite quantum algebra based on
$\alg{d}(2,1;\epsilon)\times\alg{sl}(2)$ to the one based on 
$\alg{sl}(2)\ltimes\alg{psu}(2|2)\ltimes\Real^{2,1}$
has already been demonstrated \cite{Beisert:2017xqx}.
However, it remains to promote the construction to loop algebras and to the quantum level
and to establish a suitable prescription for the subsequent reduction. 
Based on that construction, one can aim to derive the universal R-matrix,
but this will conceivably constitute an elaborate challenge in complexity. 
Therefore, lifting the notions of contraction and reduction to quantum algebra
at the level of the concrete representation developed in this article
will be a very useful next step \cite{BeisertIm3}.

Another useful aspect of the classical integrability algebras 
which we have not discussed in this article is their affine extension.
In the standard cases, the additional symmetry induced by the derivation element of a Kac--Moody algebra 
implies the difference form of the classical r-matrix. 
As we have seen, here the difference form is lost due to the reduction procedure,
nonetheless, there exists a deformation of twist of the derivation
whose co-bracket effectively describes the deviation of the r-matrix from the difference form, 
see \cite{Gomez:2007zr,Beisert:2010kk}.
Furthermore this derivation may act as a deformed Lorentz boost 
for the three-dimensional momenta \cite{Beisert:2005tm,Arutyunov:2006ak}.
In \cite{Borsato:2017lpf} the boost generator at the quantum level is proposed to be 
a Lorentz generator of the quantum-deformed Poincaré superalgebra,
and therefore the boost generator has non-trivial co-bracket in the classical limit. 
Importantly, the derivation symmetry constrains the form of the phase.
We will return to this question in \cite{BeisertIm2}. 
Furthermore, the integrability algebras admit a novel second spectral parameter 
for each evaluation representation. 
So far such a second degree of freedom has not appeared in physical applications,
and thus it may be interesting to study its potential implications.

It is also interesting to interpret the symmetry algebras
discussed in this paper from the worldsheet point of view.
In \cite{Hofman:2006xt} the momentum of the 3D Poincaré superalgebra
is associated with the stretching of a piece of string.
However, the Lorentz generators do not form a symmetry in this setting.
Therefore, it is important to understand how the extended symmetry
might fit within the worldsheet excitation picture
and whether the procedure of contraction and reduction can be realised for strings.

Whereas we now have the semi-simple algebra as the starting point, 
for which there exist the standard construction of the rational and trigonometric r-matrix, 
consistency with the reduction requires a non-standard twist of the r-matrix. 
Moreover, as we have seen in~\secref{sec:reduction} and~\secref{sec:trig}, the reduction is not unique. 
It amounts to identifying a two-dimensional ideal within the momentum subalgebra,
which, in fact, can be done in other ways. 
The twist of the r-matrix and the choice of the ideal subalgebra turn out to be closely related. 
It would be interesting to consider generic reduction schemes in order to understand the twist of the r-matrix, 
the resulting deformed loop algebras and possible models that correspond to the symmetry algebras.

Finally, the full symmetry algebra of the AdS/CFT integrability is (a quantum extension of) $\alg{psu}(2,2|4)$. 
Therefore, it is important to understand how to extend the constructions 
developed in this article to the full symmetry algebra.

%%%%%%%%%%%%%%%%%%%%%%%%%%%%%%%%%%%%%%%%%%%%%%%%%%%%%%%%%%%%%%%%%%%%%%%%%%%%%%%%
\pdfbookmark[1]{Acknowledgements}{ack}
\section*{Acknowledgements}

%We thank \remark{\ldots{}who} for discussions related to the work.
The work of NB and EI is partially supported 
by the Swiss National Science Foundation through the NCCR SwissMAP.

%%%%%%%%%%%%%%%%%%%%%%%%%%%%%%%%%%%%%%%%%%%%%%%%%%%%%%%%%%%%%%%%%%%%%%%%%%%%%%%%
%%%%%%%%%%%%%%%%%%%%%%%%%%%%%%%%%%%%%%%%%%%%%%%%%%%%%%%%%%%%%%%%%%%%%%%%%%%%%%%%
\ifarxiv\else
\begin{bibtex}[\jobname]

@article{Drinfel'd:1985,
    author = "Drinfel'd, Vladimir Gershonovich",
    title = "Hopf algebras and the quantum Yang–Baxter equation",
    journal   = "Sov. Math. Dokl.",
    volume    = "32",
    pages     = "254-258",
    year      = "1985"
}
%    journal = "Dokl. Akad. Nauk SSSR",
%    volume = "283",
%    issue = "5",
%    pages = "1060--1064",
%    year = "1985"

@article{Drinfel'd:1988,
    author = "Drinfel'd, Vladimir Gershonovich",
    title = "Quantum groups",
    doi = "10.1007/BF01247086",
    journal = "J. Sov. Math.",
    volume = "41",
    issue = "2",
    pages = "898--915",
    year = "1988"
}

@article{Hubbard:1963,
    ISSN = {00804630},
    URL = {http://www.jstor.org/stable/2414761},
    author = {J. Hubbard},
    journal = {Proc. R. Soc. London A},
    number = {1365},
    pages = {238--257},
    publisher = {The Royal Society},
    title = {Electron Correlations in Narrow Energy Bands},
    urldate = {2022-09-21},
    volume = {276},
    year = {1963}
}

@Book{Essler:2005aa,
     author    = "Essler, F. H. L. and Frahm, H. and Göhmann, F. and Klümper, A. and Korepin, V. E.",
     title     = "The one-dimensional Hubbard model",
     address   = "Cambridge, UK",
     year      = "2005",
     pages     = "690",
     publisher = "Cambridge University Press"
}

@Article{Shastry:1986bb,
    author = "Shastry, B. Sriram",
    title     = "Exact Integrability of the One-Dimensional Hubbard Model",
    journal   = "Phys. Rev. Lett.",
    volume    = "56",
    year      = "1986",
    pages     = "2453-2455",
    doi = "10.1103/PhysRevLett.56.2453",
}

@article{Beisert:2006qh,
    author = "Beisert, Niklas",
    title = "{The Analytic Bethe Ansatz for a Chain with Centrally Extended su(2$/$2) Symmetry}",
    eprint = "nlin/0610017",
    archivePrefix = "arXiv",
    reportNumber = "AEI-2006-074, PUTP-2211",
    doi = "10.1088/1742-5468/2007/01/P01017",
    journal = "J. Stat. Mech.",
    volume = "0701",
    pages = "P01017",
    year = "2007"
}

@article{Beisert:2005tm,
    author = "Beisert, Niklas",
    title = "{The SU(2$/$2) dynamic S-matrix}",
    eprint = "hep-th/0511082",
    archivePrefix = "arXiv",
    reportNumber = "PUTP-2181, NSF-KITP-05-92",
    doi = "10.4310/ATMP.2008.v12.n5.a1",
    journal = "Adv. Theor. Math. Phys.",
    volume = "12",
    pages = "945--979",
    year = "2008"
}

@article{Beisert:2010jr,
    author = "Beisert, Niklas and others",
    title = "{Review of AdS/CFT Integrability: An Overview}",
    eprint = "1012.3982",
    archivePrefix = "arXiv",
    primaryClass = "hep-th",
    reportNumber = "AEI-2010-175, CERN-PH-TH-2010-306, HU-EP-10-87, HU-MATH-2010-22, KCL-MTH-10-10, UMTG-270, UUITP-41-10",
    doi = "10.1007/s11005-011-0529-2",
    journal = "Lett. Math. Phys.",
    volume = "99",
    pages = "3--32",
    year = "2012"
}

@article{Gomez:2006va,
    author = "Gómez, César and Hernández, Rafael",
    title = "{The Magnon kinematics of the AdS/CFT correspondence}",
    eprint = "hep-th/0608029",
    archivePrefix = "arXiv",
    reportNumber = "CERN-PH-TH-2006-140, IFT-UAM-CSIC-06-37",
    doi = "10.1088/1126-6708/2006/11/021",
    journal = "JHEP",
    volume = "11",
    pages = "021",
    year = "2006"
}

@article{Plefka:2006ze,
    author = "Plefka, Jan and Spill, Fabian and Torrielli, Alessandro",
    title = "{On the Hopf algebra structure of the AdS/CFT S-matrix}",
    eprint = "hep-th/0608038",
    archivePrefix = "arXiv",
    reportNumber = "HU-EP-06-22",
    doi = "10.1103/PhysRevD.74.066008",
    journal = "Phys. Rev. D",
    volume = "74",
    pages = "066008",
    year = "2006"
}

@article{Beisert:2006fmy,
    author = "Beisert, Niklas",
    editor = "Faddeev, L. and Henneaux, M. and Kashaev, R. and Volkov, A. and Lambert, F.",
    title = "{The S-matrix of AdS/CFT and Yangian symmetry}",
    eprint = "0704.0400",
    archivePrefix = "arXiv",
    primaryClass = "nlin.SI",
    reportNumber = "AEI-2007-019",
    doi = "10.22323/1.038.0002",
    journal = "PoS",
    volume = "SOLVAY",
    pages = "002",
    year = "2006"
}

@article{Dorey:2006dq,
    author = "Dorey, Nick",
    title = "{Magnon Bound States and the AdS/CFT Correspondence}",
    eprint = "hep-th/0604175",
    archivePrefix = "arXiv",
    doi = "10.1088/0305-4470/39/41/S18",
    journal = "J. Phys. A",
    volume = "39",
    pages = "13119--13128",
    year = "2006"
}

@article{Chen:2006gp,
    author = "Chen, Heng-Yu and Dorey, Nick and Okamura, Keisuke",
    title = "{The Asymptotic spectrum of the N = 4 super Yang-Mills spin chain}",
    eprint = "hep-th/0610295",
    archivePrefix = "arXiv",
    reportNumber = "DAMTP-06-64, UY-06-17",
    doi = "10.1088/1126-6708/2007/03/005",
    journal = "JHEP",
    volume = "03",
    pages = "005",
    year = "2007"
}

@article{Matsumoto:2014cka,
    author = "Matsumoto, Takuya and Molev, Alexander",
    title = "{Representations of centrally extended Lie superalgebra psl(2$/$2)}",
    eprint = "1405.3420",
    archivePrefix = "arXiv",
    primaryClass = "math.RT",
    doi = "10.1063/1.4896396",
    journal = "J. Math. Phys.",
    volume = "55",
    pages = "091704",
    year = "2014"
}

@article{Arutyunov:2008zt,
    author = "Arutyunov, Gleb and Frolov, Sergey",
    title = "{The S-matrix of String Bound States}",
    eprint = "0803.4323",
    archivePrefix = "arXiv",
    primaryClass = "hep-th",
    reportNumber = "ITP-UU-08-15, SPIN-08-14, TCDMATH-08-03",
    doi = "10.1016/j.nuclphysb.2008.06.005",
    journal = "Nucl. Phys. B",
    volume = "804",
    pages = "90--143",
    year = "2008"
}

@article{deLeeuw:2008dp,
    author = "de Leeuw, Marius",
    title = "{Bound States, Yangian Symmetry and Classical r-matrix for the AdS$_5$ $\times$ S$^5$ Superstring}",
    eprint = "0804.1047",
    archivePrefix = "arXiv",
    primaryClass = "hep-th",
    reportNumber = "ITP-UU-08-18, SPIN-08-17",
    doi = "10.1088/1126-6708/2008/06/085",
    journal = "JHEP",
    volume = "06",
    pages = "085",
    year = "2008"
}

@article{Arutyunov:2009mi,
    author = "Arutyunov, Gleb and de Leeuw, Marius and Torrielli, Alessandro",
    title = "{The Bound State S-Matrix for AdS$_5$ $\times$ S$^5$ Superstring}",
    eprint = "0902.0183",
    archivePrefix = "arXiv",
    primaryClass = "hep-th",
    reportNumber = "ITP-UU-09-06, SPIN-09-06",
    doi = "10.1016/j.nuclphysb.2009.03.024",
    journal = "Nucl. Phys. B",
    volume = "819",
    pages = "319--350",
    year = "2009"
}

@article{Janik:2006dc,
    author = "Janik, Romuald A.",
    title = "{The AdS$_5$ $\times$ S$^5$ superstring worldsheet S-matrix and crossing symmetry}",
    eprint = "hep-th/0603038",
    archivePrefix = "arXiv",
    doi = "10.1103/PhysRevD.73.086006",
    journal = "Phys. Rev. D",
    volume = "73",
    pages = "086006",
    year = "2006"
}

@article{Hernandez:2006tk,
    author = "Hernández, Rafael and López, Esperanza",
    title = "{Quantum corrections to the string Bethe ansatz}",
    eprint = "hep-th/0603204",
    archivePrefix = "arXiv",
    reportNumber = "CERN-PH-TH-2006-048, IFT-UAM-CSIC-06-14",
    doi = "10.1088/1126-6708/2006/07/004",
    journal = "JHEP",
    volume = "07",
    pages = "004",
    year = "2006"
}

@article{Arutyunov:2006iu,
    author = "Arutyunov, G. and Frolov, S.",
    title = "{On AdS$_5$ $\times$ S$^5$ String S-matrix}",
    eprint = "hep-th/0604043",
    archivePrefix = "arXiv",
    reportNumber = "ITP-UU-06-15, SPIN-06-13",
    doi = "10.1016/j.physletb.2006.06.064",
    journal = "Phys. Lett. B",
    volume = "639",
    pages = "378--382",
    year = "2006"
}

@article{Beisert:2006ib,
    author = "Beisert, Niklas and Hernández, Rafael and López, Esperanza",
    title = "{A Crossing-symmetric phase for AdS$_5$ $\times$ S$^5$ strings}",
    eprint = "hep-th/0609044",
    archivePrefix = "arXiv",
    reportNumber = "AEI-2006-068, CERN-PH-TH-2006-176, IFT-UAM-CSIC-06-44, PUTP-2208",
    doi = "10.1088/1126-6708/2006/11/070",
    journal = "JHEP",
    volume = "11",
    pages = "070",
    year = "2006"
}

@article{Beisert:2006ez,
    author = "Beisert, Niklas and Eden, Burkhard and Staudacher, Matthias",
    title = "{Transcendentality and Crossing}",
    eprint = "hep-th/0610251",
    archivePrefix = "arXiv",
    reportNumber = "AEI-2006-079, ITP-UU-06-44, SPIN-06-34",
    doi = "10.1088/1742-5468/2007/01/P01021",
    journal = "J. Stat. Mech.",
    volume = "0701",
    pages = "P01021",
    year = "2007"
}

@article{Dorey:2007xn,
    author = "Dorey, Nick and Hofman, Diego M. and Maldacena, Juan Martin",
    title = "{On the Singularities of the Magnon S-matrix}",
    eprint = "hep-th/0703104",
    archivePrefix = "arXiv",
    doi = "10.1103/PhysRevD.76.025011",
    journal = "Phys. Rev. D",
    volume = "76",
    pages = "025011",
    year = "2007"
}

@article{Spill:2008tp,
    author = "Spill, Fabian and Torrielli, Alessandro",
    title = "{On Drinfeld's second realization of the AdS/CFT su(2$/$2) Yangian}",
    eprint = "0803.3194",
    archivePrefix = "arXiv",
    primaryClass = "hep-th",
    reportNumber = "MIT-CTP-3935, IMPERIAL-TP-08-FS-01, HU-EP-08-05",
    doi = "10.1016/j.geomphys.2009.01.001",
    journal = "J. Geom. Phys.",
    volume = "59",
    pages = "489--502",
    year = "2009"
}

@article{Beisert:2014hya,
    author = "Beisert, Niklas and de Leeuw, Marius",
    title = "{The RTT realization for the deformed gl(2$/$2) Yangian}",
    eprint = "1401.7691",
    archivePrefix = "arXiv",
    primaryClass = "math-ph",
    doi = "10.1088/1751-8113/47/30/305201",
    journal = "J. Phys. A",
    volume = "47",
    pages = "305201",
    year = "2014"
}

@article{Beisert:2016qei,
    author = "Beisert, Niklas and de Leeuw, Marius and Hecht, Reimar",
    title = "{Maximally extended sl(2$/$2) as a quantum double}",
    eprint = "1602.04988",
    archivePrefix = "arXiv",
    primaryClass = "math-ph",
    doi = "10.1088/1751-8113/49/43/434005",
    journal = "J. Phys. A",
    volume = "49",
    number = "43",
    pages = "434005",
    year = "2016"
}

@article{Matsumoto:2022nrk,
    author = "Matsumoto, Takuya",
    title = "{Drinfeld realization of the centrally extended psl(2$/$2) Yangian algebra with the manifest coproducts}",
    eprint = "2208.11889",
    archivePrefix = "arXiv",
    primaryClass = "math.QA",
    month = "8",
    year = "2022"
}

@article{Klose:2006zd,
    author = "Klose, Thomas and McLoughlin, Tristan and Roiban, Radu and Zarembo, Konstantin",
    title = "{Worldsheet scattering in AdS$_5$ $\times$ S$^5$}",
    eprint = "hep-th/0611169",
    archivePrefix = "arXiv",
    reportNumber = "ITEP-TH-61-06, UUITP-15-06",
    doi = "10.1088/1126-6708/2007/03/094",
    journal = "JHEP",
    volume = "03",
    pages = "094",
    year = "2007"
}

@article{Torrielli:2007mc,
    author = "Torrielli, Alessandro",
    title = "{Classical r-matrix of the su(2$/$2) SYM spin-chain}",
    eprint = "hep-th/0701281",
    archivePrefix = "arXiv",
    reportNumber = "MIT-CTP-3809",
    doi = "10.1103/PhysRevD.75.105020",
    journal = "Phys. Rev. D",
    volume = "75",
    pages = "105020",
    year = "2007"
}

@article{Moriyama:2007jt,
    author = "Moriyama, Sanefumi and Torrielli, Alessandro",
    title = "{A Yangian double for the AdS/CFT classical r-matrix}",
    eprint = "0706.0884",
    archivePrefix = "arXiv",
    primaryClass = "hep-th",
    reportNumber = "MIT-CTP-3843",
    doi = "10.1088/1126-6708/2007/06/083",
    journal = "JHEP",
    volume = "06",
    pages = "083",
    year = "2007"
}

@article{Beisert:2007ty,
    author = "Beisert, Niklas and Spill, Fabian",
    title = "The Classical r-matrix of AdS/CFT and its Lie Bialgebra Structure",
    eprint = "0708.1762",
    archivePrefix = "arXiv",
    primaryClass = "hep-th",
    reportNumber = "AEI-2007-116, HU-EP-07-31",
    doi = "10.1007/s00220-008-0578-2",
    journal = "Commun. Math. Phys.",
    volume = "285",
    pages = "537--565",
    year = "2009"
}

@article{Matsumoto:2007rh,
    author = "Matsumoto, Takuya and Moriyama, Sanefumi and Torrielli, Alessandro",
    title = "{A Secret Symmetry of the AdS/CFT S-matrix}",
    eprint = "0708.1285",
    archivePrefix = "arXiv",
    primaryClass = "hep-th",
    reportNumber = "MIT-CTP-3853",
    doi = "10.1088/1126-6708/2007/09/099",
    journal = "JHEP",
    volume = "09",
    pages = "099",
    year = "2007"
}

@article{Beisert:2017xqx,
    author = "Beisert, Niklas and Hecht, Reimar and Hoare, Ben",
    title = "{Maximally extended sl(2$/$2), q-deformed d(2,1;$\epsilon$) and 3D kappa-Poincaré}",
    eprint = "1704.05093",
    archivePrefix = "arXiv",
    primaryClass = "math-ph",
    doi = "10.1088/1751-8121/aa7a2f",
    journal = "J. Phys. A",
    volume = "50",
    number = "31",
    pages = "314003",
    year = "2017"
}

@article{Matsumoto:2008ww,
    author = "Matsumoto, Takuya and Moriyama, Sanefumi",
    title = "{An Exceptional Algebraic Origin of the AdS/CFT Yangian Symmetry}",
    eprint = "0803.1212",
    archivePrefix = "arXiv",
    primaryClass = "hep-th",
    doi = "10.1088/1126-6708/2008/04/022",
    journal = "JHEP",
    volume = "04",
    pages = "022",
    year = "2008"
}

@Book{Chari:1994pz,
     author    = "Chari, V. and Pressley, A.",
     title     = "A guide to quantum groups",
     address   = "Cambridge, UK",
     year      = "1994",
     pages     = "651",
     publisher = "Cambridge University Press"
}

@article{Belavin:1982,
    author = "Drinfel'd, Vladimir Gershonovich and Belavin, Aleksandr Abramovich",
    title = "Solutions of the classical Yang-Baxter equation for simple Lie algebras",
    journal = "Func. Anal. Appl.",
    volume = "16",
    issue = "3",
    pages = "159--180",
    year = "1982",
    doi = "10.1007/BF01081585"
}

@Article{Reshetikhin:1990ep,
     author    = "Reshetikhin, N.",
     title     = "Multiparameter quantum groups and twisted quasitriangular
                  Hopf algebras",
     journal   = "Lett. Math. Phys.",
     volume    = "20",
     year      = "1990",
     pages     = "331-335",
     doi       = "10.1007/BF00626530",
}

@article{Arutyunov:2006ak,
    author = "Arutyunov, Gleb and Frolov, Sergey and Plefka, Jan and Zamaklar, Marija",
    title = "{The Off-shell Symmetry Algebra of the Light-cone AdS$_5$ $\times$ S$^5$ Superstring}",
    eprint = "hep-th/0609157",
    archivePrefix = "arXiv",
    reportNumber = "AEI-2006-071, HU-EP-06-31, ITP-UU-06-39, SPIN-06-33, TCDMATH-06-13",
    doi = "10.1088/1751-8113/40/13/018",
    journal = "J. Phys. A",
    volume = "40",
    pages = "3583--3606",
    year = "2007"
}

@article{Hofman:2006xt,
    author = "Hofman, Diego M. and Maldacena, Juan Martin",
    title = "{Giant Magnons}",
    eprint = "hep-th/0604135",
    archivePrefix = "arXiv",
    doi = "10.1088/0305-4470/39/41/S17",
    journal = "J. Phys. A",
    volume = "39",
    pages = "13095--13118",
    year = "2006"
}

@article{Beisert:2010kk,
    author = "Beisert, Niklas",
    title = "The Classical Trigonometric r-Matrix for the Quantum-Deformed Hubbard Chain",
    eprint = "1002.1097",
    archivePrefix = "arXiv",
    primaryClass = "math-ph",
    reportNumber = "AEI-2010-016",
    doi = "10.1088/1751-8113/44/26/265202",
    journal = "J. Phys. A",
    volume = "44",
    pages = "265202",
    year = "2011"
}

@article{Beisert:2008tw,
    author = "Beisert, Niklas and Koroteev, Peter",
    title = "{Quantum Deformations of the One-Dimensional Hubbard Model}",
    eprint = "0802.0777",
    archivePrefix = "arXiv",
    primaryClass = "hep-th",
    reportNumber = "AEI-2008-003, ITEP-TH-06-08",
    doi = "10.1088/1751-8113/41/25/255204",
    journal = "J. Phys. A",
    volume = "41",
    pages = "255204",
    year = "2008"
}

@article{Beisert:2011wq,
    author = "Beisert, Niklas and Galleas, Wellington and Matsumoto, Takuya",
    title = "{A Quantum Affine Algebra for the Deformed Hubbard Chain}",
    eprint = "1102.5700",
    archivePrefix = "arXiv",
    primaryClass = "math-ph",
    reportNumber = "AEI-2011-005",
    doi = "10.1088/1751-8113/45/36/365206",
    journal = "J. Phys. A",
    volume = "45",
    pages = "365206",
    year = "2012"
}

@article{Delduc:2013qra,
    author = "Delduc, Francois and Magro, Marc and Vicedo, Benoit",
    title = "{An integrable deformation of the AdS$_5$ $\times$ S$^5$ superstring action}",
    eprint = "1309.5850",
    archivePrefix = "arXiv",
    primaryClass = "hep-th",
    doi = "10.1103/PhysRevLett.112.051601",
    journal = "Phys. Rev. Lett.",
    volume = "112",
    number = "5",
    pages = "051601",
    year = "2014"
}

@article{Arutyunov:2013ega,
    author = "Arutyunov, Gleb and Borsato, Riccardo and Frolov, Sergey",
    title = "{S-matrix for strings on $\eta$-deformed AdS$_5$ $\times$ S$^5$}",
    eprint = "1312.3542",
    archivePrefix = "arXiv",
    primaryClass = "hep-th",
    reportNumber = "ITP-UU-13-31, SPIN-13-23, HU-MATHEMATIK-2013-24, TCD-MATH-13-16",
    doi = "10.1007/JHEP04(2014)002",
    journal = "JHEP",
    volume = "04",
    pages = "002",
    year = "2014"
}

@article{Delduc:2014kha,
    author = "Delduc, Francois and Magro, Marc and Vicedo, Benoit",
    title = "{Derivation of the action and symmetries of the q-deformed AdS$_5$ $\times$ S$^5$ superstring}",
    eprint = "1406.6286",
    archivePrefix = "arXiv",
    primaryClass = "hep-th",
    doi = "10.1007/JHEP10(2014)132",
    journal = "JHEP",
    volume = "10",
    pages = "132",
    year = "2014"
}

@article{Nahm:1977tg,
    author = "Nahm, W.",
    title = "{Supersymmetries and their Representations}",
    reportNumber = "CERN-TH-2341",
    doi = "10.1016/0550-3213(78)90218-3",
    journal = "Nucl. Phys. B",
    volume = "135",
    pages = "149",
    year = "1978"
}

@article{Eberhardt:2019niq,
    author = "Eberhardt, Lorenz and Gaberdiel, Matthias R.",
    title = "{Strings on AdS$_3$ $\times$ S$^3$ $\times$ S$^3$ $\times$ S$^1$}",
    eprint = "1904.01585",
    archivePrefix = "arXiv",
    primaryClass = "hep-th",
    doi = "10.1007/JHEP06(2019)035",
    journal = "JHEP",
    volume = "06",
    pages = "035",
    year = "2019"
}

@article{Arutyunov:2004vx,
    author = "Arutyunov, Gleb and Frolov, Sergey and Staudacher, Matthias",
    title = "{Bethe ansatz for quantum strings}",
    eprint = "hep-th/0406256",
    archivePrefix = "arXiv",
    reportNumber = "AEI-2004-046",
    doi = "10.1088/1126-6708/2004/10/016",
    journal = "JHEP",
    volume = "10",
    pages = "016",
    year = "2004"
}

@Article{BeisertIm2,
     author    = "Beisert, N. and Im, E.",
     title     = "Affine Classical Lie Bialgebras for AdS/CFT Integrability",
     note      = "in preparation"
}

@article{Beisert:2004ry,
    author = "Beisert, Niklas",
    title = "{The Dilatation operator of N = 4 super Yang-Mills theory and integrability}",
    eprint = "hep-th/0407277",
    archivePrefix = "arXiv",
    reportNumber = "AEI-2004-057",
    doi = "10.1016/j.physrep.2004.09.007",
    journal = "Phys. Rept.",
    volume = "405",
    pages = "1--202",
    year = "2004"
}

@Article{BeisertIm3,
     author    = "Beisert, N. and Im, E.",
     note      = "work in progress"
}

@article{Gomez:2007zr,
    author = "Gómez, César and Hernández, Rafael",
    title = "{Quantum deformed magnon kinematics}",
    eprint = "hep-th/0701200",
    archivePrefix = "arXiv",
    reportNumber = "IFT-UAM-CSIC-07-03",
    doi = "10.1088/1126-6708/2007/03/108",
    journal = "JHEP",
    volume = "03",
    pages = "108",
    year = "2007"
}

@article{Borsato:2017lpf,
    author = "Borsato, Riccardo and Torrielli, Alessandro",
    title = "{$q$-Poincaré supersymmetry in AdS$_5$/CFT$_4$}",
    eprint = "1706.10265",
    archivePrefix = "arXiv",
    primaryClass = "hep-th",
    reportNumber = "DMUS-MP-17-07, NORDITA-2017-066",
    doi = "10.1016/j.nuclphysb.2018.01.017",
    journal = "Nucl. Phys. B",
    volume = "928",
    pages = "321--355",
    year = "2018"
}

@article{VanderJeugt:1985,
    author = {Van der Jeugt,J. },
    title = {Irreducible representations of the exceptional Lie superalgebras D(2,1;$\alpha$)},
    journal = {J. Math. Phys.},
    volume = {26},
    number = {5},
    pages = {913-924},
    year = {1985},
    doi = {10.1063/1.526547}
}

@article{OhlssonSax:2011ms,
    author = "Ohlsson Sax, Olof and Stefanski, Jr., B.",
    title = "{Integrability, spin-chains and the AdS$_3$/CFT$_2$ correspondence}",
    eprint = "1106.2558",
    archivePrefix = "arXiv",
    primaryClass = "hep-th",
    reportNumber = "UUITP-17-11",
    doi = "10.1007/JHEP08(2011)029",
    journal = "JHEP",
    volume = "08",
    pages = "029",
    year = "2011"
}

\end{bibtex}
\fi

\bibliographystyle{nb}
\bibliography{\jobname}

\end{document}